\documentclass[twocolumn]{aastex63}

\received{--}
\revised{--}
\accepted{--}

\submitjournal{ApJ}

\shorttitle{Poynting flux of MHD modes in magnetic solar vortex tubes}
\shortauthors{Skirvin et al.}

\usepackage{graphicx}
\usepackage{amsmath}
\usepackage{bm}

\begin{document}

\title{Poynting flux of MHD modes in magnetic solar vortex tubes}

\correspondingauthor{Samuel Skirvin}
\email{s.skirvin@sheffield.ac.uk}
\author[0000-0002-3814-4232]{Samuel J. Skirvin}
\affiliation{Plasma Dynamics Group, School of Electrical and Electronic Engineering, University of Sheffield, Sheffield, S1 3JD, UK}

\author[0000-0002-0893-7346]{Viktor Fedun}
\affiliation{Plasma Dynamics Group, School of Electrical and Electronic Engineering, University of Sheffield, Sheffield, S1 3JD, UK}

\author[0000-0002-7830-7147]{Marcel Goossens}
\affiliation{Centre for Mathematical Plasma Astrophysics, Department of Mathematics, KU Leuven, Celestijnenlaan 200B bus 2400, B-3001 Leuven, Belgium}

\author[0000-0001-5414-0197]{Suzana S. A. Silva}
\affiliation{Plasma Dynamics Group, School of Electrical and Electronic Engineering, University of Sheffield, Sheffield, S1 3JD, UK}

\author[0000-0002-9546-2368]{Gary Verth}
\affiliation{Plasma Dynamics Group, School of Mathematical and Physical Sciences, University of Sheffield, Sheffield S3 7RH, UK}

\begin{abstract}
Magnetic flux tubes in the presence of background rotational flows, known as solar vortex tubes, are abundant throughout the solar atmosphere and may act as conduits for MHD waves to transport magnetic energy to the upper solar atmosphere. We aim to investigate the Poynting flux associated with these waves within solar vortex tubes. We model a solar vortex tube as a straight magnetic flux tube with a background azimuthal velocity component. The MHD wave solutions in the equilibrium configuration of a vortex tube are obtained using the SESAME code and we derive an expression for the vertical component of the Poynting flux, $S_z$, associated with MHD modes. In addition, we present 2D visualisations of the spatial structure of $S_z$ for different MHD modes under different background flow strengths. We show that $S_z$ increases in the presence of a background rotational flow when compared to a flux tube with no rotational flow. When the strength of the background flow is greater than $100$ times the strength of the perturbation, the $S_z$ associated with non-axisymmetric ($|m|>0$) modes increases by over $1000\%$ when compared to a magnetic flux tube in the absence of a background rotational flow. Furthermore, we present a fundamental property of solar vortices that they cannot solely produce an upwards Poynting flux in an untwisted tube, meaning that any observed $S_z$ in straight flux tubes must arise from perturbations, such as MHD waves.
\end{abstract}


\section{Introduction} \label{sec:intro}

The mechanisms which maintain the temperature of the solar corona remain elusive despite considerable progress over recent years in the understanding of both AC \citep{vanDoorsselaere2020SSRv} and DC \citep{Pontin2022} heating mechanisms. However, it is evident that the magnetic field plays a dominating role in the supply and transport of energy available for the heating of solar plasma. 

One method by which energy can be transported from the photosphere to the corona is through magnetohydrodynamic (MHD) waves generated at the photosphere. However, MHD waves face many challenges in making it to the corona where their energy can be available for heating. Transverse (cross-field) structuring is essential for magnetic waveguides to exist in the solar atmosphere to support the propagation of MHD waves. Classically, features acting as magnetic waveguides such as coronal loops, sunspots, pores, spicules and prominences, to name a few, have been modelled as static, straight magnetic flux tubes to study their feasibility in supporting MHD wave propagation. However, it has recently become evident that magnetic vortex tubes, such as solar tornadoes and spinning solar jets, may also act as MHD waveguides, channeling increased amounts of energy flux to the corona \citep{Yadav2020, Yadav2021, Finley2022, Kuniyoshi2023, Silva2024}. Furthermore, rotational motions of solar jets are present in both numerical simulations \citep{gon2019, Skirvin2023_JET} and observations \citep{Suematsu_et_al_2008, Sharma_2017, Sharma2018} making it difficult to diagnose the properties of MHD waves guided by these structures using the model of a static magnetic flux tube. Moreover, rotational motions are frequently reported in both observations and numerical studies in relation to MHD waves and/or untwisting of the magnetic field \citep[e.g.][]{Wedemeyer2012, Tzi2018, Tzi2020,  Murabito2020, Aljohani2022, Liakh2023,Petrova2024}. Magnetic vortex tubes can be modelled as straight magnetic flux tubes but in the presence of background rotational plasma flows, with implications for the spectrum of MHD waves that can be supported. Features resembling vortex tubes are ubiquitous throughout the solar atmosphere \citep{Tziotziou2023SSRv}, therefore, it is vital to develop the theory and understanding of MHD wave properties in the presence of rotational plasma motions to correctly diagnose them in solar observations and numerical simulations.

The stability of MHD modes in an equilibrium containing background rotational flows has been studied in a solar \citep{sol2010, zaq2015, Cher2018} and astrophysical \citep{Keppens2002, Brughmans2024} context. However, literature on the broad spectrum of MHD modes and their observability in rotating magnetic flux tubes is limited. Although, the spectrum of MHD waves in magnetic flux tubes in the presence of background rotational flows has been explored in both an adiabatic \citep{Wang2004, Skirvin2023rotflow} and non-adiabatic context \citep{Hermans2024}. In both cases, it is found that the modes are altered by the background flow providing a Doppler shift to the continua, modifying the shape of the eigenfunctions from the case of a static flux tube, with implications for interpreting observational data of MHD waves in rotating structures.

The Poynting flux vector is an important quantity to understand the magnetic energy transfer by waves in the structured solar atmosphere. A study by \citet{Goossens2013} derived expressions for the energy fluxes associated with transverse ($m=1$) kink waves in pressureless flux tubes with a piece-wise constant density. The authors demonstrated how calculating the vertical component of the Poynting flux when using classic, volume filling bulk Alfv\'{e}n waves can significantly overestimate the real energy flux carried by transverse MHD kink waves in the presence of spatial structuring. Moreover, the wave energy fluxes associated with axisymmetric MHD waves ($m=0$) in photospheric magnetic flux tubes was investigated by \citet{moreels2015}. This study built on the work by \citet{Goossens2013} through the inclusion of plasma pressure and these expressions have been exploited in numerous observational results over recent years to determine the energy fluxes from observations of MHD waves in the solar atmosphere \citep[e.g.][]{grant2015, keys2018}. However, these derivations assume a static magnetic flux tube, and it is unclear how the presence of background flows may affect the Poynting flux carried by MHD waves.

This paper is structured as follows: in Section \ref{sec:methods} we layout the equations necessary to conduct an analytical investigation into the Poynting flux associated with MHD modes in rotating flux tubes. In section \ref{sec:Results} we obtain the wave solutions in a rotating equilibrium using the SESAME code \citet{Skirvin2021, Skirvin2022, Skirvin2023rotflow}, and derive an expression for the vertical component of the Poynting flux, $S_z$, associated with MHD waves in such an equilibrium configuration. We present 2D visualisations demonstrating how $S_z$ may appear for different modes under varying strengths of the background flow, and explore how the presence of a background rotational flow affects the magnitude of the Poynting flux when compared with the static case. In section \ref{sec:conclusions} we summarise our findings and outline future work.

\section{Methods} \label{sec:methods}
To investigate the Poynting flux in solar vortex tubes, we work in a cylindrical geometry $(r, \varphi, z)$ and consider a straight and untwisted flux tube, such that the background magnetic field vector can be written as $\mathbf{B}_0 = (0,0,B_{0,z})$. The flux tube under investigation exhibits rotational behaviour and the radially dependent background velocity field vector is given by $\mathbf{v}_0 = (0, v_{0,\varphi}(r), 0)$. For simplicity, we will consider a solid body rotation, such that the background rotational flow is described as:
\begin{equation}\label{flow_eqn}
v_{0,\varphi}(r) = Ar, 
\end{equation}
where $A$ is the amplitude of the rotational flow.

The ideal MHD equations used in this study are:

 \begin{equation}\label{continuity}
     \frac{d\rho}{dt}+\rho\nabla\cdot\textbf{v}=0,
 \end{equation}
\begin{equation}\label{momentum}
     \rho\left(\frac{d\textbf{v}}{dt}\right)=-\nabla p+\frac{1}{\mu_0}\left(\nabla\times\textbf{B}\right)\times\textbf{B},
 \end{equation}
 \begin{equation}\label{induction}
     \frac{\partial\textbf{B}}{\partial t}=\nabla\times\left(\textbf{v}\times\textbf{B}\right),
 \end{equation}
 \begin{equation}\label{divB}
     \nabla \cdot \textbf{B} = 0,
 \end{equation}
 \begin{equation}\label{energy}
         \frac{d}{dt}\left(\frac{p}{\rho^\gamma}\right)=0,
 \end{equation}
where $\rho$, $\textbf{v}$, $p$, $\textbf{B}$, $\gamma$ and $\mu$ denote plasma density, plasma velocity, plasma pressure, magnetic field, the ratio of specific heats (taken $\gamma = 5/3$) and the magnetic permeability respectively. Physical effects such as gravity, flux tube expansion and non-ideal terms are neglected in the current study. In the lower solar atmosphere it should be noted that magnetic flux tubes possess significant non-vertical magnetic field as a result of flux tube expansion to maintain pressure balance due to gravitational stratification, however, this effect can be considered to be negligible in the current study due to the analysis in the local plasma environment. Since the equilibrium quantities depend on $r$ only, the perturbed quantities can be Fourier-analysed with respect to the ignorable coordinates $\varphi, z$ and time $t$ and put proportional to: 
\begin{equation*}
\text{exp}\left[i\left(m\varphi + kz - \omega t\right)\right],
\end{equation*}
where $m$ is the azimuthal wave number, $k$ is the vertical (parallel to the magnetic field) wavenumber and $\omega$ is the wave frequency.

To obtain the wave solutions for the flux tube equilibrium under consideration, Equations (\ref{continuity})-(\ref{divB}) are linearised resulting in a system of two differential equations containing the total pressure perturbation $\hat{P}_T$ and the Lagrangian displacement perturbation in the radial direction $r\hat{\xi}_r$ \cite[see e.g.][]{sak1991, goo1992}, which can be written as:

\begin{align}
    D\frac{d}{dr}\left(r\hat{\xi}_r\right)&=C_1 r\hat{\xi}_r - C_2 r \hat{P}_T, \label{rxi_r_diff} \\
    D\frac{d\hat{P}_T}{dr}&=C_3 \hat{\xi}_r - C_1 \hat{P}_T, \label{P_diff}
\end{align}
where,
\begin{align}
    D &= \rho_0\left(c_s^2+v_A^2 \right)\left(\Omega^2-k^2v_A^2\right)\left(\Omega^2-k^2c_T^2\right), \label{D} \\
    \Omega &= \omega - \frac{m}{r}v_{0,\varphi}, \label{Omega} \\
    c^2 &= \frac{\gamma p}{\rho_0},\ \ \ \ v_A^2 = \frac{B_{0,z}^2}{\mu\rho_0}, \ \ \ \ c_T^2 = \frac{v_A^2 c_s^2}{\left(c_s^2 + v_A^2\right)}, \\
    C_1 &= Q\Omega^2 - 2m\left(c_s^2+v_A^2\right)\left(\Omega^2-k^2c_T^2\right)\frac{T^2}{r^2}, \label{C_1} \\
    C_2 &= \Omega^4 - \left(c_s^2+v_A^2\right)\left(\frac{m^2}{r^2}+k^2\right)\left(\Omega^2-k^2c_T^2\right), \label{C_2} \\
    C_3 &= D\left\{\rho_0\left(\Omega^2-k^2v_A^2 \right) + r\frac{d}{dr}\left[- \rho\left(\frac{v_{0,\varphi}}{r}\right)^2\right]\right\} + \\ \nonumber &+ Q^2 - 4\left(c_s^2+v_A^2 \right)\left(\Omega^2-k^2c_T^2\right)\frac{T^2}{r^2}, \\
    Q &= -\left(\Omega^2-k^2v_A^2\right)\frac{\rho_0 v_{0,\varphi}^2}{r}, \label{Q} \\
    T &= \rho_0 \Omega v_{0,\varphi}. \label{T}
\end{align}
Quantities $c_s^2$, $v_A^2$, and $c_T^2$ define the squares of the equilibrium local sound, Alfv\'{e}n and cusp (tube) speeds respectively. The quantity $\Omega$ represents the Doppler shifted frequency as a result of the background plasma flow.

Equations (\ref{rxi_r_diff})-(\ref{P_diff}) can be combined to create a single, second-order differential equation in either $r\hat{\xi}_r$:
\begin{equation}\label{rxi_r_diff_eqn}
    \frac{d}{dr}\left[ f(r)\frac{d}{dr}\left(r\hat{\xi}_r\right)\right] - g(r)\left(r\hat{\xi}_r\right) = 0,
\end{equation}
where,
\begin{equation}
f(r) = \frac{D}{rC_2},
\end{equation}
\begin{equation}
g(r) = \frac{d}{dr}\left(\frac{C_1}{rC_2}\right) - \frac{1}{rD}\left(C_3 - \frac{C_1^2}{C_2}\right),
\end{equation}
or $\hat{P}_T$:
\begin{equation}\label{P_diff_eqn}
    \frac{d}{dr}\left[ \Tilde{f}(r)\frac{d\hat{P}_T}{dr}\right] - \Tilde{g}(r)\hat{P}_T = 0,
\end{equation}
where,
\begin{equation}
\Tilde{f}(r) = \frac{rD}{C_3},
\end{equation}
\begin{equation}
\Tilde{g}(r) = -\frac{d}{dr}\left(\frac{rC_1}{C_3}\right) - \frac{r}{D}\left(C_2 - \frac{C_1^2}{C_3}\right).
\end{equation}
Equations (\ref{rxi_r_diff_eqn}) and (\ref{P_diff_eqn}) reduce to the well-known Bessel equations presented in \citet{edrob1983} when the plasma inside and outside the flux tube is uniform and does not depend on the spatial coordinate $r$. 

For a non-uniform plasma, the governing Equations (\ref{rxi_r_diff})-(\ref{P_diff}) possess regular singularities where the wave frequency matches the local characteristic frequencies at:
\begin{equation}\label{alfven_flow_continuum}
    \omega = \frac{m}{r}v_{0,\varphi}(r) \pm k v_A,
\end{equation}
\begin{equation}\label{cusp_flow_continuum}
    \omega = \frac{m}{r}v_{0,\varphi}(r) \pm k c_T.
\end{equation}
Equations (\ref{alfven_flow_continuum}) and (\ref{cusp_flow_continuum}) define the flow continua modified by the local Alfv\'{e}n ($\Omega_A$) and slow ($\Omega_T$) frequencies, respectively. In ideal MHD, the wave solutions existing inside the continua, with positions given by Equations (\ref{alfven_flow_continuum}) and (\ref{cusp_flow_continuum}), are known as `quasi-modes' where the wave frequency becomes a complex quantity \citep{DeGroof2000, goe2004, Geer2022} and the waves may be resonantly damped or become unstable. However, given that the rotational flow profile in the present study depends linearly on the radial coordinate, the continua in this case reduce to single point values.

Obtaining an equilibrium in a rotating magnetic flux tube is achieved in the same manner outlined in \citet{Skirvin2023rotflow}. In order to maintain total pressure balance across the waveguide the following expression must be satisfied \citep{goo2011}:
\begin{equation}\label{rot_flow_pressure_balance}
    \frac{d}{dr}\left( p + \frac{B_{0z}^2}{2\mu} \right) = \frac{\rho v_{\varphi}^2(r)}{r} = \rho A^2 r.
\end{equation}
Integration of Equation (\ref{rot_flow_pressure_balance}) yields:
\begin{equation}\label{integrate_pressure_balance}
    p + \frac{B_{0z}^2}{2\mu} = \rho A^2 \frac{r^2}{2},
\end{equation}
where the constant of integration is absorbed into the gas pressure term, $p$, and corresponds to the plasma pressure on the axis of the cylinder where the amplitude of the flow is zero \citep[see e.g.][]{Cher2018}. Under the photospheric conditions considered in this work, the total pressure balance is achieved by an increase in temperature to balance the increase in azimuthal flow amplitude towards the boundary of the flux tube.

The presence of a background rotational flow not only modifies the equilibrium pressure balance relationship, but also affects the continuity conditions on the boundary of the waveguide. Considering a magnetic flux tube in the presence of a background rotational flow, the resulting boundary continuity conditions state, for the Lagrangian displacement in the radial direction and the total pressure perturbation:
\begin{align}
    \label{rotflow_boundary_condition_xi} \hat{\xi}_{re}\Bigr\rvert_{r = a} &= \hat{\xi}_{ri}\Bigr\rvert_{r = a}, \\
    \label{rotflow_boundary_condition_PT} \hat{P}_{Te}\Bigr\rvert_{r = a} &= \left(\hat{P}_{Ti} + \frac{\rho_{0i} v_{0,\varphi}^2}{a}\hat{\xi}_{ri} \right)\Bigr\rvert_{r = a}.
\end{align}
Here, plasma variables inside the flux tube are denoted by subscript $i$, and outside the flux tube are denoted with subscript $e$. The change in boundary conditions are accounted for in the numerical eigensolver, and a pair of eigenvalues will only be retrieved for values satisfying the above conditions.

\section{Results}\label{sec:Results}
\subsection{MHD wave solutions in a rotating magnetic flux tube}\label{sec:wave_solutions}

Consider a magnetic flux tube under photospheric conditions where the characteristic speeds both inside and outside the tube can be defined as: $c_i=1$, $c_e=1.5c_i$, $v_{Ai}=2c_i$, $v_{Ae}=0.5c_i$ with internal density $\rho_i=1$. This choice of equilibrium parameters results in a density contrast between the internal and external plasma to be roughly $\rho_i/\rho_e = 0.567$ such that the flux tube is under-dense with respect to the external plasma. The amplitude of the background rotational flow is chosen as $A=0.1$, such that $v_{0,\varphi} = 0.1r$ and the flow is both subsonic and subalfv\'{e}nic throughout the domain. This choice of flow amplitude allows us to investigate the effect of the rotating tube on the possible trapped MHD waves within and study their magnetic energy transport without entering the regime of any flow-driven instabilities such as Kelvin-Helmholtz or encounter the nonlinear regime. 

Firstly, it is necessary to obtain the MHD wave solutions to Equations (\ref{rxi_r_diff_eqn}) and (\ref{P_diff_eqn}) with the associated boundary conditions provided by Equations (\ref{rotflow_boundary_condition_xi}) and (\ref{rotflow_boundary_condition_PT}). To do so, a numerical eigensolver is implemented as both Equations (\ref{rxi_r_diff_eqn}) and (\ref{P_diff_eqn}) have no known closed form analytical solutions, without making assumptions that somehow reduce the mathematical complexity. Therefore, investigating the properties of wave modes propagating within an equilibrium which is non-uniform must be done numerically. The numerical approach used in this study utilises the numerical eigensolver SESAME (Shooting Eigensolver for SolAr Magnetohydrostatic Equilibria) developed and applied in \citet{Skirvin2021, Skirvin2022} for non-uniform magnetic slabs and non-uniform flux tubes,  respectively, in addition to the study of MHD wave modes in rotating photospheric tubes by \citet{Skirvin2023rotflow}. The SESAME code obtains the wave solutions, $k$ and $\omega$, for a given azimuthal wavenumber $m$ and matches the necessary boundary conditions of $\xi_r$ and $P_T$ for a provided equilibrium configuration.

\begin{figure*}
\centering
\includegraphics[width=0.99\textwidth]{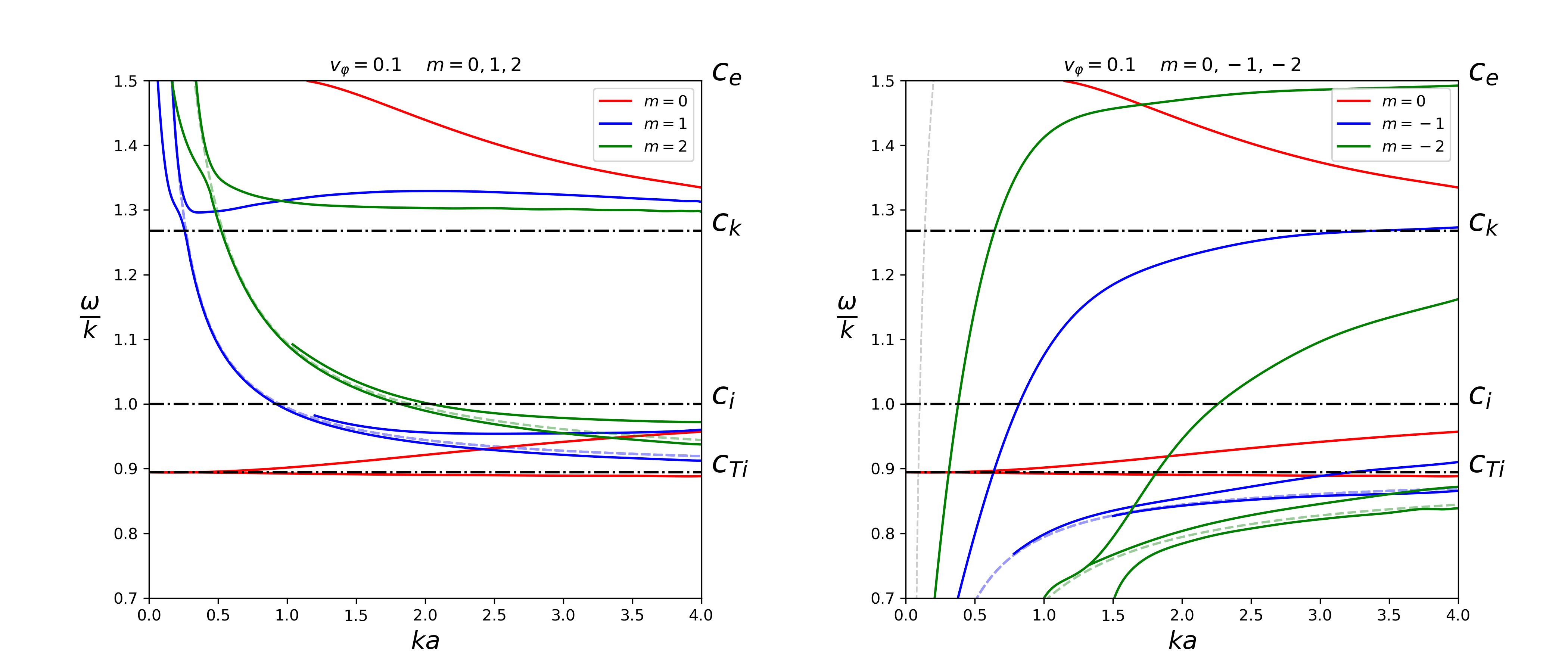}
\caption{The dispersion diagram for a rotating photospheric flux tube with $v_{0,\varphi} = 0.1r$ indicating the wave solutions obtained with SESAME. Red curves denote the sausage $m=0$ mode, blue curves highlight the $m=\pm 1$ kink mode and green curves denote fluting $m=\pm 2$ modes. The dashed blue and green curves indicate the flow modified slow continuum for the $m=1$ kink mode and $m=2$ fluting mode, respectively. The dashed grey curve indicates the flow modified Alfv\'{e}n continuum for the $m=-1$ kink mode.
\label{fig:disp_diag_v01_m_comparison}}
\end{figure*}

The SESAME solutions are shown on the dispersion diagram in Figure \ref{fig:disp_diag_v01_m_comparison} for sausage, kink and fluting modes with azimuthal wave numbers $m=0, \pm 1, \pm 2$, respectively, in a rotating photospheric flux tube with amplitude of rotation $v_{0,\varphi} = 0.1r$. As reported by \citet{Skirvin2023rotflow} the presence of a background rotational flow has no considerable effect here on the solutions for the axisymmetric $m=0$ sausage mode, however, there may be some modification of the solutions in the thin-tube limit. This is expected because, although in Equations (\ref{rxi_r_diff})-(\ref{T}) the background azimuthal flow is usually multiplied by the azimuthal wavenumber $m$, there are some instances where $v_{0,\varphi}$ is present independent of $m$, for instance in the expressions of the variables $Q$ and $T$ in Equations (\ref{Q}) and (\ref{T}), respectively.  

Figure \ref{fig:disp_diag_v01_m_comparison} highlights the effect of the rotational flow on non-axisymmetric modes (e.g. $m=\pm 1, \pm 2$) and indicates the position of the flow modified slow continua. It is evidently clear that the eigenvalues of the slow body and slow surface modes follow the respective continua which depends on azimuthal wavenumber $m$. The phase speed of the positive modes (e.g. $m > 0$) is increased compared to the scenario of no background rotational flow, as these modes rotate in the same direction as the background flow and their propagation is supported and enhanced by the flow. On the other hand, the phase speed of the negative modes ($m < 0$) is reduced, as these modes rotate against the direction of the flow and their propagation is suppressed. Moreover, for the negative modes rotating against the flow, the slow modes (slow body and slow surface) do not appear to exist in the thin-tube (long-wavelength) limit, instead they become absorbed into the flow modified slow continuum where they may become resonantly damped or unstable. For the negative modes, the fast surface kink mode follows the trajectory of the flow modified Alfv\'{e}n continuum and is absorbed by this in the long-wavelength limit regime. The fast surface fluting $m=-2$ modes also follow their respective Doppler shifted Alfv\'{e}n continuum, however, due to the greater shift to their continuum resulting from the higher order azimuthal mode, solutions which typically exist in the leaky regime for photospheric conditions ($\omega/k > c_e$), are now shifted into the trapped regime. In a uniform static magnetic flux tube, the phase speed of the fast surface fluting mode tends towards the kink speed in the long-wavelength limit. However, in the presence of a background rotational flow, it appears to follow the trajectory of the flow modified Alfv\'{e}n continuum with $m=-2$ where it then appears to merge with the slow body mode branch. This merging may be an interesting focus of future studies and the location of the sharing of properties between the two modes will be dependent on a case-by-case basis of particular flow amplitudes and background plasma equilibrium values.

\subsection{Derivation of expression for $S_z$ for MHD modes in a rotating flux tube}

Once the MHD wave solutions are obtained for a given equilibrium configuration of a magnetic flux tube, it is possible to investigate the Poynting flux associated with these modes. The Poynting flux, $\mathbf{S}$ can be written as:
\begin{equation}\label{eqn:PoyntingFlux1}
    \mathbf{S} = \frac{1}{4\pi}\mathbf{E}\times\mathbf{B}.
\end{equation}
The Poynting vector $\mathbf{S}$ describes the direction in which magnetic energy is flowing and its magnitude is expressed in W/m$^2$. From this point, we will drop the $1/4\pi$ term, representing the magnetic permeability, and absorb it into magnetic field variable $\mathbf{B}$. Using the MHD expression for the definition of the electric field $\mathbf{E}$, Equation (\ref{eqn:PoyntingFlux1}) becomes:
\begin{equation}\label{eqn:PoyntingFlux}
    \mathbf{S} = -(\mathbf{v}\times\mathbf{B})\times\mathbf{B}.
\end{equation}
Similar to the analysis procedure undertaken in Section \ref{sec:wave_solutions}, the Poynting flux vector can be linearised to study the contribution from different wave modes. However, typically after linearising, when considering wave energy fluxes the first order terms average out over a wavelength (due to the equal contribution from positive and negative values). Therefore, to study the energy fluxes associated with MHD waves, we should also consider the second order terms which are inherently nonlinear in nature. To do so, we follow an approach similar to that described in Section 9.3 of the book by \citet{Walker2005} which allows us to use quantities that are averaged over a complete
cycle of the wave \citep{Goossens2013}. This approach demonstrates how treating the variables as products of complex conjugates can result in expressions for the energy flux which are quadratic in nature, however, arise from only first-order reduced MHD equations \citep{Goossens2013, moreels2015}. Given the fact that we are only interested in trapped MHD waves in this study, which can be expressed solely in terms of real-valued variables (as opposed to complex quantities), we will linearise the Poynting flux vector but include second order terms in the expansion. Now, let us use subscript $0$ to denote equilibrium quantities and subscript $1$ to denote the small perturbations. The \textit{linearised} Poynting vector can be expressed in the following terms:
\begin{multline}\label{linearised}
    \mathbf{S} = - \underbrace{(\mathbf{v_0}\times\mathbf{B_0})\times\mathbf{B_0}}_{\text{T1}} \ - \ \underbrace{(\mathbf{v_1}\times\mathbf{B_0})\times\mathbf{B_0}}_{\text{T2}} \ - \ \\ -\underbrace{(\mathbf{v_0}\times\mathbf{B_0})\times\mathbf{B_1}}_{\text{T3}} \ -  \underbrace{(\mathbf{v_0}\times\mathbf{B_1})\times\mathbf{B_0}}_{\text{T4}} \ - \ \underbrace{(\mathbf{v_1}\times\mathbf{B_0})\times\mathbf{B_1}}_{\text{T5}}- \\ - \ \underbrace{(\mathbf{v_0}\times\mathbf{B_1})\times\mathbf{B_1}}_{\text{T6}}  - \underbrace{(\mathbf{v_1}\times\mathbf{B_1})\times\mathbf{B_0}}_{\text{T7}}.
\end{multline}
It can be seen that there are seven terms describing the total Poynting flux in Equation (\ref{linearised}) comprising the interaction between the background rotating plasma and the perturbations resembling the MHD waves. The MHD Poynting flux vector can be decomposed into two separate contributions, one relating to the vertical transport of horizontal magnetic field and another connected with the horizontal buffeting of vertical magnetic fields \citep{shelyag2012}. In this study, the equilibrium magnetic field is purely vertical, such that we only model the contribution from horizontal motions of the vertical magnetic field as a mechanism for producing vertical Poynting flux. From this point, all variables can be assumed to be perturbations unless sub-scripted with $0$.

The background equilibrium configuration assumed in this study consists of a straight magnetic flux tube with a background $v_{0,\varphi}$ component which is linear in the radial direction. The term in Equation (\ref{linearised}) labelled T$1$ represents the Poynting flux solely from the background plasma, therefore, this term is of zero order and represents the Poynting flux associated with the magnetic vortex tube itself. This term contributes only to the azimuthal component of the Poynting vector and does not produce any net upwards magnetic energy flux. Explicitly, this can be written as:
\begin{equation}        
[(\mathbf{v_0}\times\mathbf{B_0})\times\mathbf{B_0}] = 0 \mathbf{\hat{r}} \ + \ v_{0,\varphi}B_{0,z}^2  \bm{\hat{\varphi}} \ + \ 0 \mathbf{\hat{z}},
\end{equation}
which is a fundamental result, because any magnetic solar vortex tube in observations or numerical simulations, must contain a significant azimuthal component of the magnetic field vector, complimented by an azimuthal or vertical velocity field component, in order to self-generate an upwards Poynting flux. Therefore, the magnitude of $S_z$ associated with solar vortices will heavily depend upon the pitch angle of the background magnetic field ($B_{0,\varphi}/B_{0,z}$), which should be incorporated and investigated in future work. The fact that solar vortex tubes with very small pitch angles ($B_{0,\varphi} \ll B_{0,z}$) produce very little vertical Poynting flux may explain why the horizontal component of the Poynting flux has suggested to be important in the solar atmosphere \citep{silva2022}. Therefore, any vertical Poynting flux produced in straight rotating magnetic flux tubes must be a result of perturbations only, and may be interpreted as the Poynting flux associated with MHD waves guided by the vortex tube. In fact, the full inclusive expression for the vertical Poynting flux $S_z$ in a twisted magnetic flux tube with background flow is given by:
\begin{equation}
[(\mathbf{v_0}\times\mathbf{B_0})\times\mathbf{B_0}]_z = (B_{0,\varphi} v_{0,z} - v_{0,\varphi}B_{0,z})B_{0,\varphi},
\end{equation}
highlighting the necessity for a twisted magnetic tube for the presence of a background Poynting flux component in the vertical direction.

In this study, we are most interested in $S_z$, as this represents the magnetic energy transported upwards along the vertical magnetic field, which would be available for heating in the upper atmosphere. The components of first order, given by terms highlighted T2, T3, T4 in Equation (\ref{linearised}), which contribute to the vertical Poynting flux can be written as:
\begin{equation}
    S_z = -v_{0,\varphi}B_{0,z}B_{\varphi}.
\end{equation}
The above expression demonstrates that the vertical component of the Poynting flux is zero, as expected for the first order terms, unless there is a background flow which can advect the flux associated with the wave. The background flow does not need to be along the axis of the tube, in this case, the flow is around the tube axis, however as the magnetic field is vertical, an upwards magnetic energy flux is generated.

The components of second order, terms $5,6,7$ in Equation (\ref{linearised}), of the vertical Poynting flux can be written as:
\begin{equation}
    S_z = B_{0z}\left(B_{\varphi}v_{\varphi} + B_rv_r\right) + v_{0,\varphi}B_{\varphi}B_z. 
\end{equation}
The above expression indicates the quadratic terms to $S_z$ which will not cancel out one another when averaged over a full cycle of the wave. Therefore, by combining the first and second order terms of $S_z$ (remembering that the zero order expression is zero for the $z$ component), the full $S_z$ expression is given by:
\begin{equation}\label{s_z_components}
    S_z = -v_{0,\varphi}B_{0,z}B_{\varphi} + B_{0z} \left(B_{\varphi}v_{\varphi} + B_rv_r\right) + v_{0,\varphi}B_{\varphi}B_z. 
\end{equation}
Equation (\ref{s_z_components}) contains both the background plasma flow and the perturbations of components of the velocity and magnetic field associated with MHD waves in a magnetic flux tube. The numerical eigensolver SESAME finds the wave solutions for a given equilibrium by matching the necessary boundary conditions for $P_T$ and $\xi_r$ provided in Equations (\ref{rotflow_boundary_condition_xi}) and (\ref{rotflow_boundary_condition_PT}). Therefore, it would be instructive to write Equation (\ref{s_z_components}) in terms of quantities relating to the plasma environment and the eigenfunctions $P_T$ and $\xi_r$. Following \citet{goo1992}, we can use the following expressions:
\begin{equation}\label{expr_vr}
    v_r = -i\Omega \xi_r,
\end{equation}
\begin{equation}
    v_{\varphi} = \frac{B_{0,z}\Omega}{\rho_0 \left( \Omega^2 - k^2v_A^2\right)}\left[\frac{m}{r}P_T - 2T\frac{\xi_r}{r} \right],
\end{equation}
\begin{equation}
    B_r = -i k B_{0,z} \xi_r,
\end{equation}
\begin{equation}
    B_{\varphi} = -\frac{kB_{0,z}^2}{\rho_0 \left( \Omega^2 - k_z^2v_A^2\right)}\left[\frac{m}{r}P_T - 2T\frac{\xi_r}{r} \right],
\end{equation}
\begin{equation}\label{expr_bz}
    B_z = B_{0,z} \left( \frac{k^2 v_{0,\varphi}^2}{\Omega^2}\frac{\xi_r}{r} - \left[\frac{\Omega^2P_T - Q\xi_r}{\rho_0\left(\Omega^2-k^2 c_T^2 \right)\left(c^2+v_A^2\right)}\right]\left[\frac{k^2 c^2}{\Omega^2}+1 \right] \right).
\end{equation}
Inserting Equations (\ref{expr_vr})-(\ref{expr_bz}) into Equation (\ref{s_z_components}) yields:

\begin{multline} \label{S_z_equation_withflow}
        S_z = k \Omega B_{0,z}^2 \xi_r^2 - \left( \frac{k v_{0,\varphi} B_{0,z}^3}{\rho_0 \left( \Omega^2 - k^2v_A^2\right)}\left[\frac{m}{r}P_T - 2T\frac{\xi_r}{r} \right] \right)\times \\ \times \left( \frac{k^2 v_{0,\varphi}^2}{\Omega^2}\frac{\xi_r}{r} - \left[\frac{\Omega^2P_T - Q\xi_r}{\rho_0\left(\Omega^2-k^2 c_T^2 \right)\left(c^2+v_A^2\right)}\right]\left[\frac{k^2 c^2}{\Omega^2}+1 \right] -1\right) - \\ - \frac{k \Omega B_{0,z}^4}{\rho_0^2\left(\Omega^2 - k^2v_A^2 \right)^2}\left(\frac{m}{r}P_T - 2T\frac{\xi_r}{r} \right)^2.
\end{multline}

Equation (\ref{S_z_equation_withflow}) provides, to the best of our knowledge, the first explicit expression describing the field-aligned Poynting flux associated with MHD waves in a solar vortex tube resembling a straight untwisted magnetic flux tube with a background rotational flow. However, Equation (\ref{S_z_equation_withflow}) is extremely complex, as the spatial structure of the eigenfunctions will also depend upon the wavenumber and specific mode under investigation. For example, in the long-wavelength limit as $k$ approaches zero, the azimuthal perturbation of displacement is dominant for the fast surface mode over the vertical displacement component, with obvious implications in determining which term is dominant in Equation (\ref{s_z_components}).

It is easy to validate Equation (\ref{S_z_equation_withflow}) against previously obtained analytical expressions for the Poynting flux in less complicated equilibrium configurations. For example, if the presence of a background flow is neglected ($v_{0,\varphi}=0$), we can see from Equations (\ref{s_z_components}) or (\ref{S_z_equation_withflow}):
\begin{equation}
    S_z = B_{0z} \left(B_{\varphi}v_{\varphi} + B_rv_r\right),
\end{equation}
which, using Equations (\ref{expr_vr})-(\ref{expr_bz}), can be written as:
\begin{equation}\label{s_z_noflow}
    S_z = \omega k B_{0,z}^2\left(\xi_{\varphi}^2 + \xi_r^2 \right).
\end{equation}
The work by \citet{Goossens2013} adopt the complex conjugate approach outlined in \citet{Walker2005}. They also assumed a pressureless plasma such that $c=0$ (sound speed) and $\xi_z=0$. In addition, they studied the $m=1$ kink mode, therefore, following the approach of \citet{Goossens2013}, it is clear that Equation (\ref{s_z_noflow}) can be written as:
\begin{equation}
    S_z = \omega k B_{0,z}^2\left( \bm{\xi}\cdot \bm{\xi^*}\right),
\end{equation}
which can be manipulated to:
\begin{equation}\label{goossens2013}
    S_z = \rho_0 \omega_A^2 \left( \bm{\xi}\cdot \bm{\xi^*}\right)v_{ph},
\end{equation}
where $v_{ph}=\omega/k$ and $\omega_A^2=k^2v_A^2$. Equation (\ref{goossens2013}) agrees with Equation (13) from \citet{Goossens2013}  when averaged over a complete wavelength.

In a magnetic flux tube with no background rotational flow, Equation (\ref{S_z_equation_withflow}) can be expressed in terms of $P_T$ and $\xi_r$ as:
\begin{equation}\label{Sz_no_flow_full}
    S_z = \rho_0 \omega k v_A^2 \left( \left[ \frac{B_{0,z}}{\rho_0\left(\omega^2- k^2 v_A^2 \right)}\frac{m}{r}P_T \right]^2 + \xi_r^2 \right).
\end{equation}
We can substitute $m=0$ into Equation (\ref{Sz_no_flow_full}) to recover Equation (5) from \citet{moreels2015} where the authors were investigating the energy fluxes associated with sausage modes in photospheric flux tubes. Substituting $m=0$ yields:
\begin{equation}
    S_z = \rho_0 \omega k v_A^2 \xi_r^2,
\end{equation}
which can be rewritten as:
\begin{equation}
    S_z = \rho_0 v_A^2 k \omega \xi_r \xi_r^* v_{ph},
\end{equation}
as derived in \citet{moreels2015} when considering averaging of the wave energy flux over a full wavelength. Therefore, we can be confident that Equation (\ref{S_z_equation_withflow}) describes the field-aligned component of the Poynting flux associated with MHD waves in rotating flux tubes.

\subsection{Amplitude ratios}
Throughout the preceding derivation for $S_z$ associated with MHD modes in rotating magnetic flux tubes, we are free to choose the relative amplitudes of the perturbations (i.e. the strength of the wave) for the eigenfunctions $\xi_r$ and $P_T$ with respect to the strength of the background plasma flow. It is likely that throughout the solar atmosphere, there are plasma dynamics generating waves with varying amplitudes with respect to the strength of background flows. Therefore, it would be instructive to study the properties and observability of MHD modes with varying amplitude ratios. In other words, to understand the change in behaviour of the modes as the amplitude of the flow and/or the amplitude of the perturbation changes with respect to each other.

\begin{figure}
\centering
\includegraphics[width=0.45\textwidth]{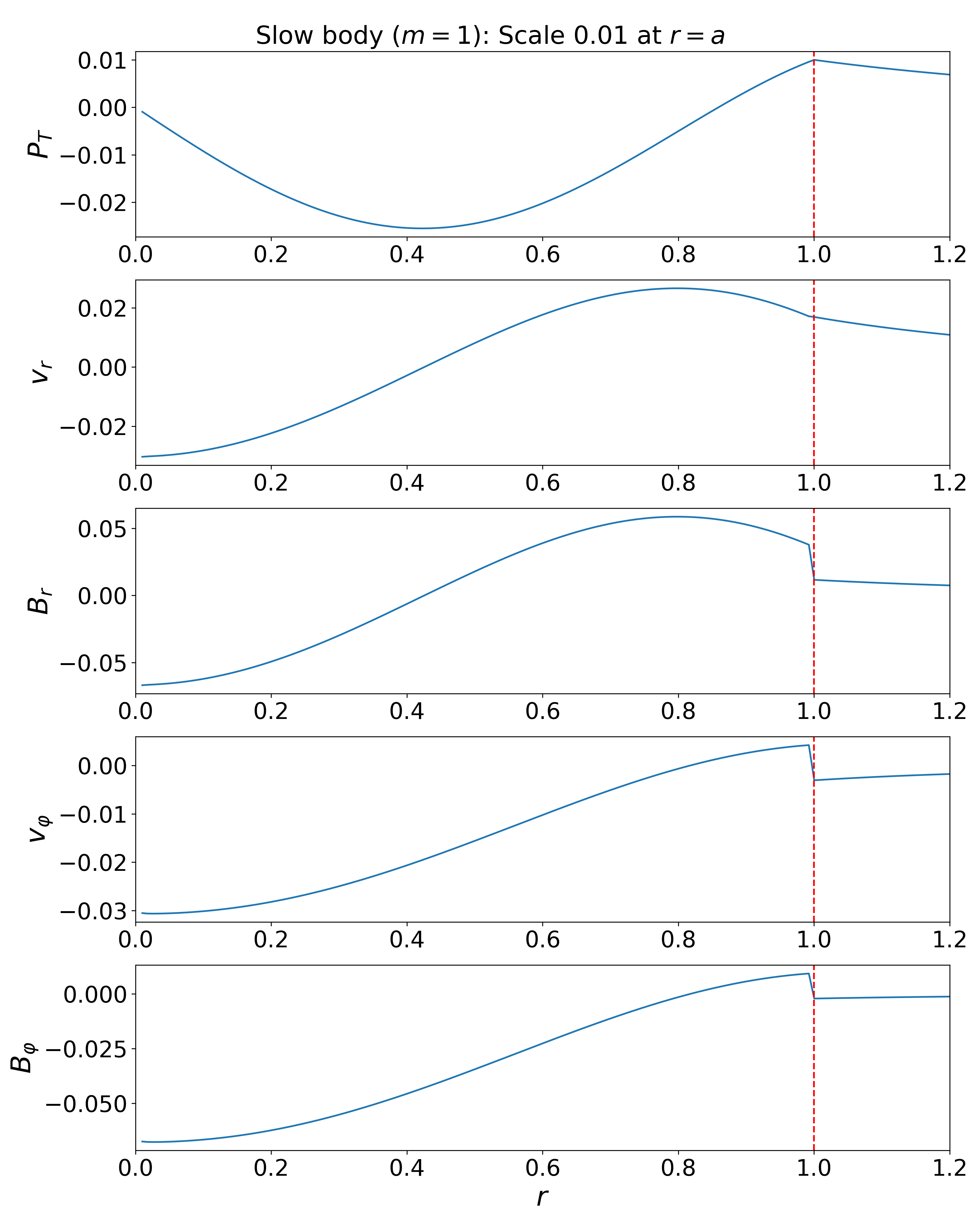}
\caption{Spatial structure of the eigenfunctions ($P_T, v_r, B_r, v_{\varphi}$ and $B_{\varphi}$) for the slow body kink mode ($m=1$) normalised to $0.01$ at $r=a$ with $k=1.76$ and $\omega=1.69$. The boundary of the magnetic flux tube with a background rotational flow is denoted by the dashed red line at $r=1$.}
\label{fig:spatial_eigenfuncs_amplituderatio}
\end{figure}

\begin{figure*}
\centering
\includegraphics[width=0.85\textwidth]{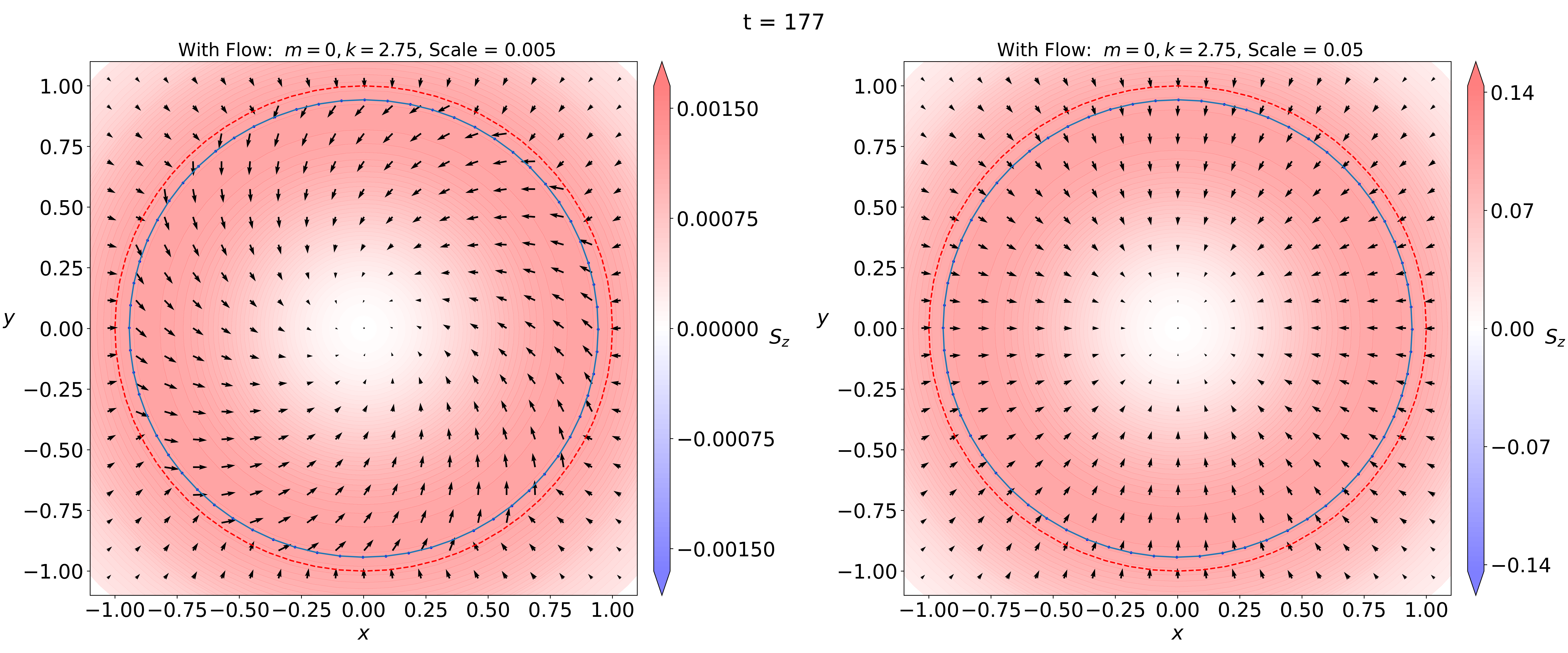}
\caption{The vertical component of the Poynting flux $S_z$ for the slow body sausage mode in the presence of background rotational flow at a specific snapshot in time ($t=177$ in arbitrary units). Panel (a) shows the case with amplitude ratio $v_{0, \varphi}=20\hat{f}$, where $\hat{f}$ denotes the strength of the perturbation, whereas panel (b) highlights the case with amplitude ratio $v_{0, \varphi}=2\hat{f}$. The arrows denote the total velocity vector (background plus perturbation) and are scaled with the background flow amplitude, respectively. The boundary of the flux tube is represented by the solid blue line whereas the equilibrium location of the boundary is shown with the dashed red line.}
\label{fig:Sz_m0_scale_comparison}
\end{figure*}

To investigate this, let us focus on one of the most frequently observed modes in the lower solar atmosphere in waveguides such as sunspots and pores; slow body modes. Their spatial behaviour is more strongly affected by the presence of a rotational flow inside the flux tube, when compared to surface waves whose strongest perturbation exists on the tube boundary. Let us fix the amplitude of the rotational flow to be $0.1$ at the flux tube boundary, as is consistent throughout this work. However, we now scale the amplitude of the perturbation such that, at $r=a$, the magnitude of the perturbation ranges from $0.001$, $0.005$, $0.01$, $0.02$, $0.05$, $0.1$ and $1$. In other words, we are investigating the regime where the amplitude of the flow is $100$ times greater than the wave (e.g. $v_{0,\varphi} = 100 \hat{f}$ where $\hat{f}$ denotes the strength of the perturbation) to the regime whereby the amplitude of the wave is $10$ times greater than the flow (e.g. $v_{0,\varphi} = 0.1 \hat{f}$) which is approaching the limit where the flow can almost be neglected, as the perturbation is much stronger than the background flow speed. It should be noted here that the wave amplitude is normalised to its value at the boundary of the flux tube, however, inside the flux tube the amplitude of the wave may be larger than the normalised value, as we are studying body modes which are oscillatory in nature inside the waveguide. An example of the eigenfunctions for the slow body mode ($m=1$) with $k=1.76$ and $\omega=1.69$, is displayed in Figure \ref{fig:spatial_eigenfuncs_amplituderatio} for the scaling of $0.01$, which corresponds to $10\%$ of the flow velocity. The spatial structure of the modes are exactly the same for different amplitude ratios considered in this work for each mode, respectively. In the solar atmosphere, rotational flows from observations have reported amplitudes of roughly $1-4$ km s$^{-1}$ for photospheric bright points \citep{Bonet2008}, $5-13$ km s$^{-1}$ for chromospheric swirls \citep{Park2016} and $0.23-0.48$ km s$^{-1}$ in convectively driven photospheric sinks \citep{vargasD2011,Requerey2017}. Therefore, the amplitude ratios studied in this work would correspond to perturbation amplitudes on the order of $0.01-0.2$ km s$^{-1}$ (for the strongest background rotational flows). These perturbation amplitudes are consistent with a range of possible phenomena in the lower solar atmosphere such as acoustic oscillations (p-modes) and granular/supergranular motions \citep{Hart1956, Priest2014, McClure2019}. For a comprehensive review into the rotational flow amplitudes of vortex motions in the lower solar atmosphere, see Table 5 from \citet{Tziotziou2023SSRv}.

Firstly, the spatial structure of the $m=0$ sausage mode is displayed in Figure \ref{fig:Sz_m0_scale_comparison} for two different amplitude ratios, one corresponding to the amplitude ratio $v_{0,\varphi} = 20 \hat{f}$ and the other for the case where $v_{0,\varphi} = 2 \hat{f}$. The presence of a background rotational flow does not significantly affect the spatial structure of $S_z$ when compared to the case of a flux tube with no background rotational flow. The amplitude of $S_z$ increases as the strength of the perturbation is increased, however, this is to be expected. For the regime when the strength of the background flow is much stronger than the amplitude of the wave perturbation, there is a rotational behaviour of the velocity vector as a result of the background flow, however, as the strength of the flow is decreased (or alternatively as the amplitude of the perturbation is increased) then the resulting velocity field resembles that of a classic sausage mode.

\begin{figure*}
\gridline{\fig{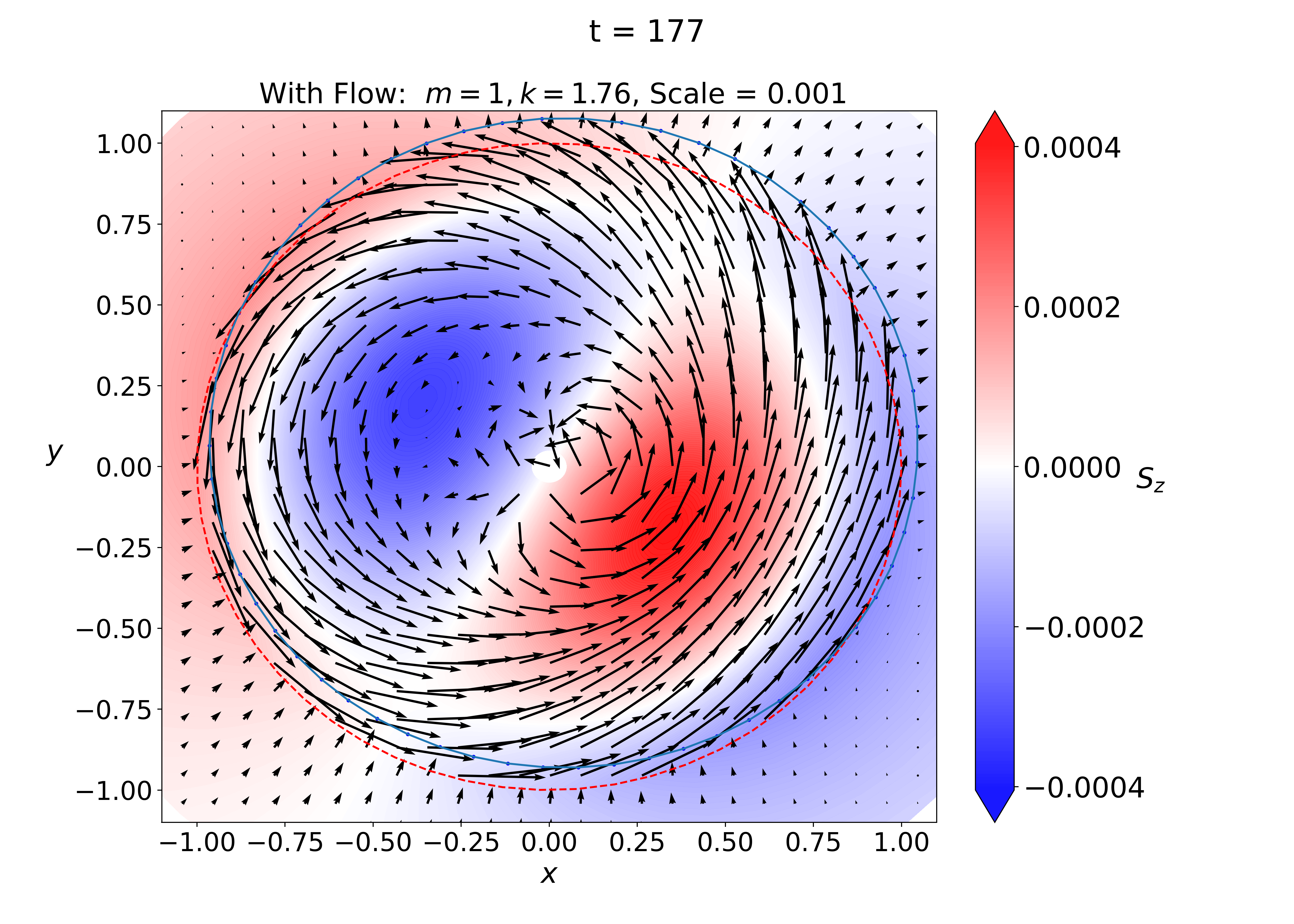}{0.33\textwidth}{(a)}
          \fig{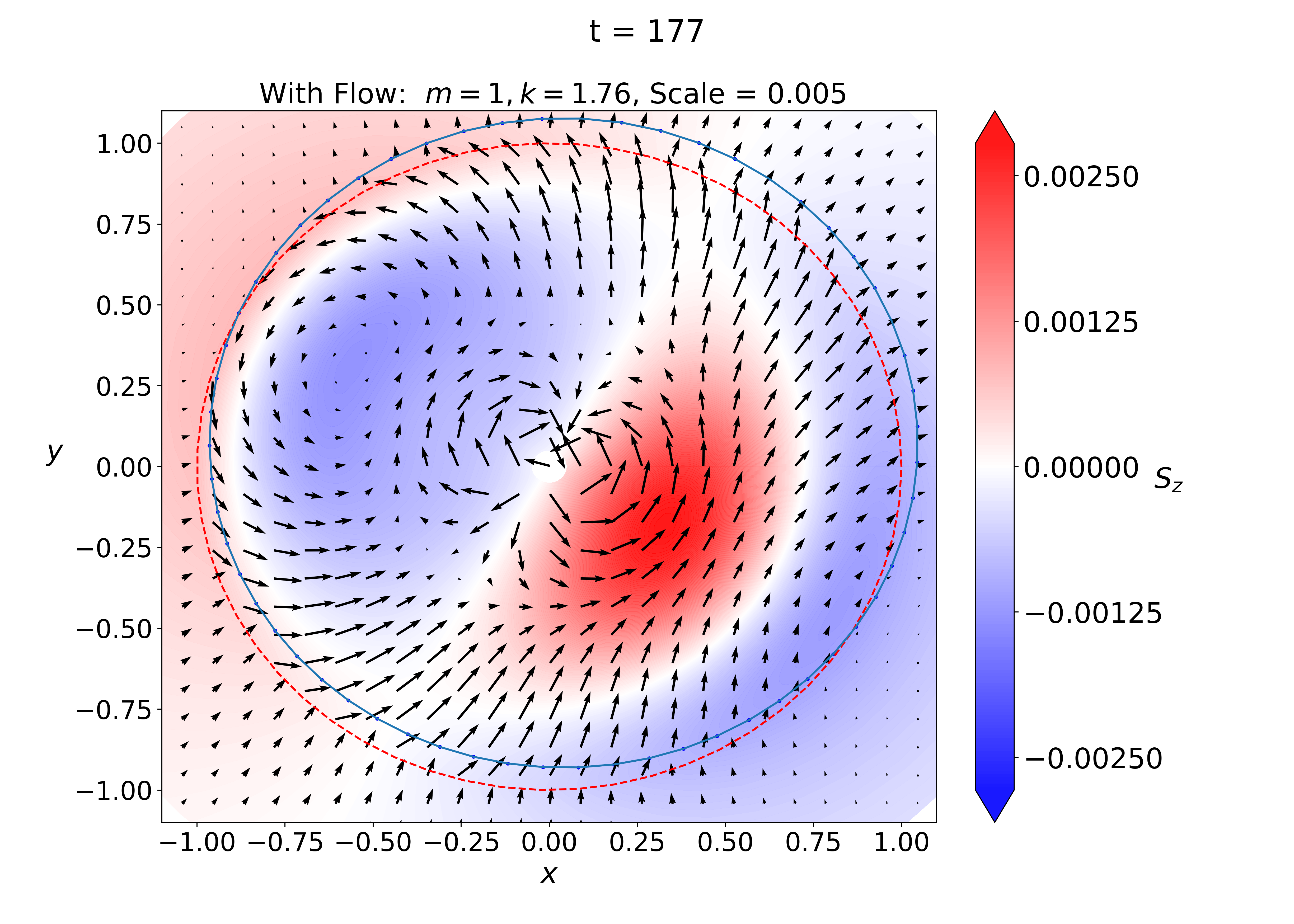}{0.33\textwidth}{(b)}          
          \fig{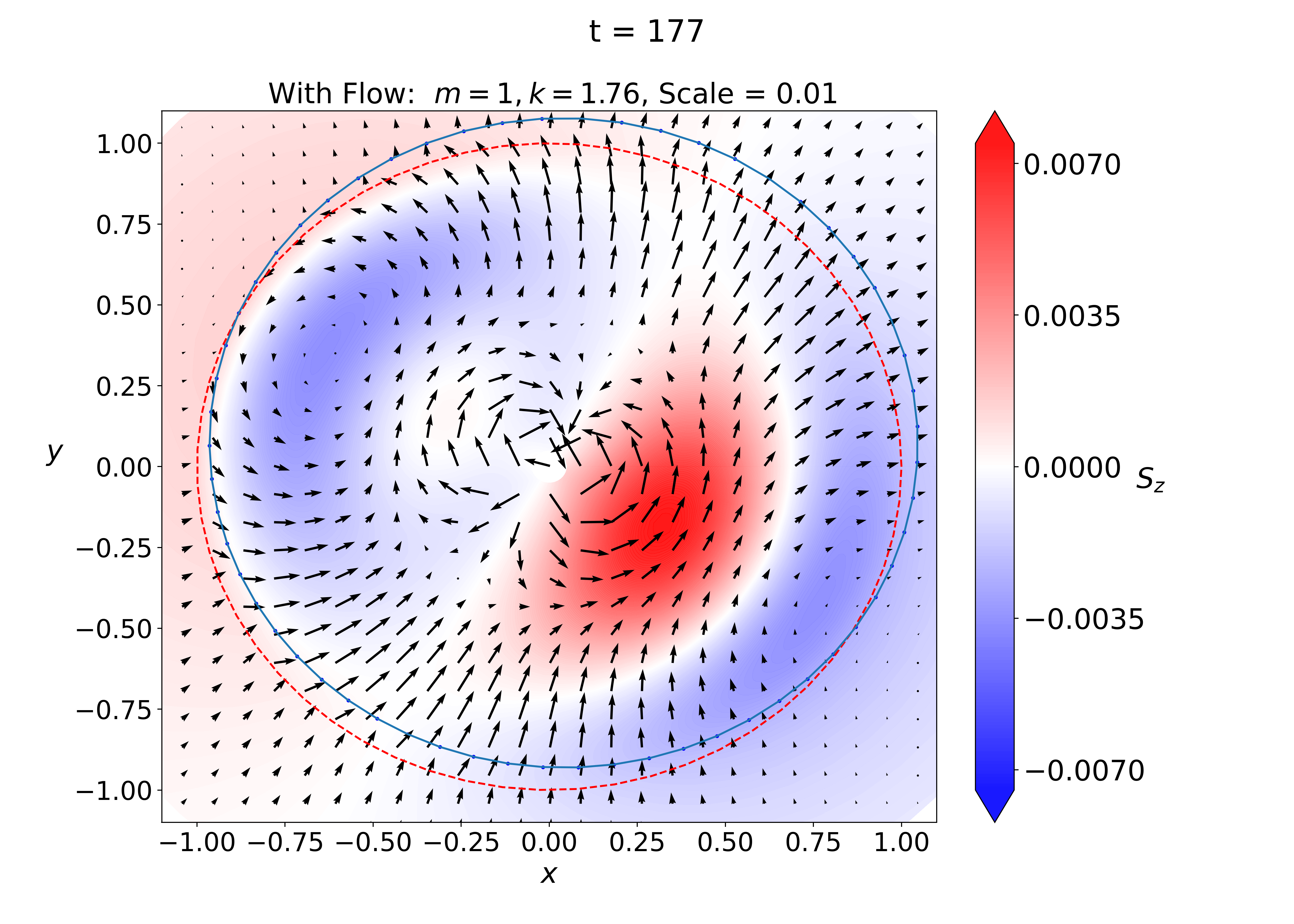}{0.33\textwidth}{(c)}
          }          
\gridline{\fig{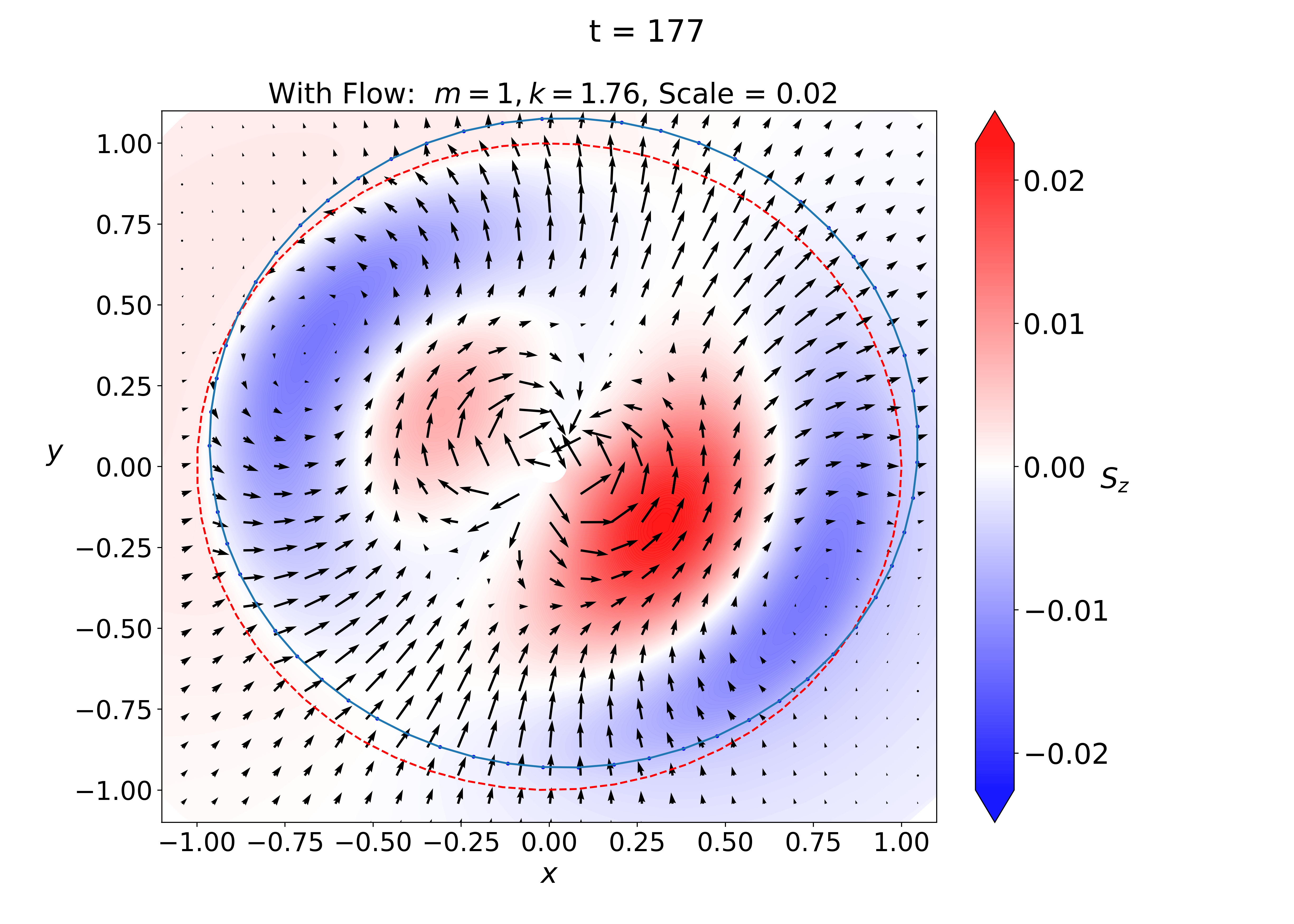}{0.33\textwidth}{(d)}
        \fig{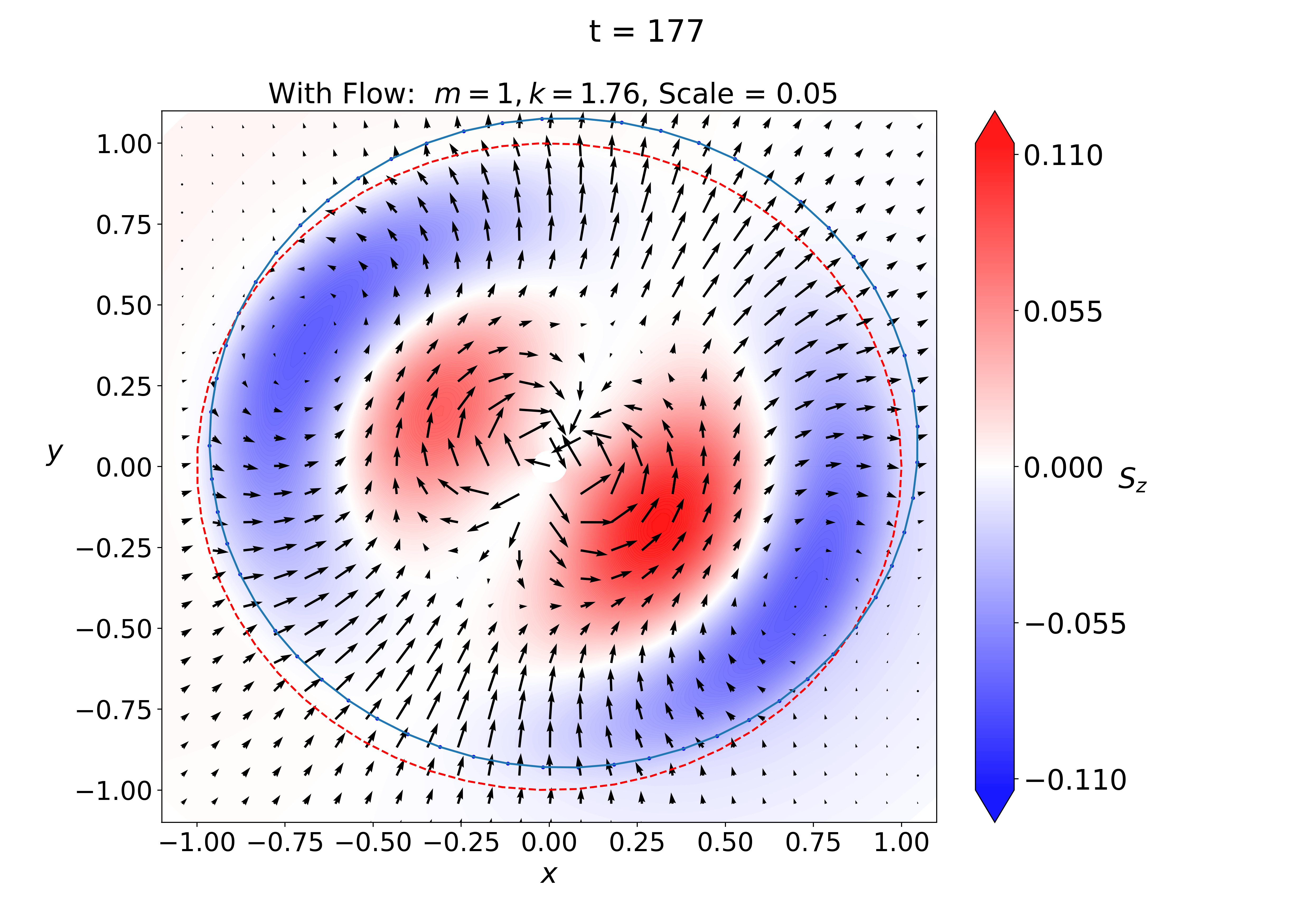}{0.33\textwidth}{(e)}
        \fig{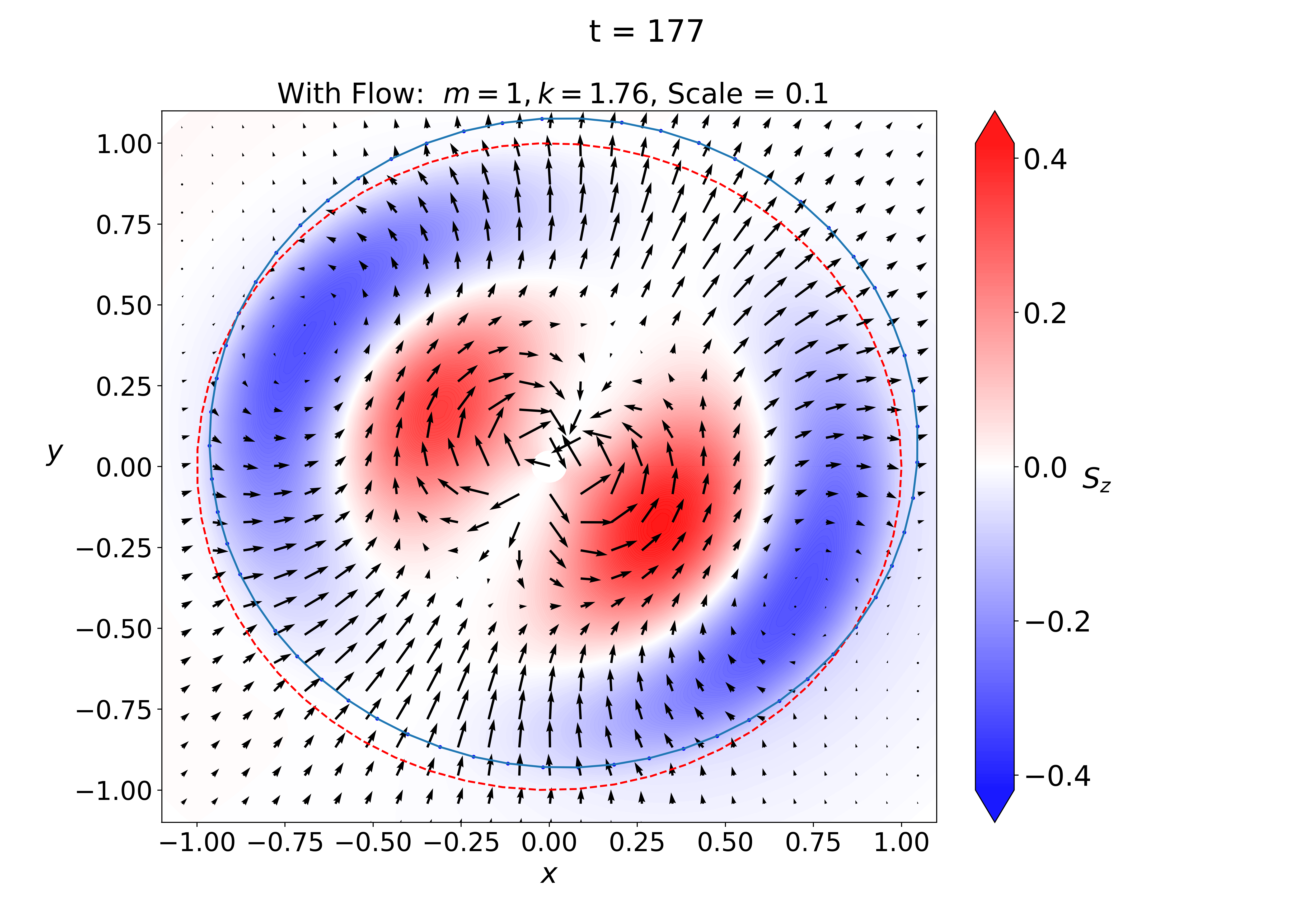}{0.33\textwidth}{(f)}
          }
\caption{Spatial profiles of the vertical component of the Poynting flux $S_z$ for the kink mode with $m=1$ in a rotating magnetic flux tube at the same snapshot in time ($t=177$ in arbitrary units). The background rotational flow profile is given by $v_{0,\varphi} = 0.1r$ such that the maximum amplitude of the rotational flow is $0.1$. We display $S_z$ for varying amplitude ratios ranging from (a) $v_{0,\varphi} = 100\hat{f}$, (b) $v_{0,\varphi} = 20\hat{f}$, (c) $v_{0,\varphi} = 10\hat{f}$, (d) $v_{0,\varphi} = 5\hat{f}$, (e) $v_{0,\varphi} = 2\hat{f}$ and (f) $v_{0,\varphi} = \hat{f}$, where $\hat{f}$ denotes the strength of the perturbation. The velocity vectors (background flow plus the perturbation) are scaled respectively for visualisation purposes. The boundary of the flux tube is represented by the solid blue line whereas the equilibrium location of the boundary is shown with the dashed red line. \label{fig:m1_Sz_spatial}}
\end{figure*}
The spatial structure of the kink mode ($m=1$) in a rotating magnetic flux tube is displayed in Figure \ref{fig:m1_Sz_spatial} for varying amplitude ratios of the wave strength with respect to the background flow. Figure \ref{fig:m1_Sz_spatial}(a) corresponds to the regime where the amplitude of the background rotational flow is much greater than the amplitude of the perturbation, and the perturbation amplitude is increased until Figure \ref{fig:m1_Sz_spatial}(f) which displays the case where the amplitude of the background flow and of the perturbation are of equal strength at the boundary. The result of decreasing the strength of the background flow (alternatively increasing the strength of the perturbation) is not only visible in the velocity streamlines, however, can also be seen to affect the value of $S_z$. As expected, when the wave amplitude is weak and the flow is dominant, the $S_z$ value is very small (as the background magnetic flux tube has no associated $S_z$). However, when the amplitude of the wave is increased, the resulting Poynting flux also increases and the spatial pattern of the slow body kink mode becomes more obvious throughout the volume of the flux tube. Interestingly, for values of the amplitude ratio when the flow is a factor of $5-20$ times greater than the strength of the wave perturbation, there is a clear asymmetry in the azimuthal direction of the vertical Poynting flux. In some spatial locations, there are even `null' points where the vertical Poynting flux becomes zero.

\begin{figure*}
\gridline{\fig{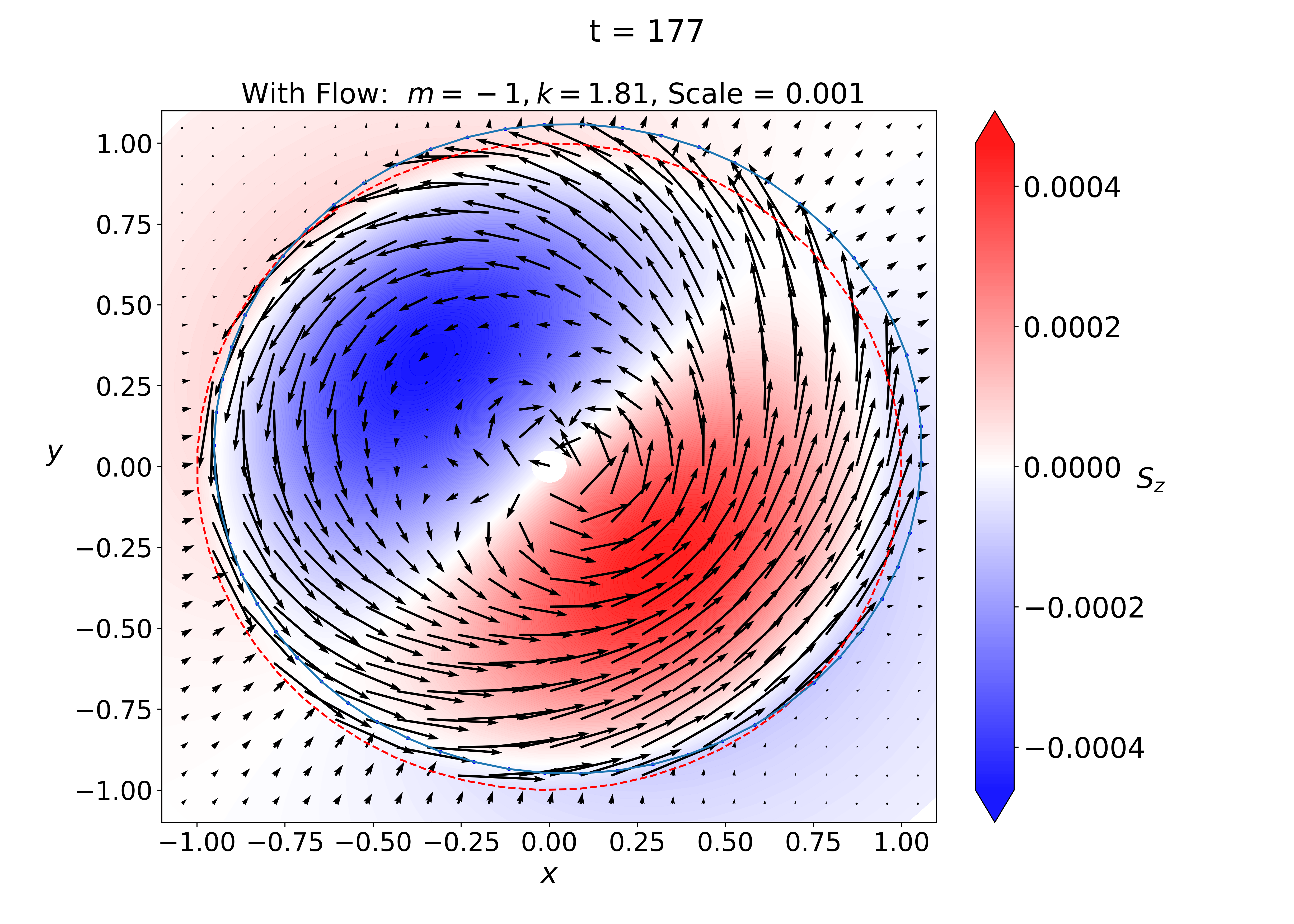}{0.33\textwidth}{(a)}
          \fig{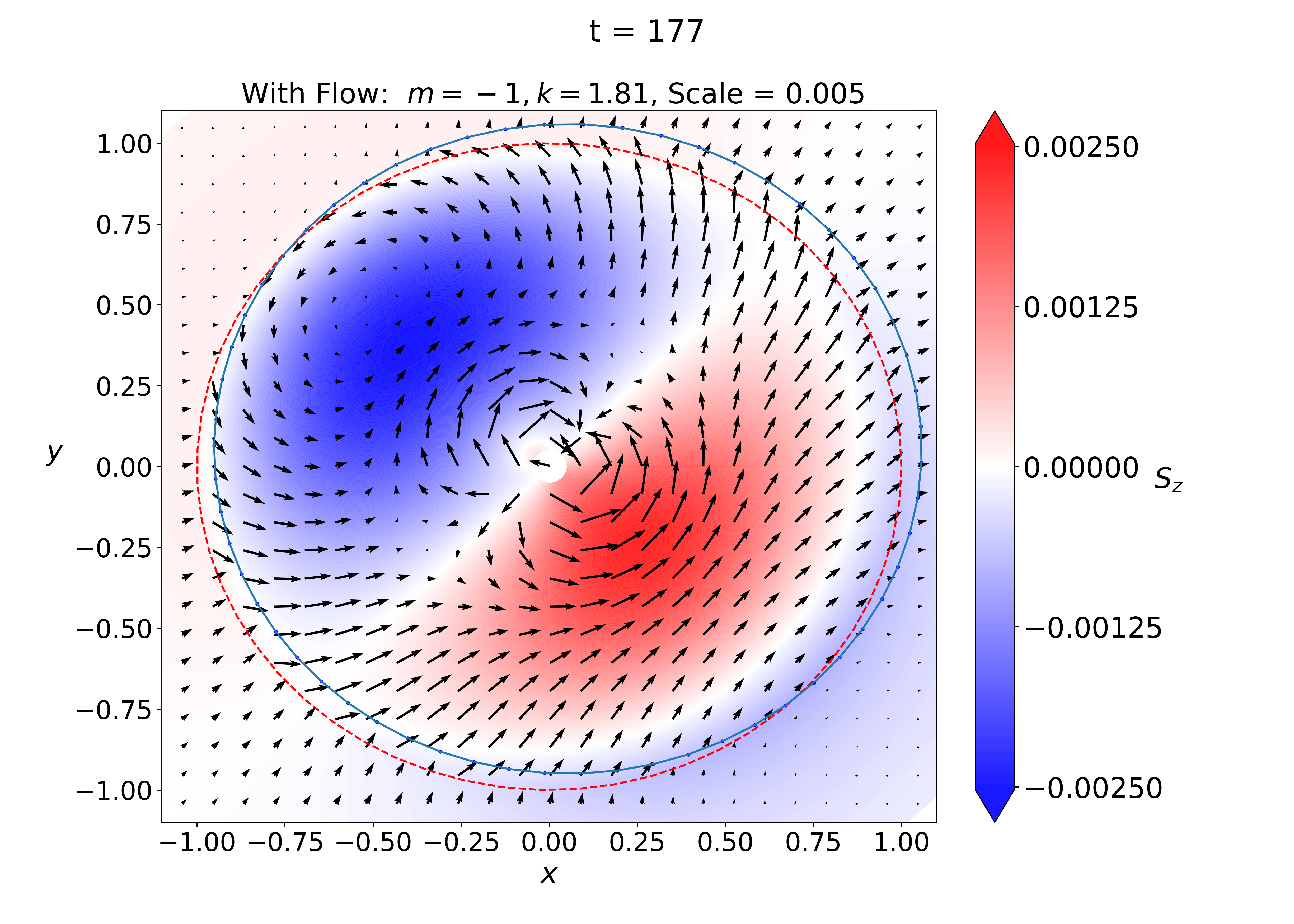}{0.33\textwidth}{(b)}          
          \fig{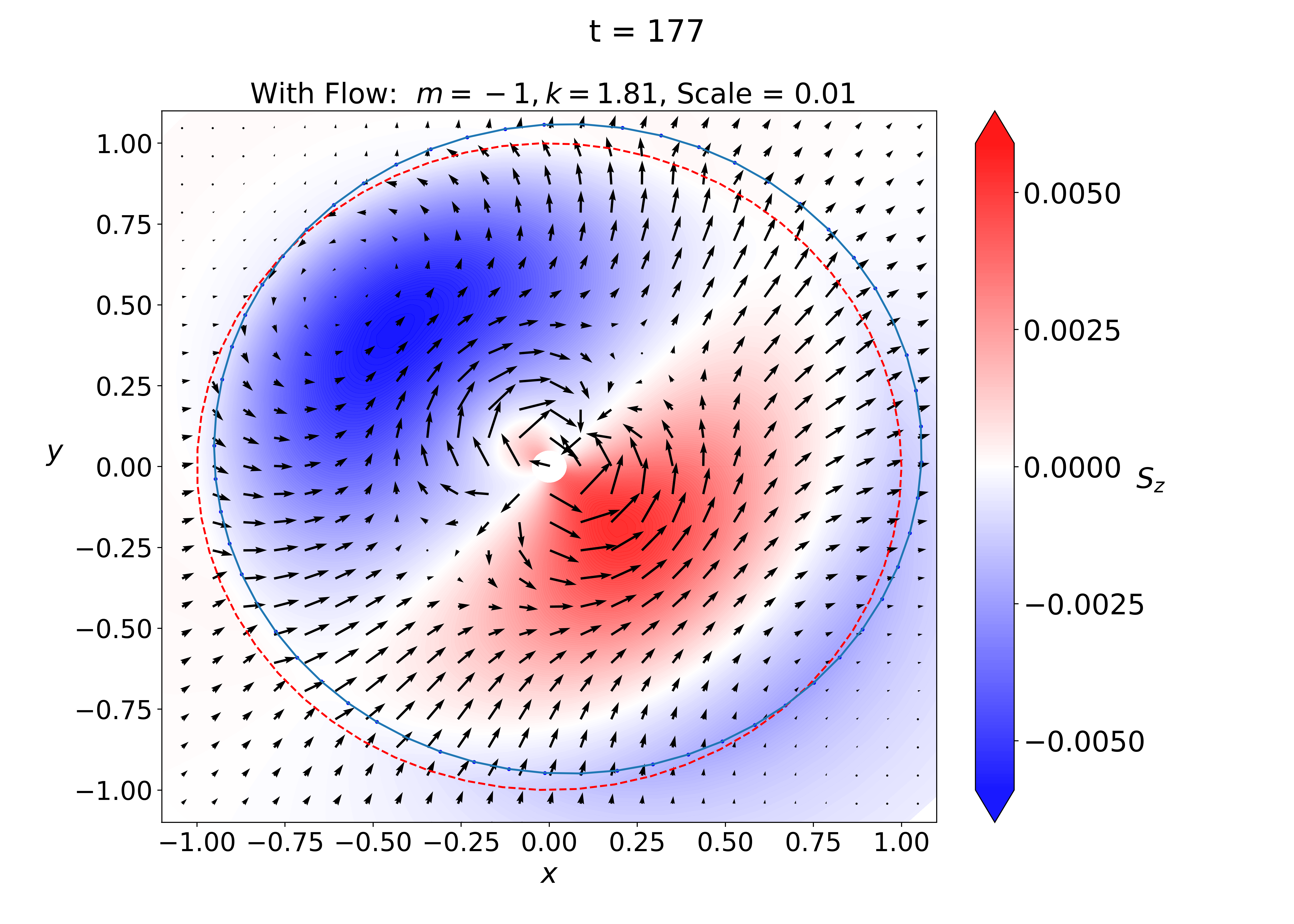}{0.33\textwidth}{(c)}
          }          
\gridline{\fig{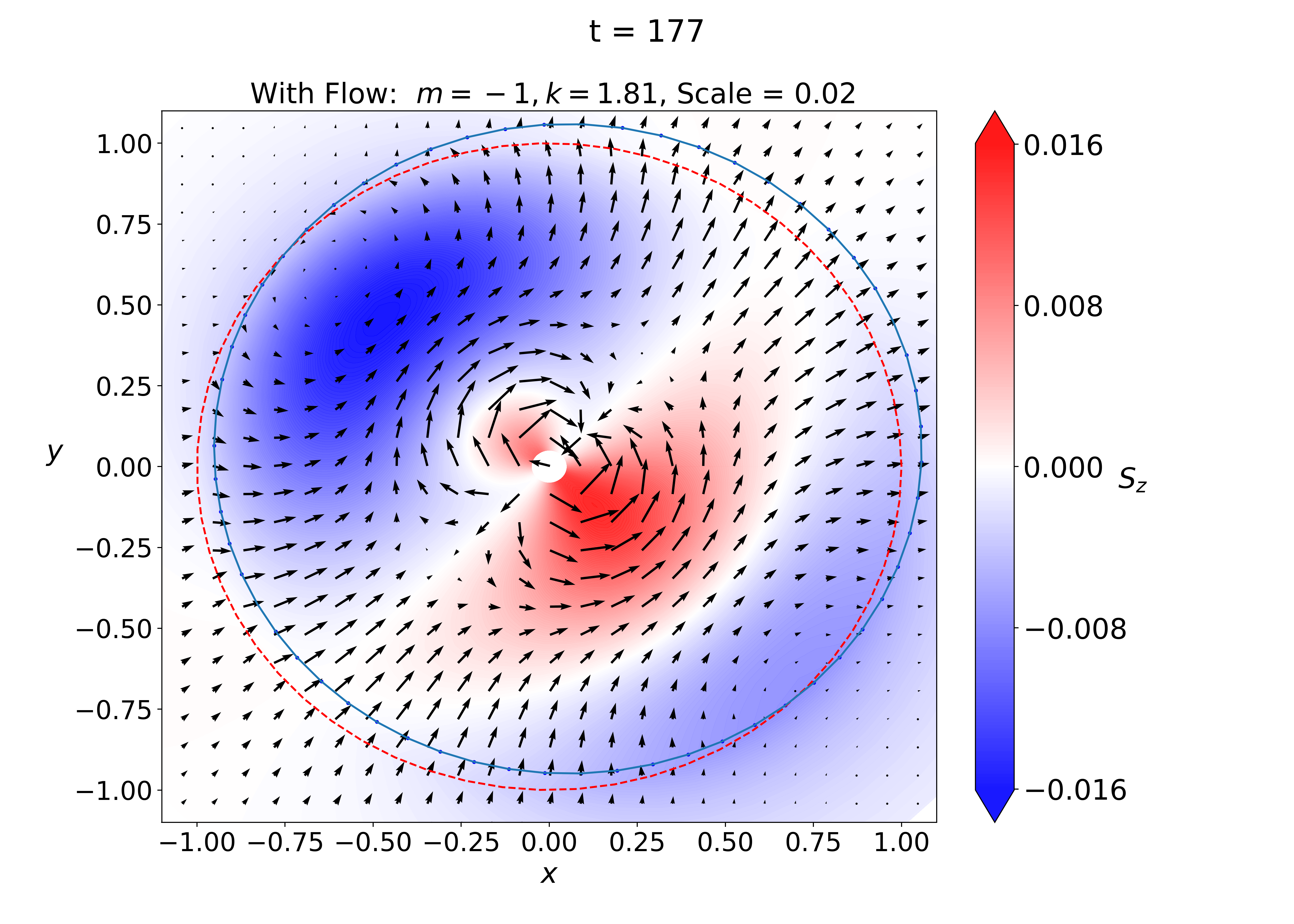}{0.33\textwidth}{(d)}
        \fig{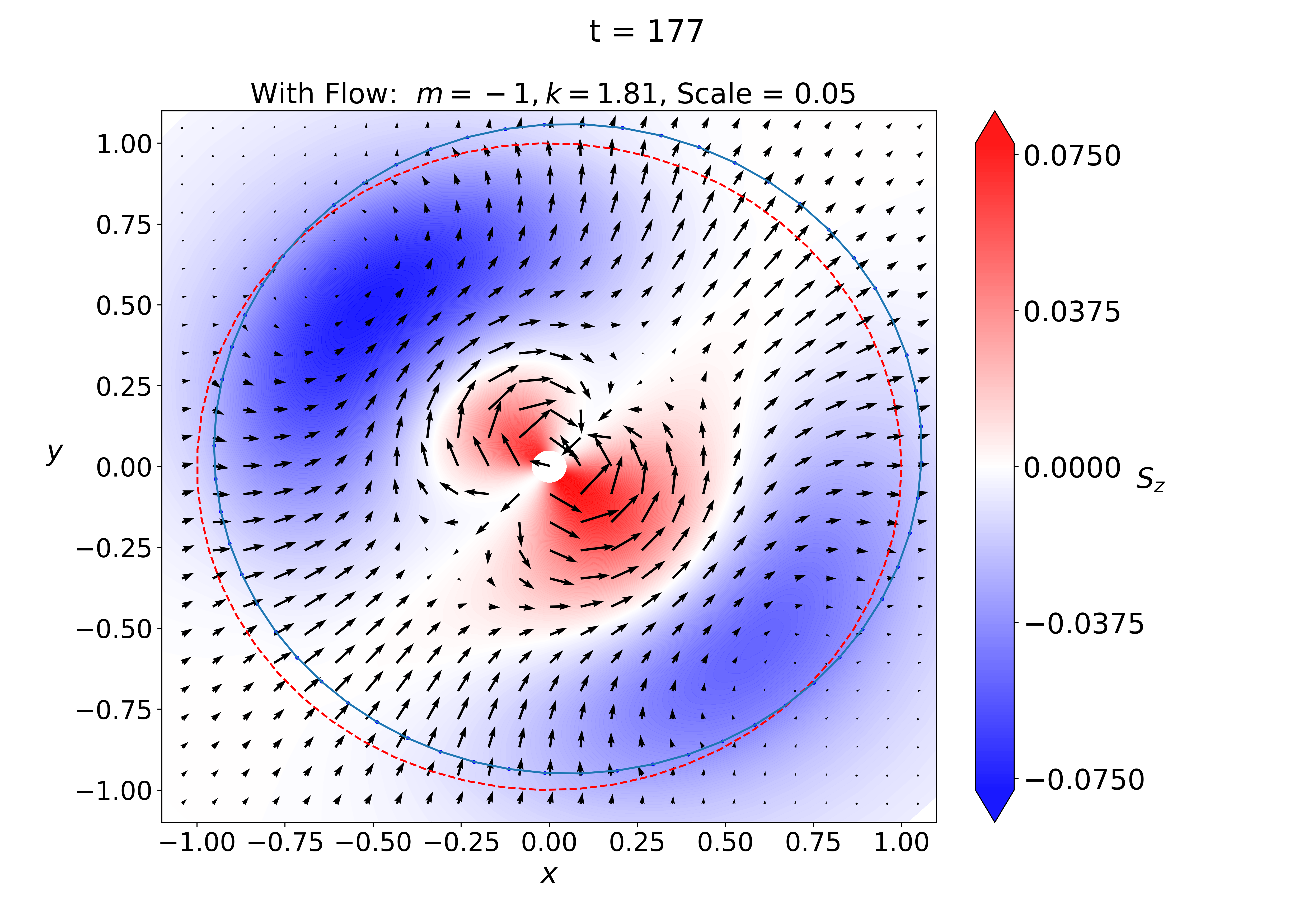}{0.33\textwidth}{(e)}
        \fig{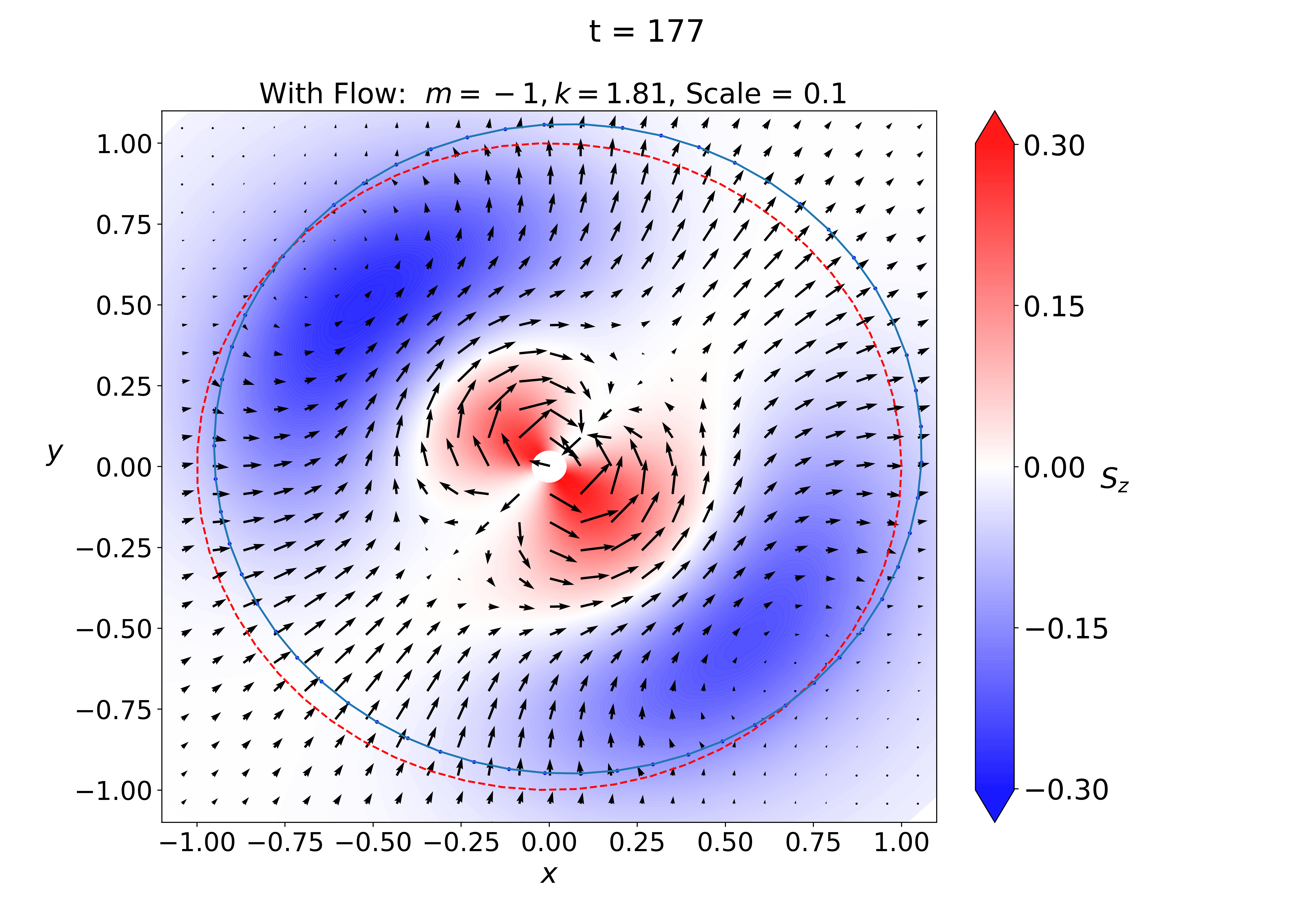}{0.33\textwidth}{(f)}
          }
\caption{Same as Figure \ref{fig:m1_Sz_spatial} but for the $m=-1$ kink mode. \label{fig:mneg1_Sz_spatial}}
\end{figure*}

Similarly, the spatial structure of $S_z$ for the kink mode ($m=-1$) is highlighted in Figure \ref{fig:mneg1_Sz_spatial} for the same values of amplitude ratio. As expected, the magnitude of the vertical Poynting flux increases as the strength of the wave perturbation with respect to the background flow is increased. Moreover, there is also a notable difference in the spatial pattern of the $S_z$ signal as the wave amplitude is increased, similar to the case when $m=1$. It is interesting to note that, even when the strength of the background rotational flow is much greater than the strength of the perturbation, there is still an asymmetry of the velocity vectors for the non-axisymmetric modes ($m=1$ and $m=-1$), whereas, the asymmetry in the $S_z$ signal only becomes visible when the flow amplitude is greater than roughly $20$ times the amplitude of the wave perturbation, highlighted in Figure \ref{fig:mneg1_Sz_spatial} panel (b). This asymmetry in the $S_z$ signal reduces when the amplitude of the wave perturbation becomes comparable to the strength of the background rotational flow shown in Figure \ref{fig:mneg1_Sz_spatial} panel (f).

\begin{figure*}
\gridline{\fig{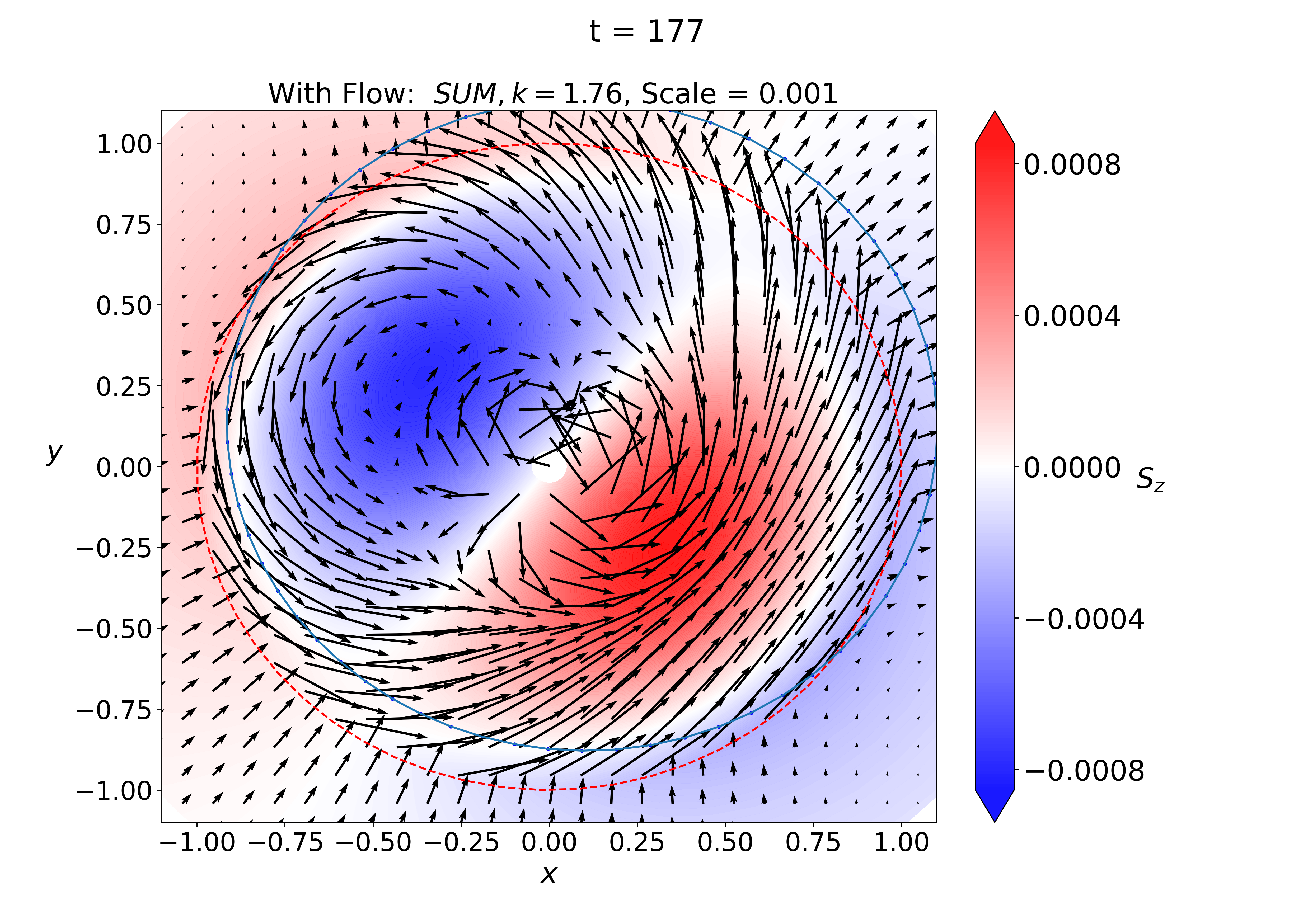}{0.33\textwidth}{(a)}
          \fig{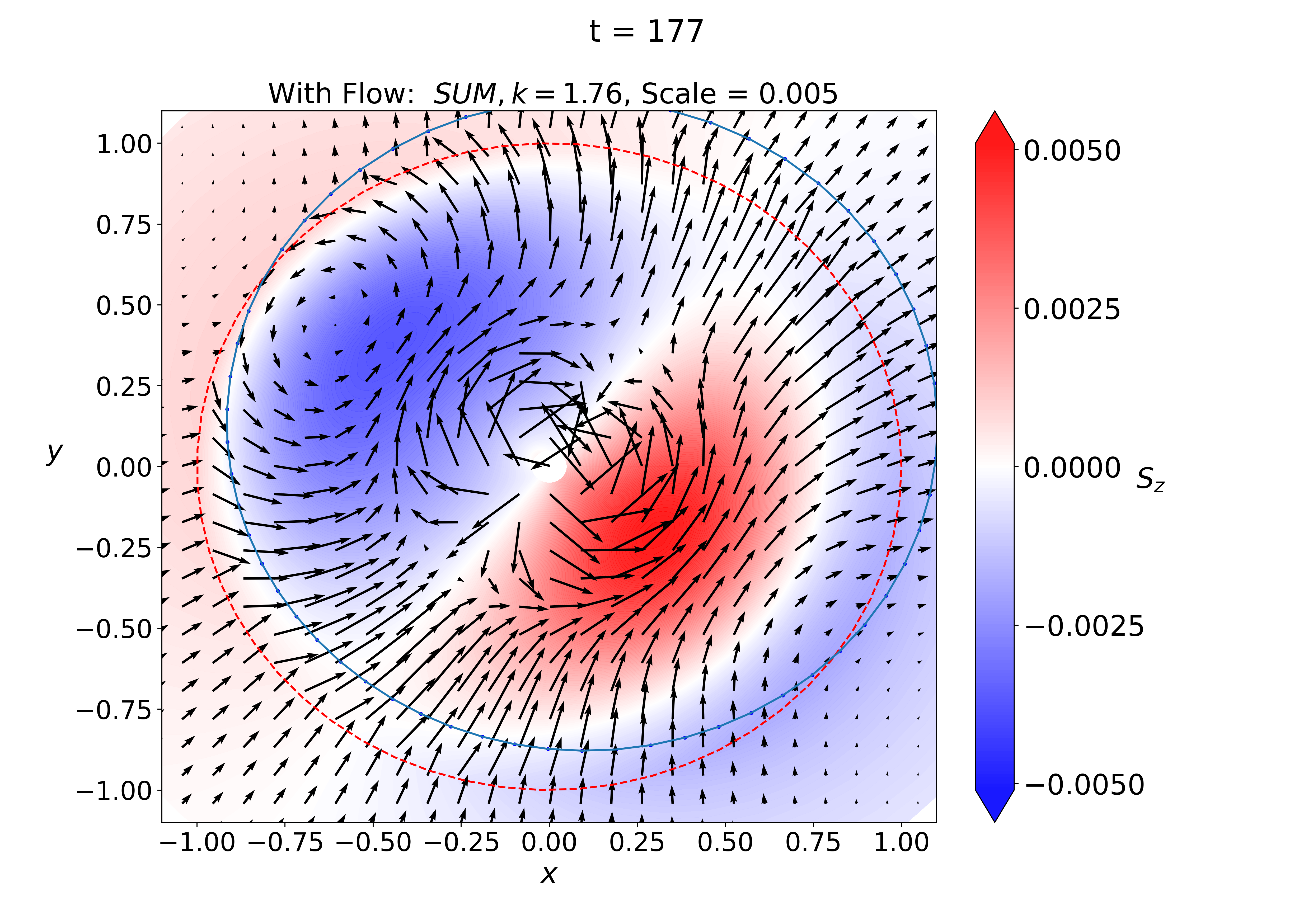}{0.33\textwidth}{(b)}          
          \fig{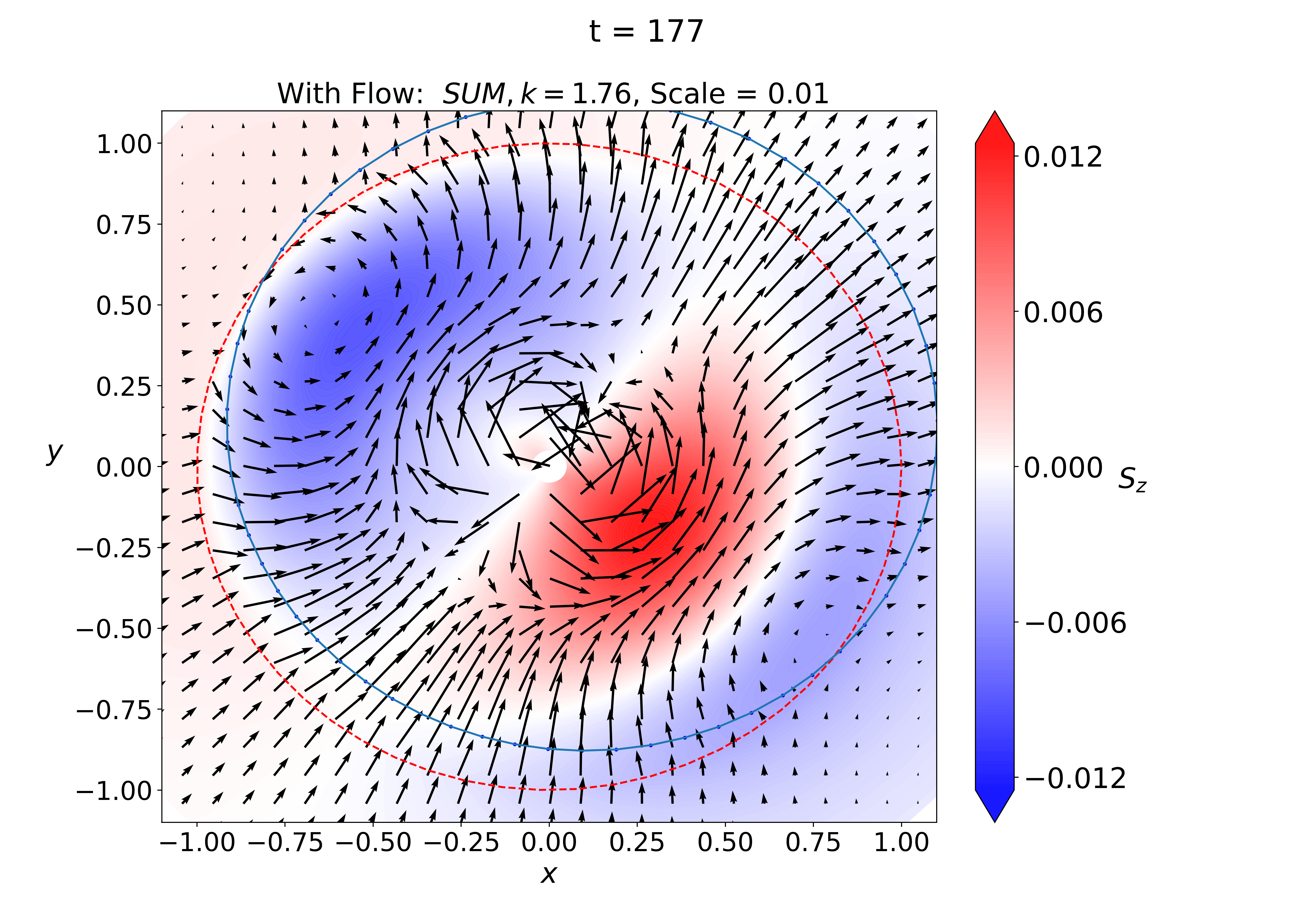}{0.33\textwidth}{(c)}
          }
          
\gridline{\fig{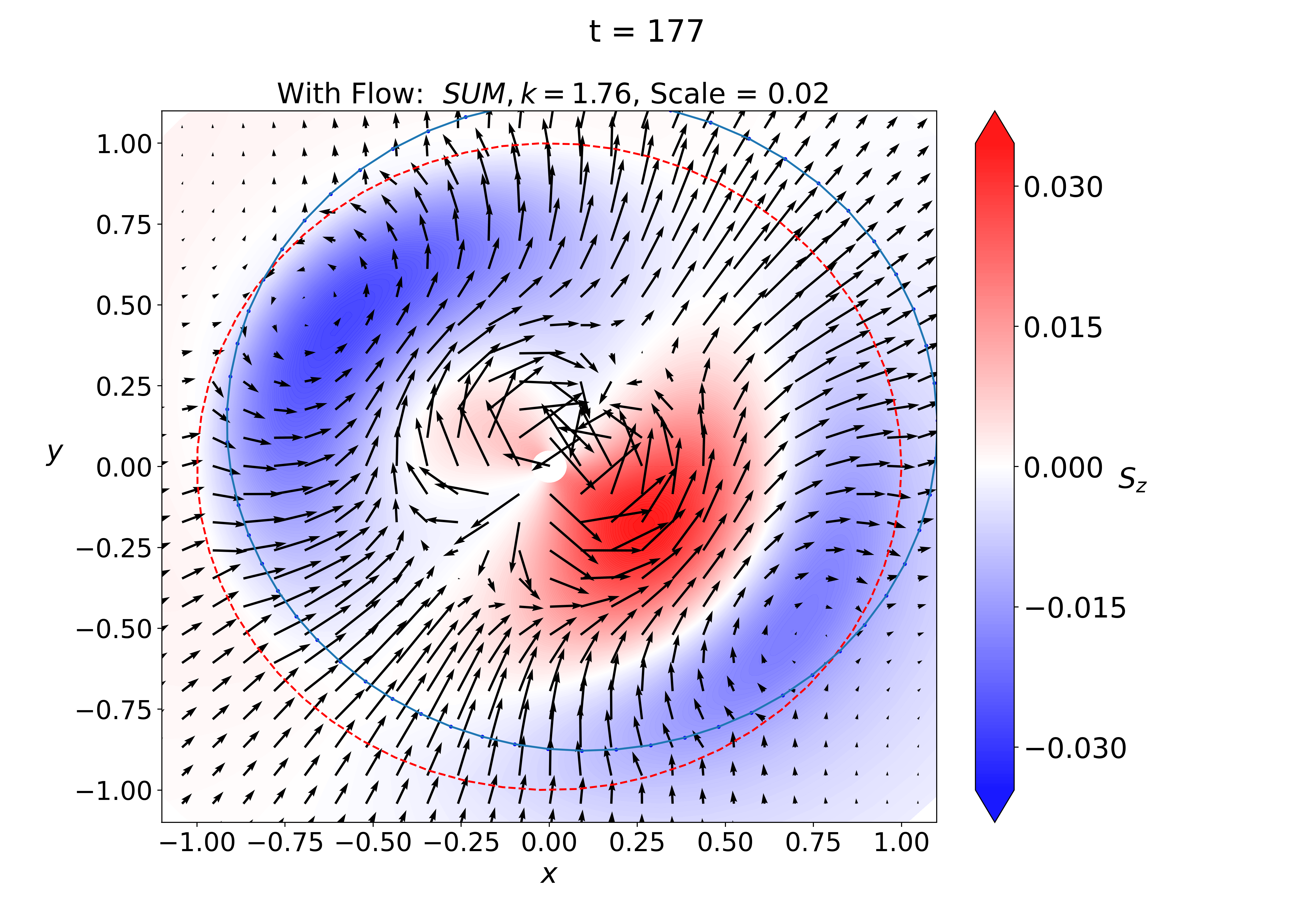}{0.33\textwidth}{(d)}
        \fig{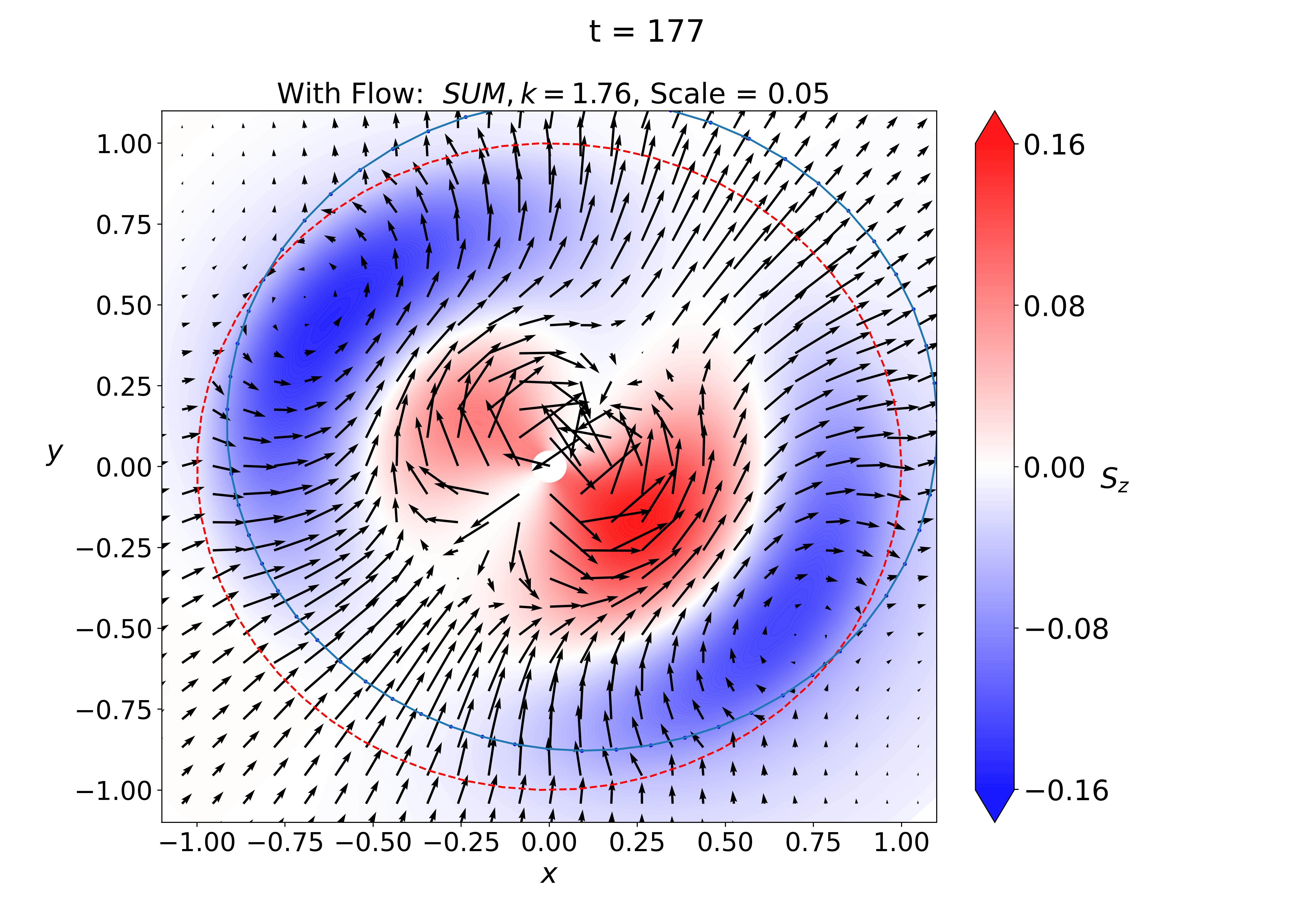}{0.33\textwidth}{(e)}
        \fig{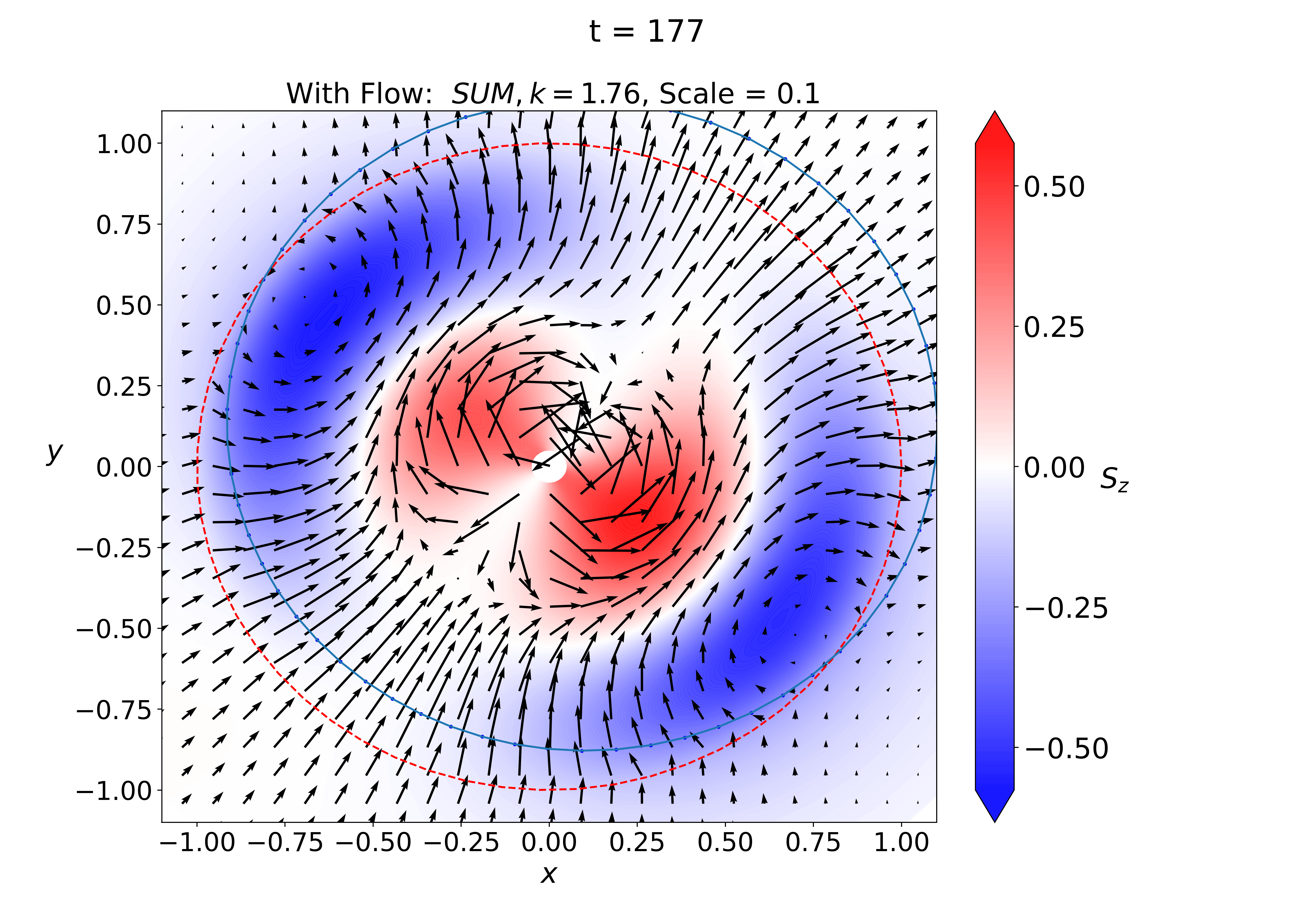}{0.33\textwidth}{(f)}
          }

\caption{Same as Figure \ref{fig:m1_Sz_spatial} but for the linear superposition of the kink modes $m=1$ and $m=-1$. Here we assume that both modes have equal strength as one another. \label{fig:m1sum_Sz_spatial}}
\end{figure*}
The superposition of the signals from both the $m=1$ and $m=-1$ modes are displayed in Figure \ref{fig:m1sum_Sz_spatial} for varying amplitude ratios. The sum of the two modes in this figure assumes that both the $m=1$ and $m=-1$ modes have the same normalisation. In other words, we assume that the strength of the two modes are equal. However, it is likely in reality that both the positive and negative modes are generated with unequal amplitudes, which will change the superposition of the signal. Nonetheless, making the assumption that the kink mode is generated with equal amplitude for both the positive and negative azimuthal wave numbers is instructive when analysing wave modes in vortices in the solar atmosphere. As discussed in \citet{Skirvin2023rotflow}, the presence of a background rotational flow does not permit a standing kink mode in the azimuthal direction. This is highlighted in Figure \ref{fig:m1sum_Sz_spatial} as the movement of the boundary of the flux tube is no longer transverse in one plane, instead, the superposition of the two modes produces a transverse displacement which itself rotates around the central axis over time, resulting in a circularly polarised kink mode.

\begin{figure*}
\gridline{\fig{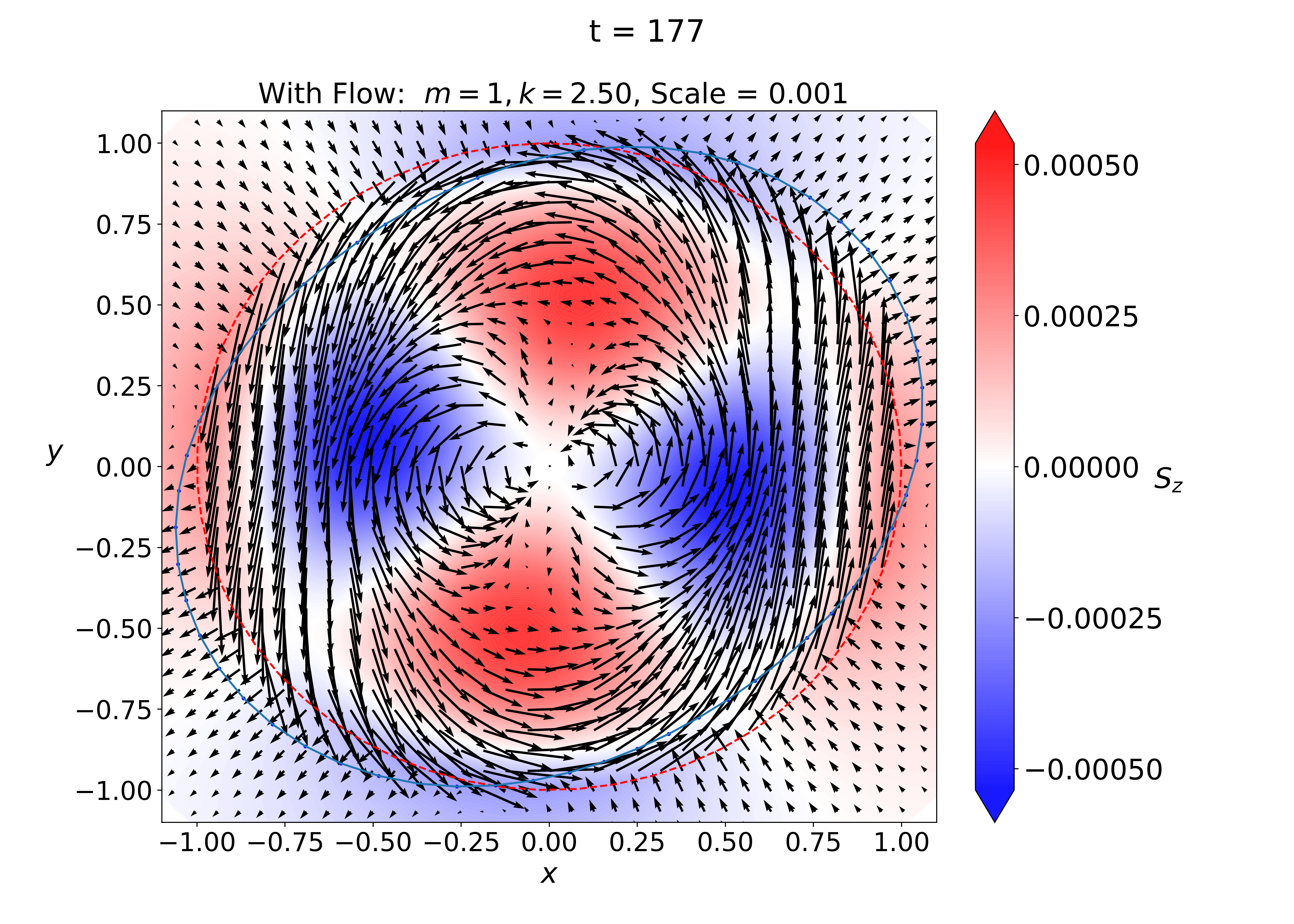}{0.33\textwidth}{(a)}
          \fig{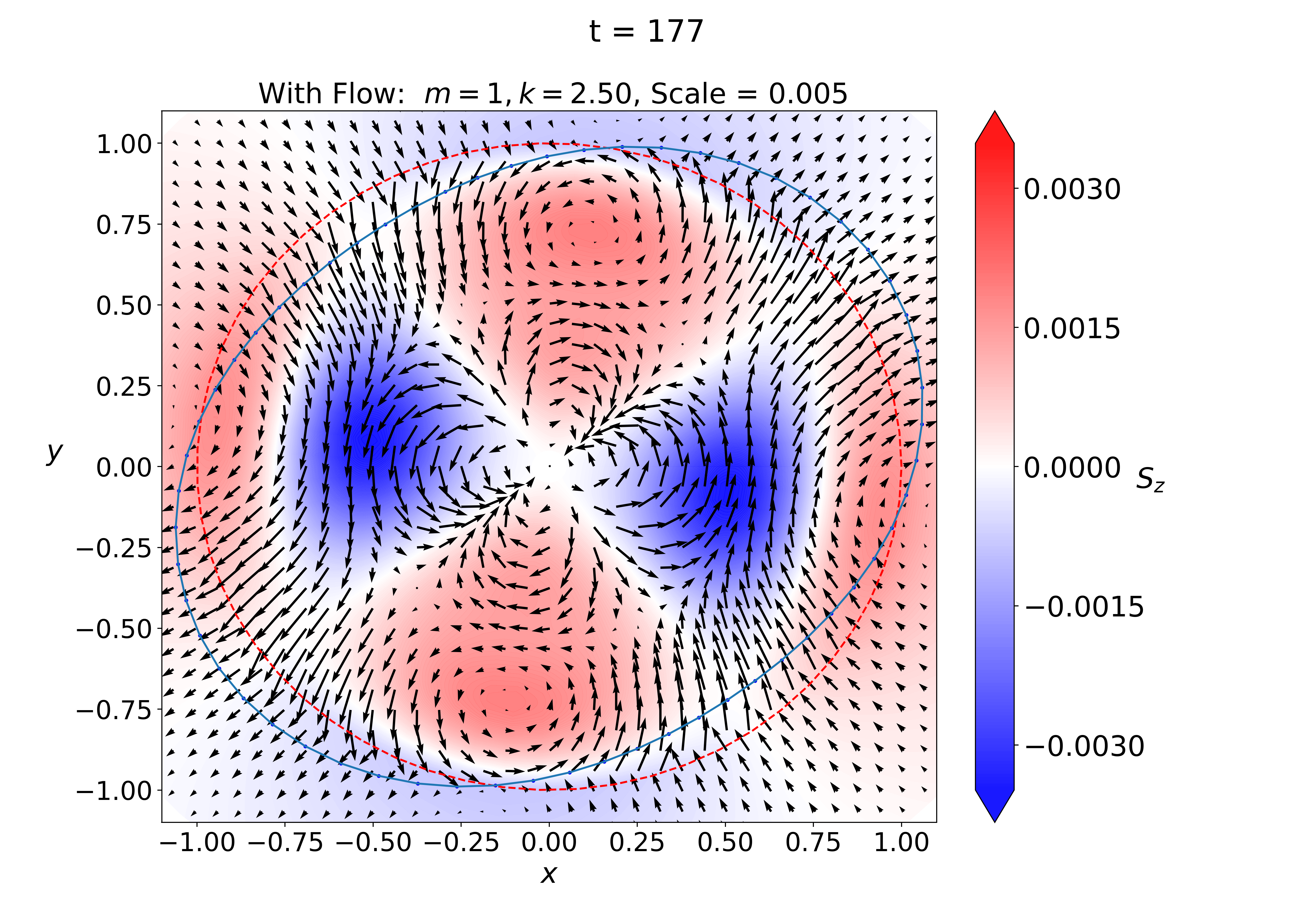}{0.33\textwidth}{(b)}          
          \fig{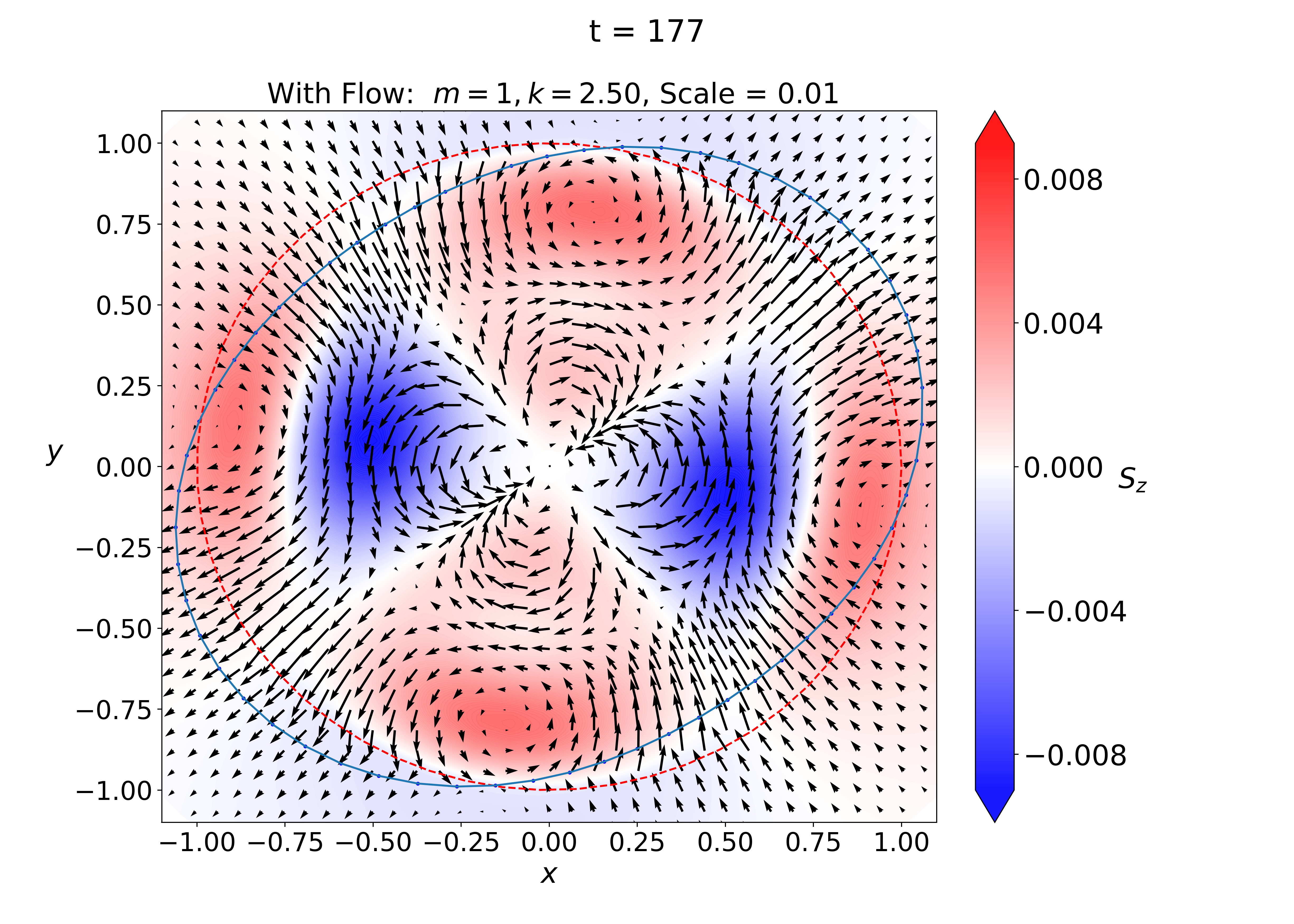}{0.33\textwidth}{(c)}
          }
          
\gridline{\fig{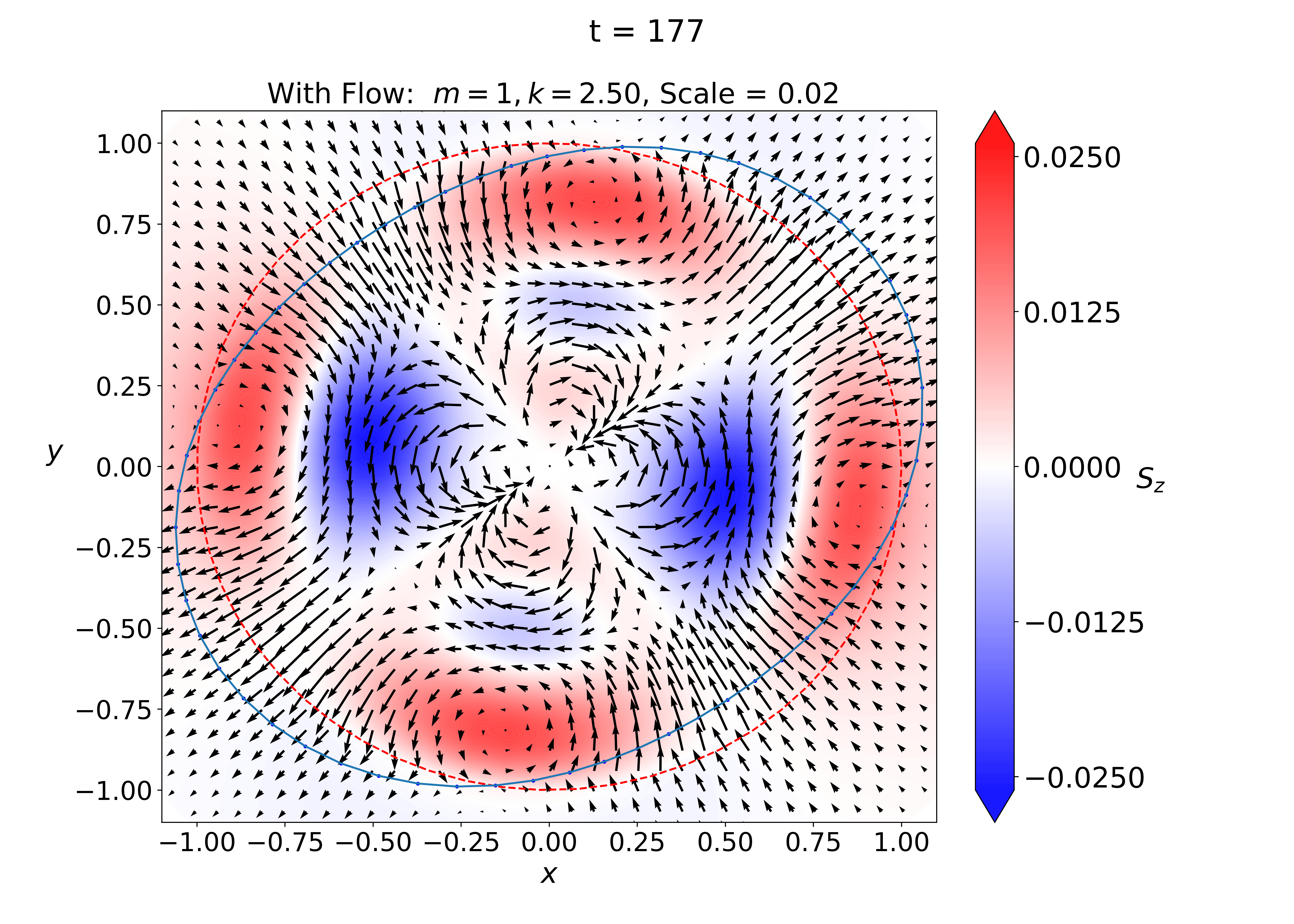}{0.33\textwidth}{(d)}
        \fig{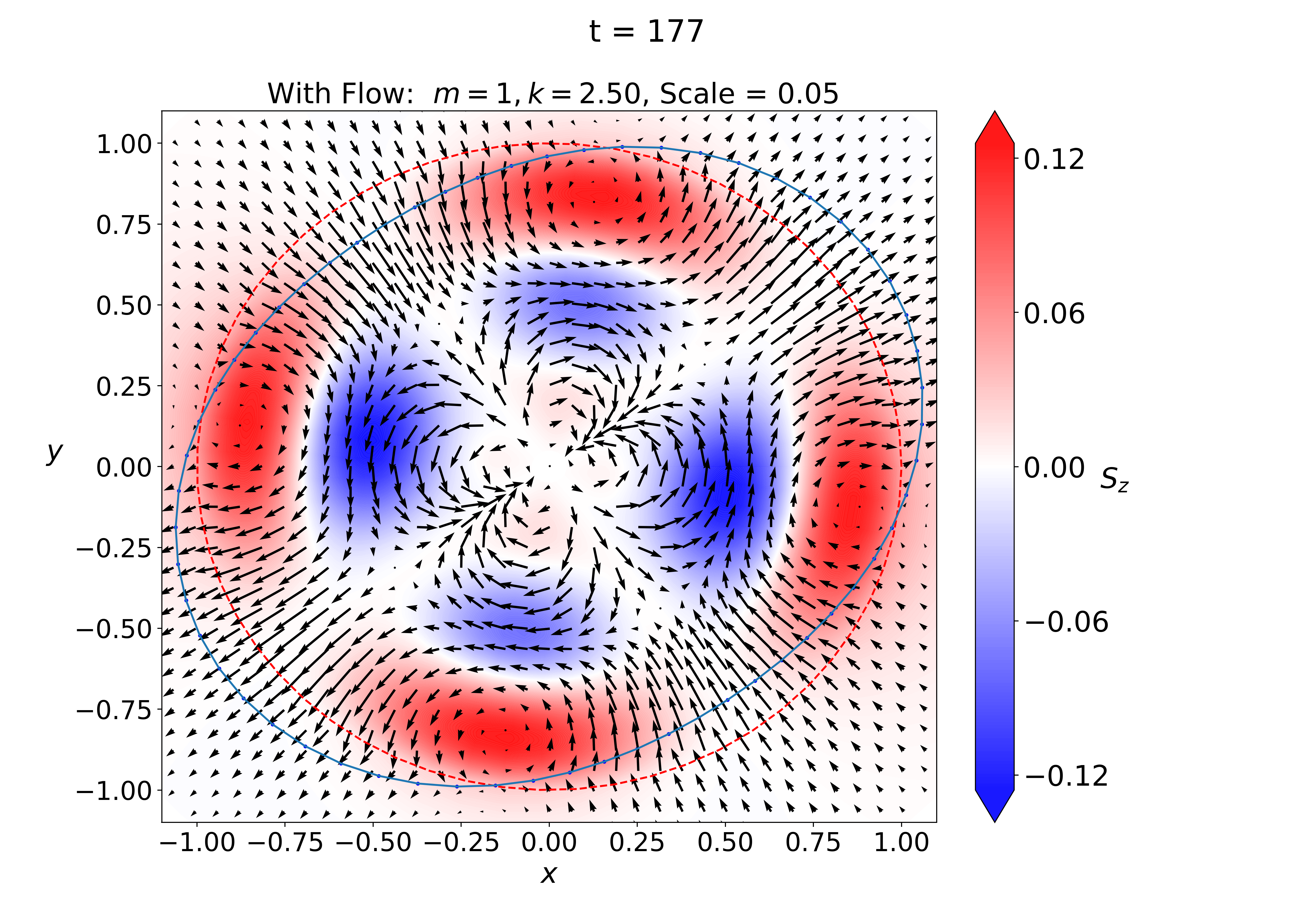}{0.33\textwidth}{(e)}
        \fig{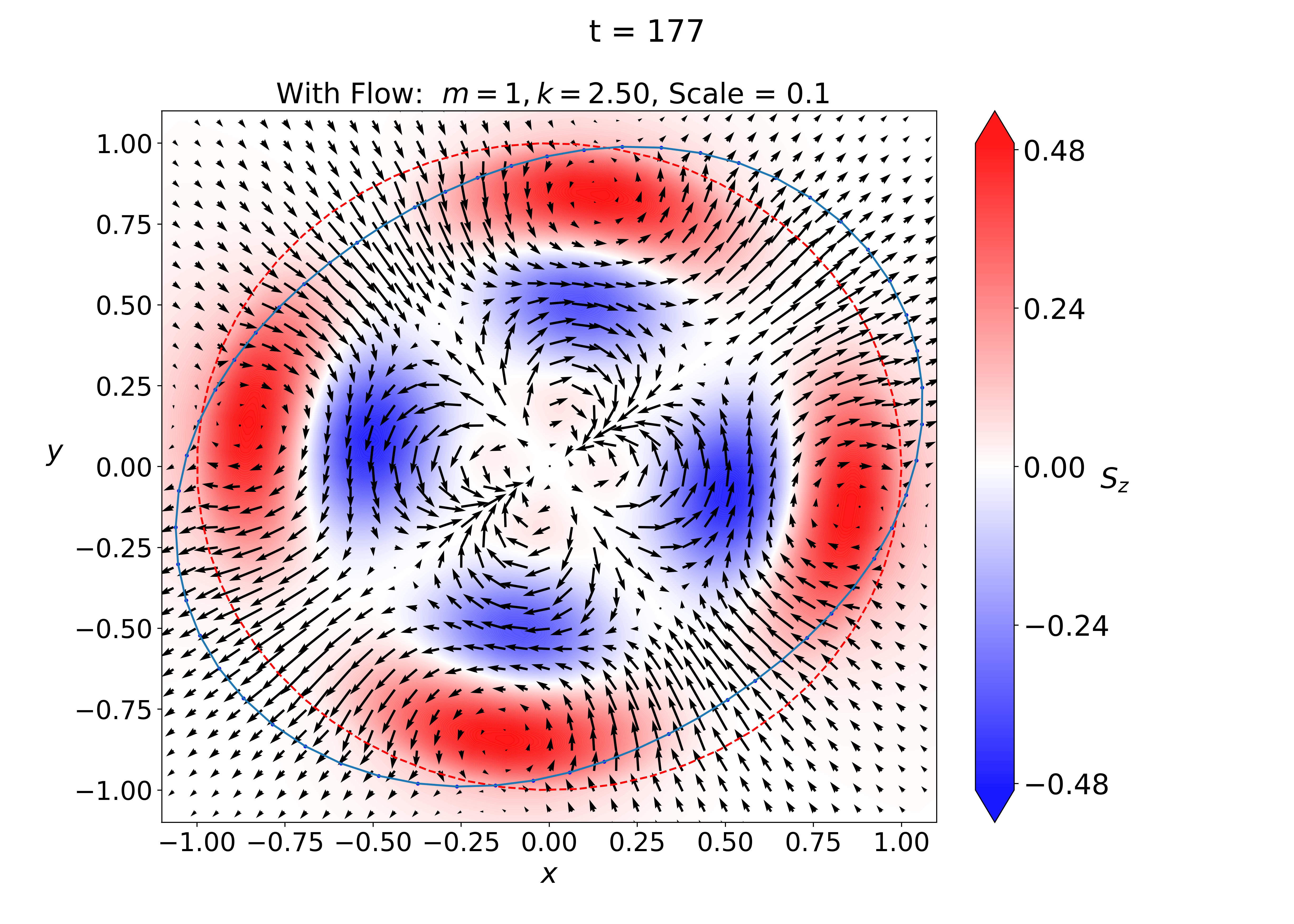}{0.33\textwidth}{(f)}
          }
\caption{Same as Figure \ref{fig:m1_Sz_spatial} but for the $m=2$ fluting mode.\label{fig:m2_Sz_spatial}}
\end{figure*}

\begin{figure*}
\gridline{\fig{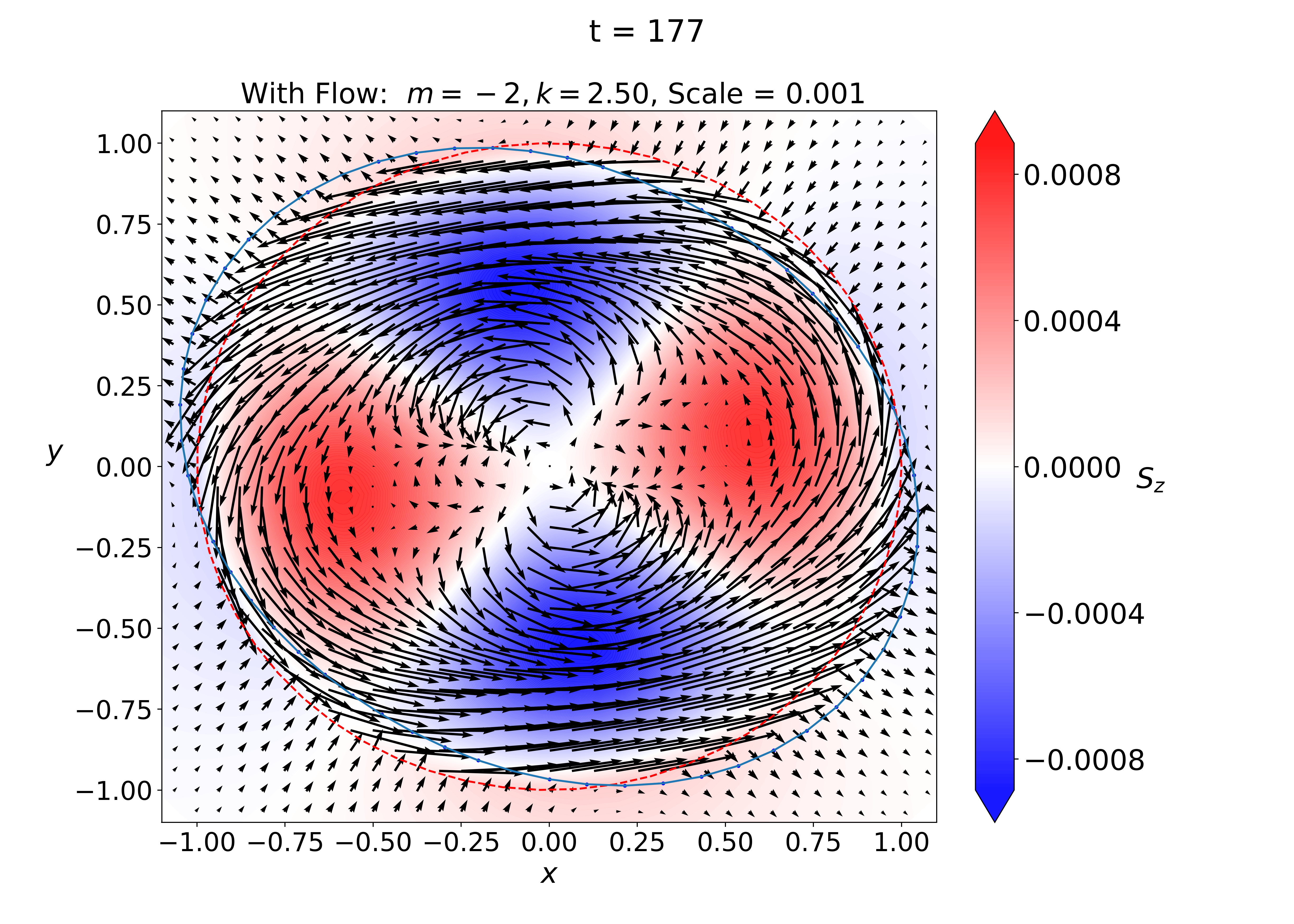}{0.33\textwidth}{(a)}
          \fig{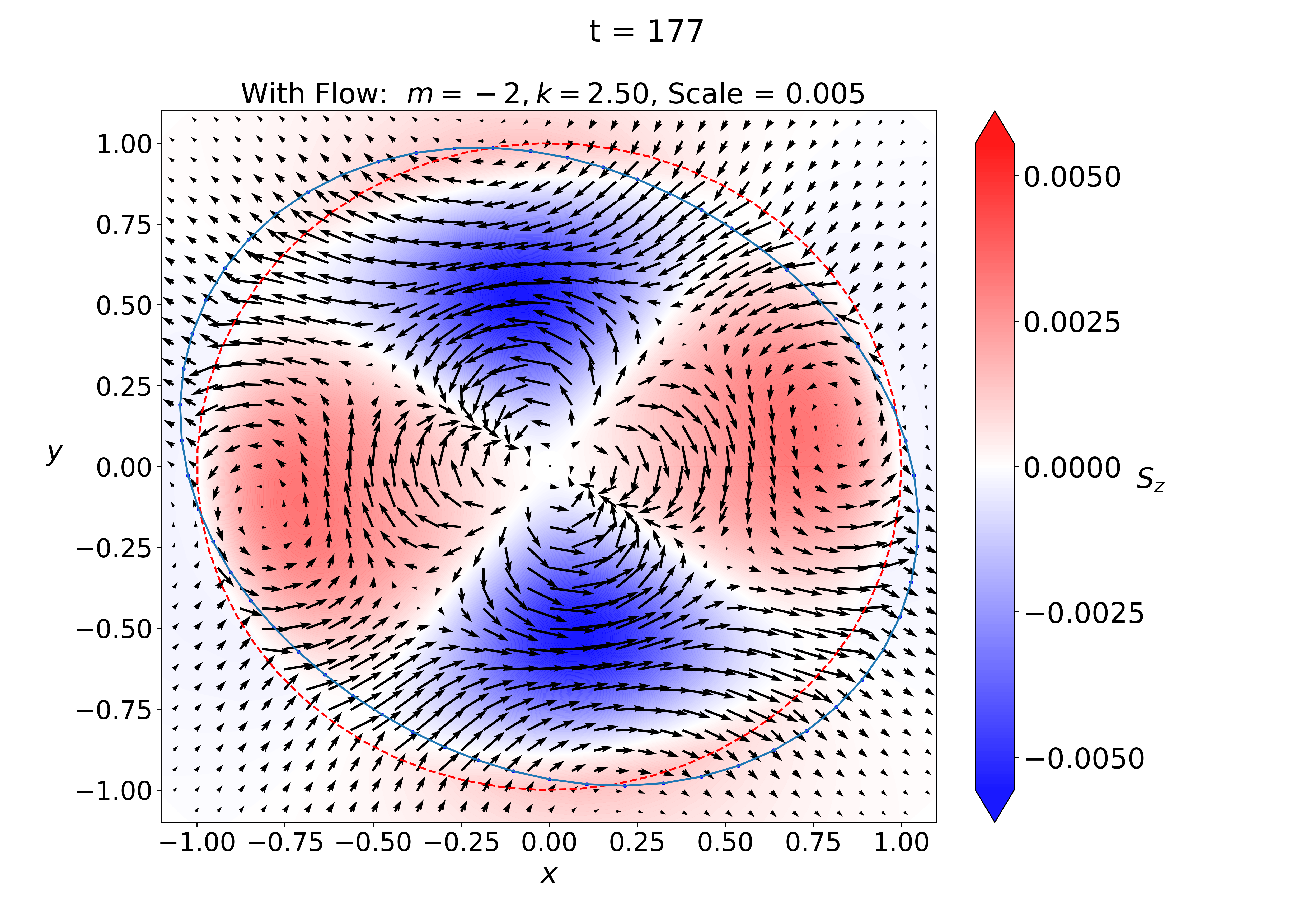}{0.33\textwidth}{(b)}          
          \fig{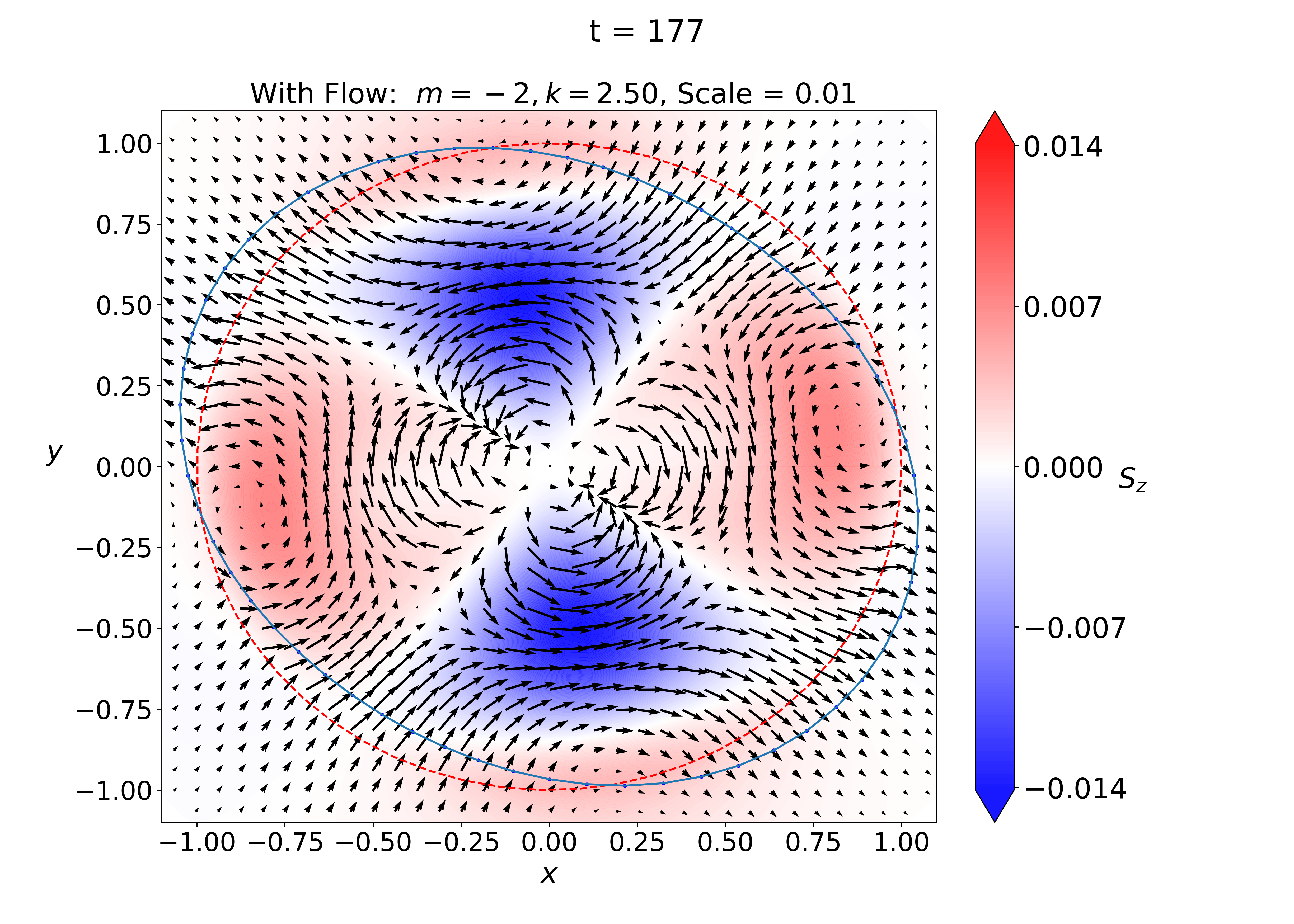}{0.33\textwidth}{(c)}
          }          
\gridline{\fig{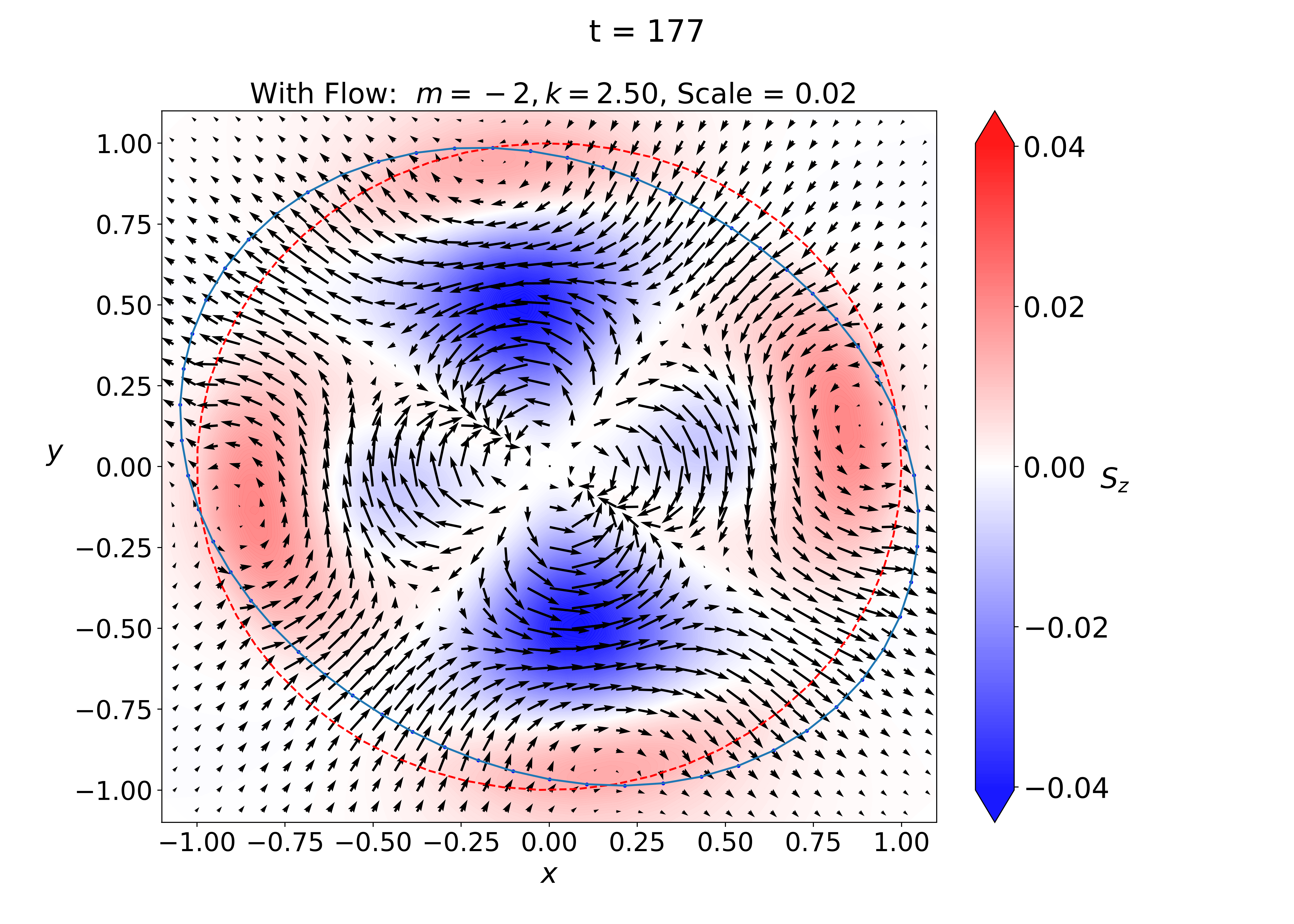}{0.33\textwidth}{(d)}
        \fig{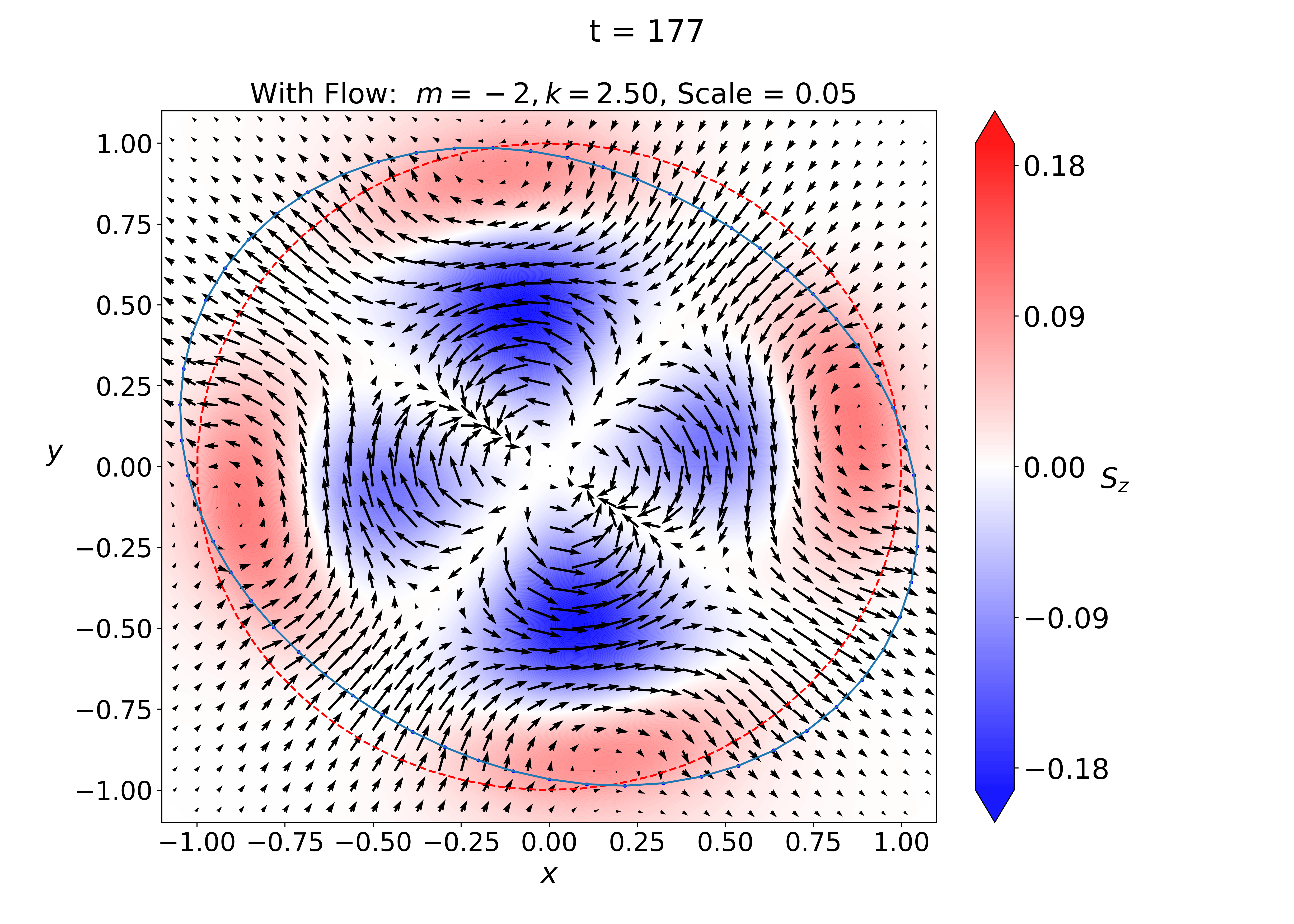}{0.33\textwidth}{(e)}
        \fig{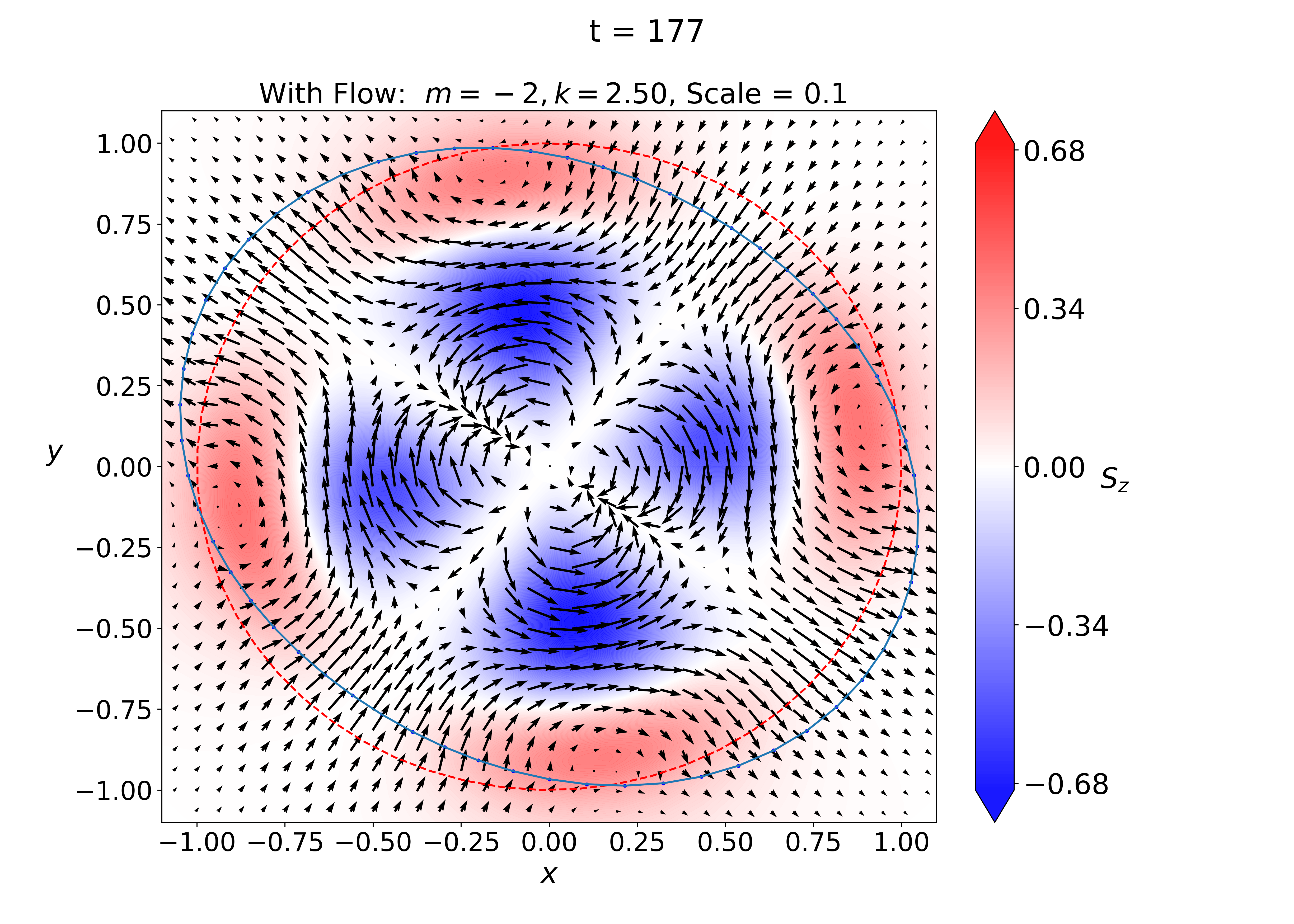}{0.33\textwidth}{(f)}
          }
\caption{Same as Figure \ref{fig:m1_Sz_spatial} but for the $m=-2$ fluting mode.
\label{fig:mneg2_Sz_spatial}}
\end{figure*}

The spatial distributions of $S_z$ for the $m=2$ 
 and $m=-2$ fluting mode are displayed in Figures \ref{fig:m2_Sz_spatial} and \ref{fig:mneg2_Sz_spatial}, respectively. Similar to the kink mode, the pattern of $S_z$ becomes more detailed with a greater range of sub-structuring as the amplitude of the flow is decreased (increased perturbation). The effect of the background rotational flow appears to decrease the complexity of the $S_z$ signal, with a small apparent swirling nature visible for both modes in the case of strong flow. Moreover, there is a visible rotational pattern in the velocity streamlines for cases when the amplitude of the flow is stronger than 10 times the perturbation, shown in Figures \ref{fig:m2_Sz_spatial} and \ref{fig:mneg2_Sz_spatial} panels (a) and (b), which would be measurable in simulations and detectable in observations.

\begin{figure*}
\gridline{\fig{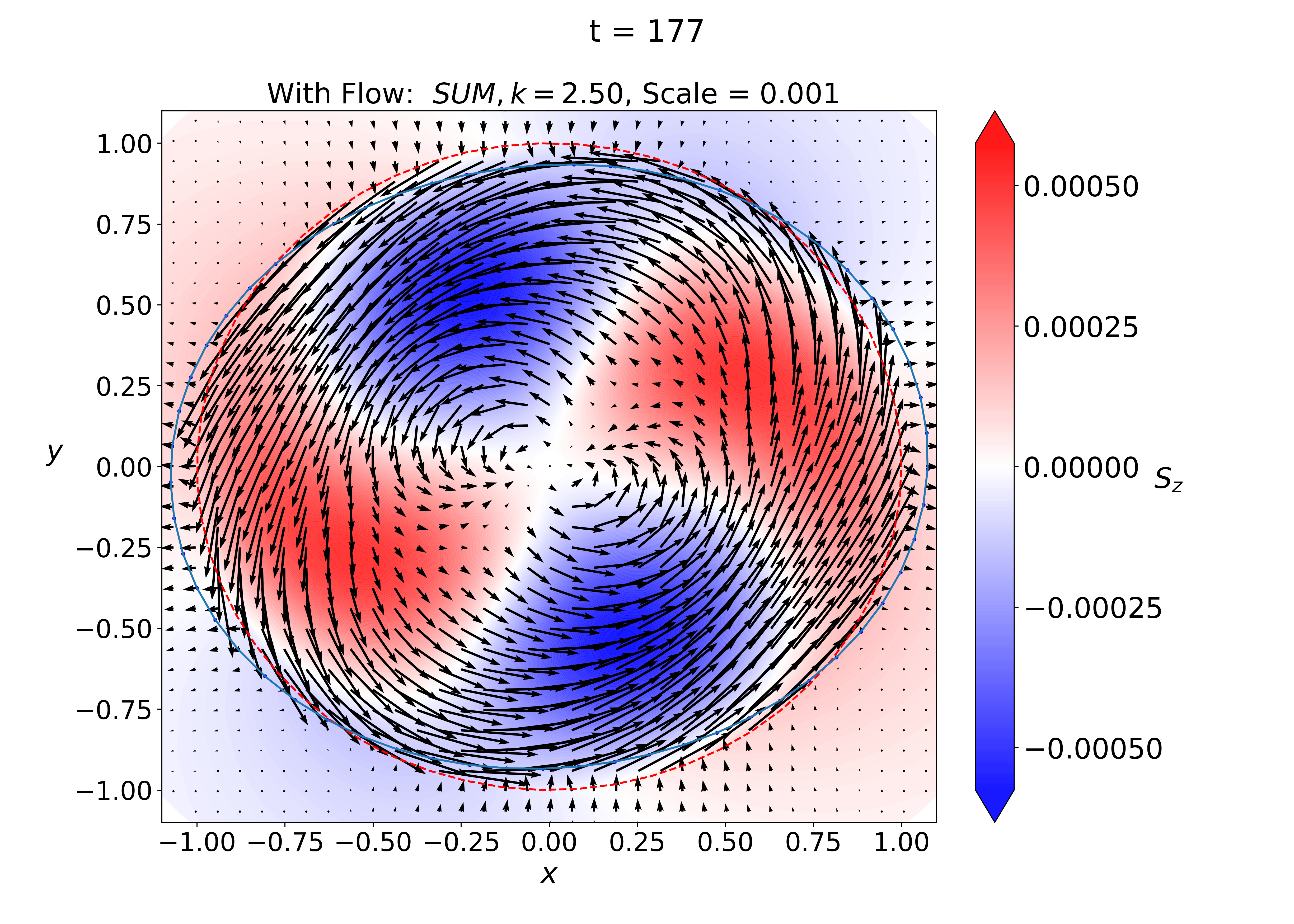}{0.33\textwidth}{(a)}
          \fig{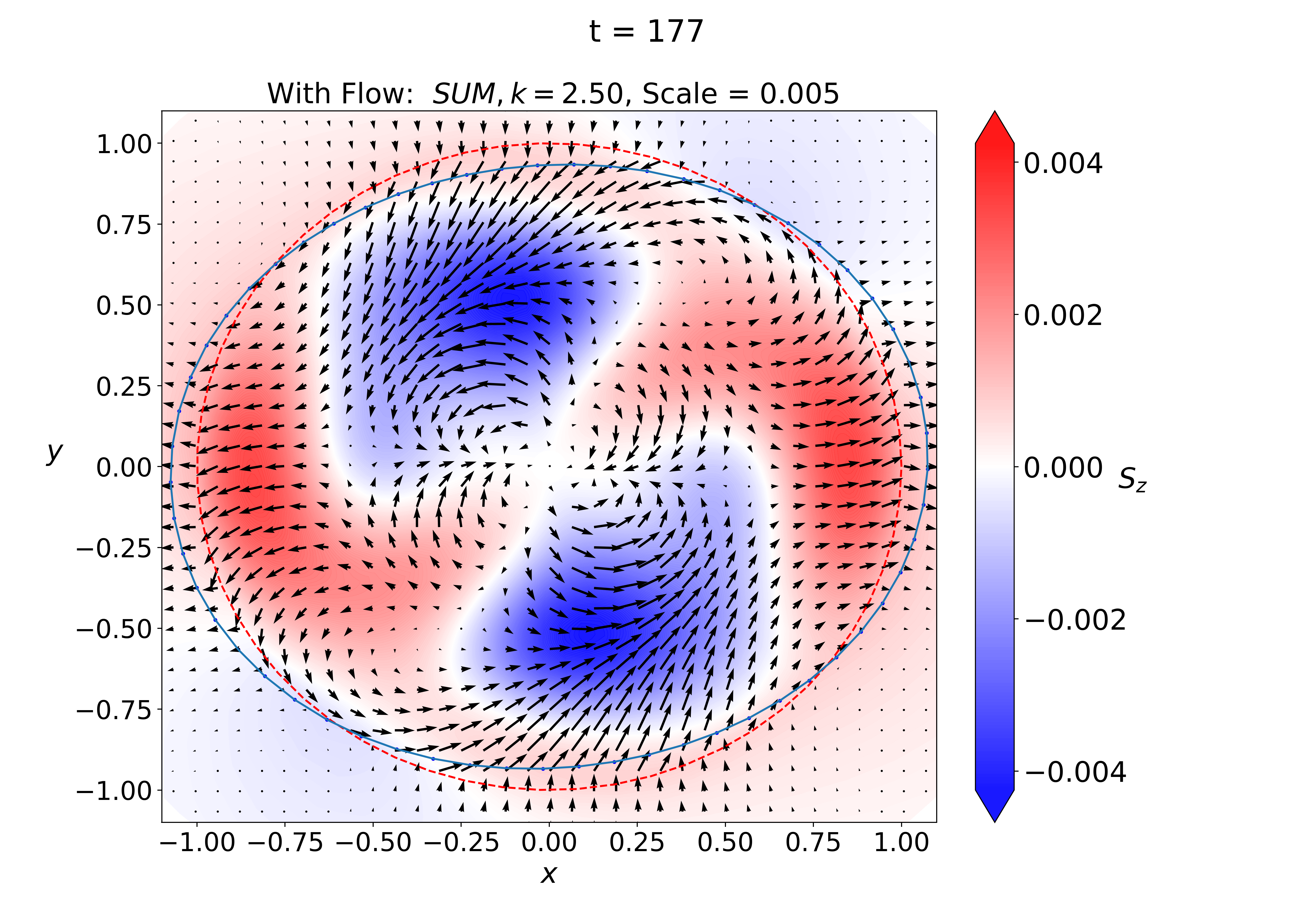}{0.33\textwidth}{(b)}          
          \fig{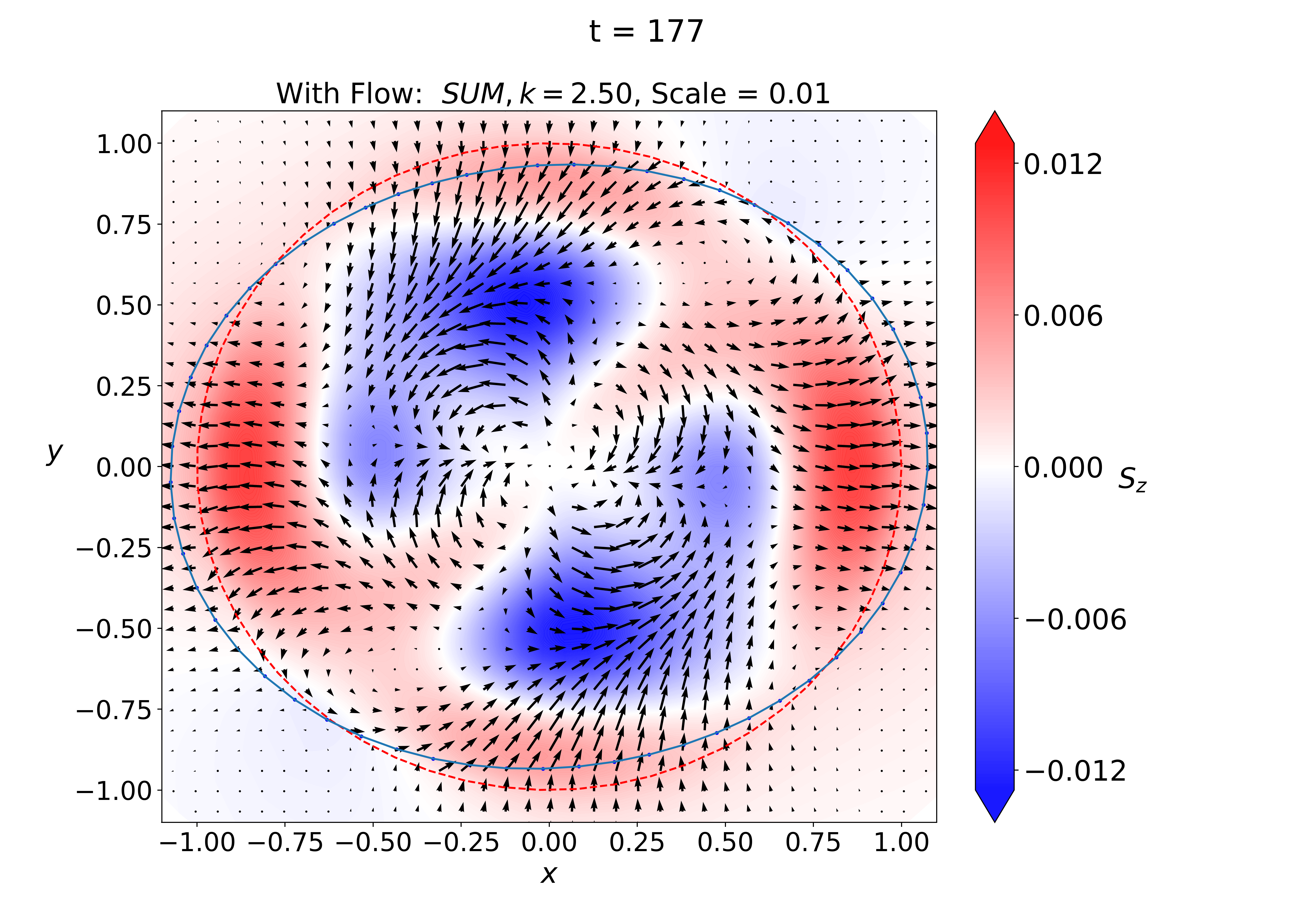}{0.33\textwidth}{(c)}
          }          
\gridline{\fig{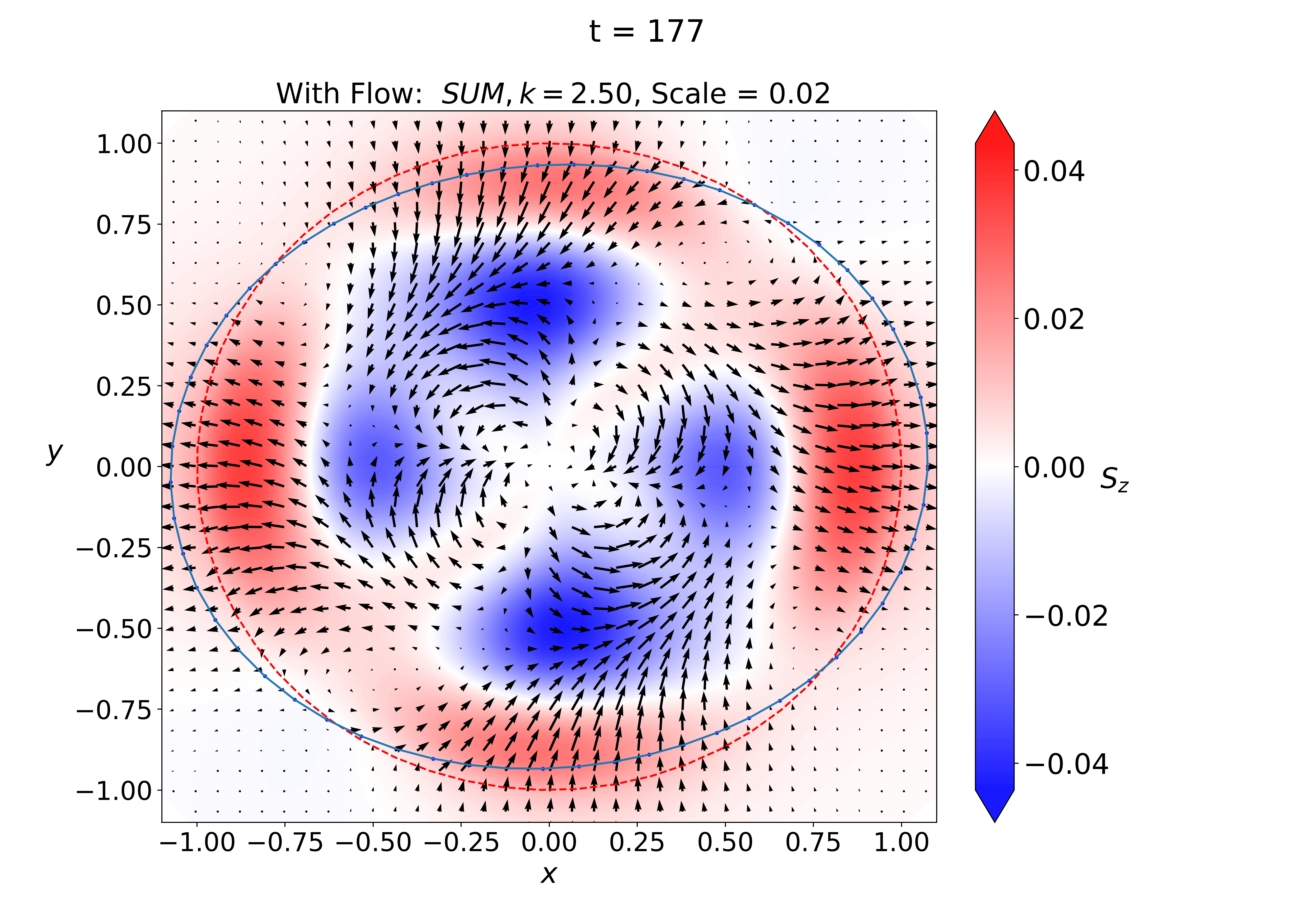}{0.33\textwidth}{(d)}
        \fig{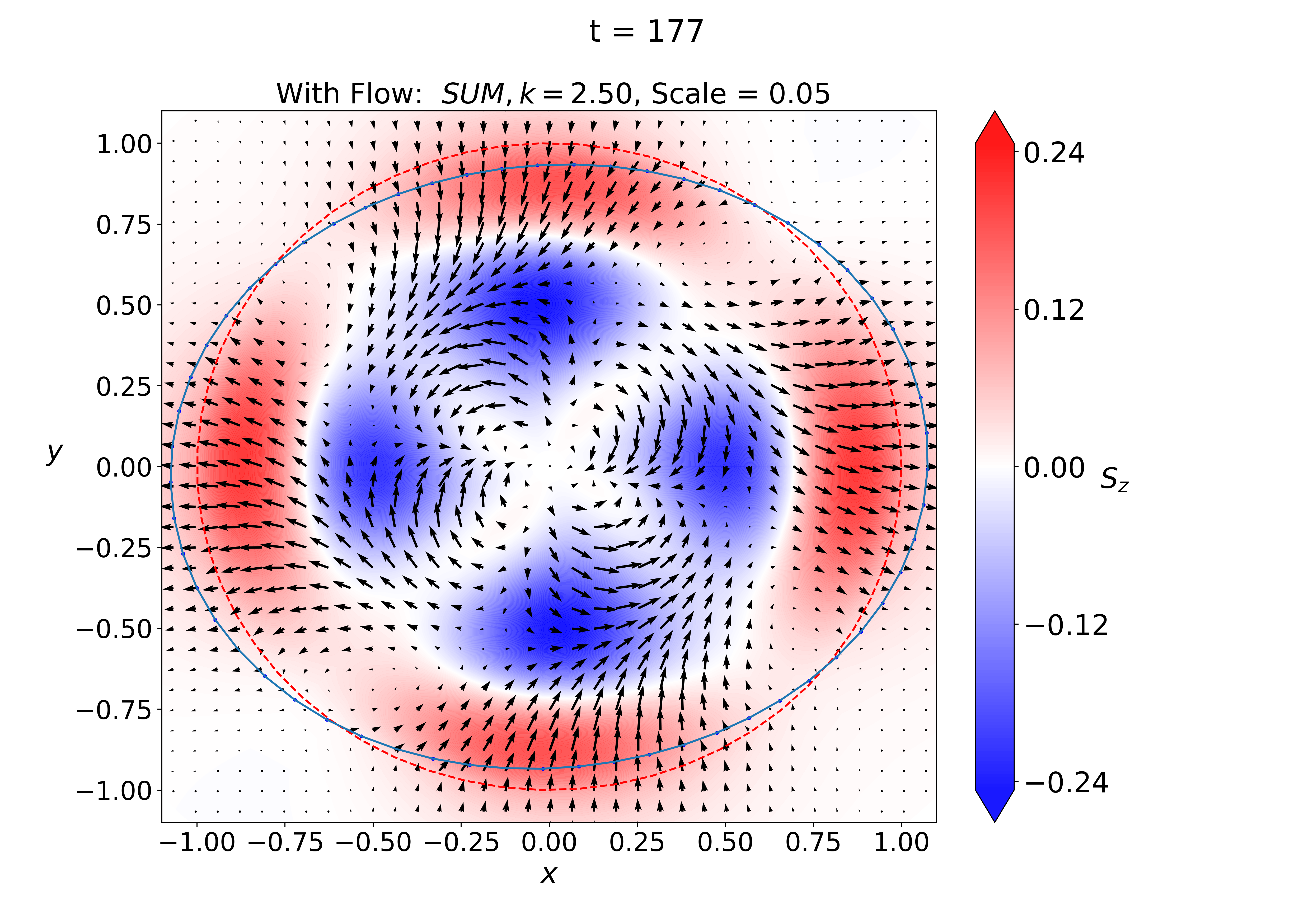}{0.33\textwidth}{(e)}
        \fig{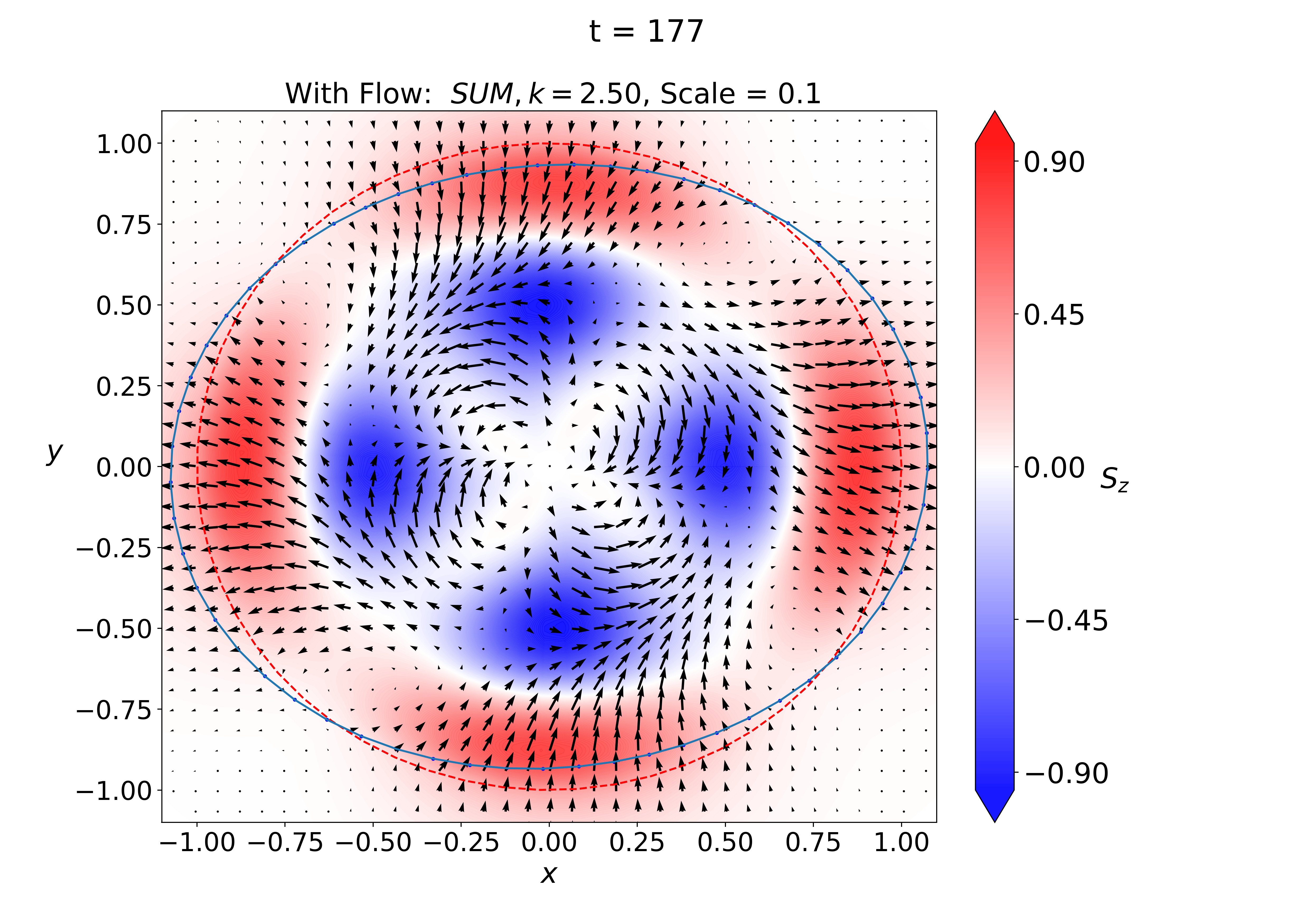}{0.33\textwidth}{(f)}
          }
\caption{Same as Figure \ref{fig:m1_Sz_spatial} but for the linear superposition of the modes $m=2$ and $m=-2$. Here we assume that both modes have equal strength as one another.
\label{fig:m2sum_Sz_spatial}}
\end{figure*}

The spatial pattern of $S_z$ for the combined superposition of the $m=2$ and $m=-2$ modes is shown in Figure \ref{fig:m2sum_Sz_spatial}. The effect of the background rotational flow is much more evident for higher order modes, and a strong swirling nature of the $S_z$ signal can be seen for the case when the flow is greater than $5$ times stronger than the perturbation, seen in Figure \ref{fig:m2sum_Sz_spatial} panels (a)-(d). The asymmetry in the azimuthal direction, caused by the presence of the rotational flow, results in the positive and negative modes propagating with different phase speeds, no longer forming a standing mode pattern in the azimuthal direction. With increasing azimuthal wavenumber, the sub-structuring becomes more complex and, as a result, the presence of the background rotational flow distorts the observed signal heavily. Only when the amplitude of the flow is weak compared to the strength of the perturbation, do we see a spatial pattern which is reminiscent of the fluting mode of azimuthal wavenumber $m=2$, in Figure \ref{fig:m2sum_Sz_spatial} panels (e)-(f).

\begin{figure*}
\gridline{\fig{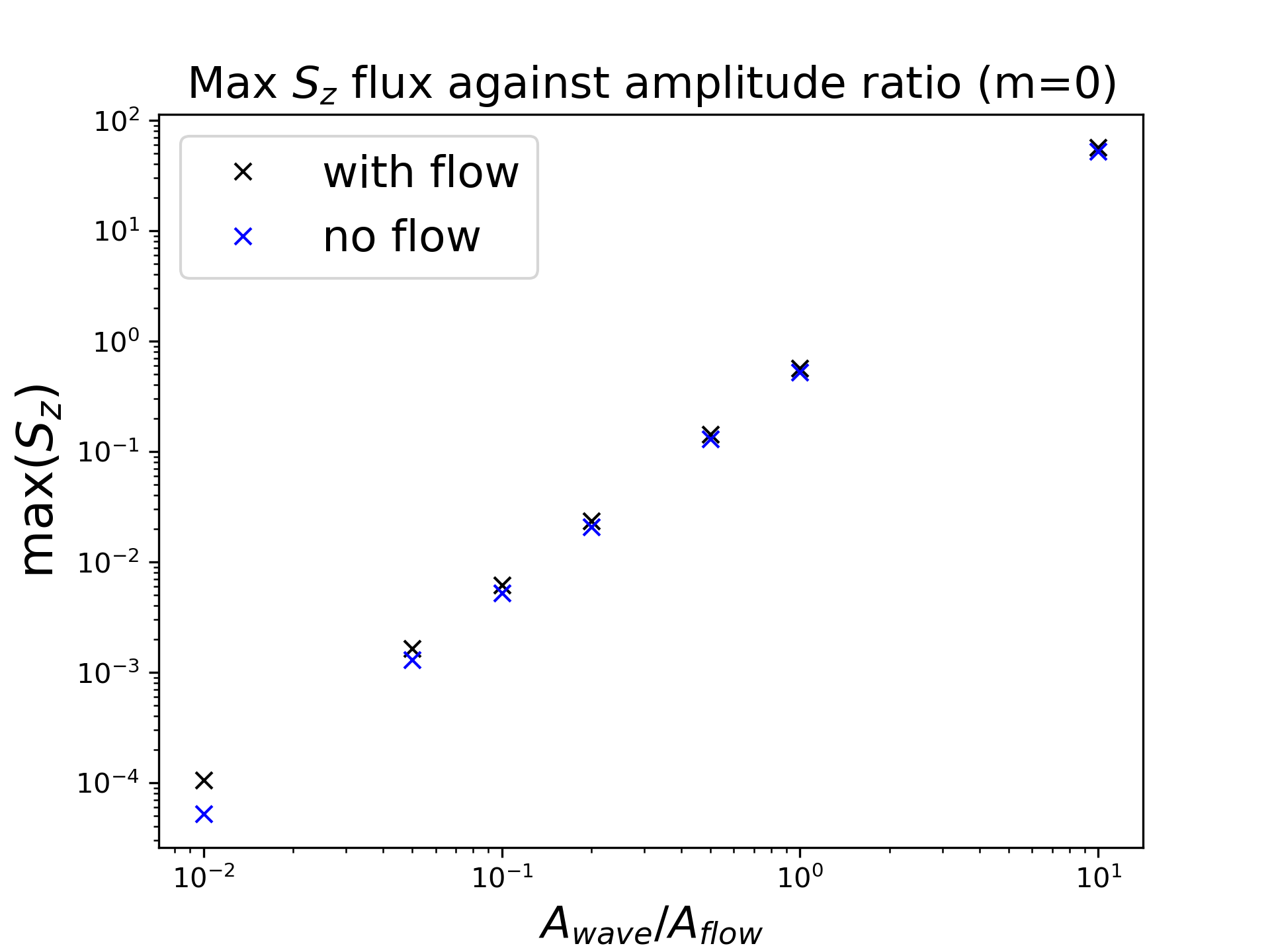}{0.33\textwidth}{(a)}
          \fig{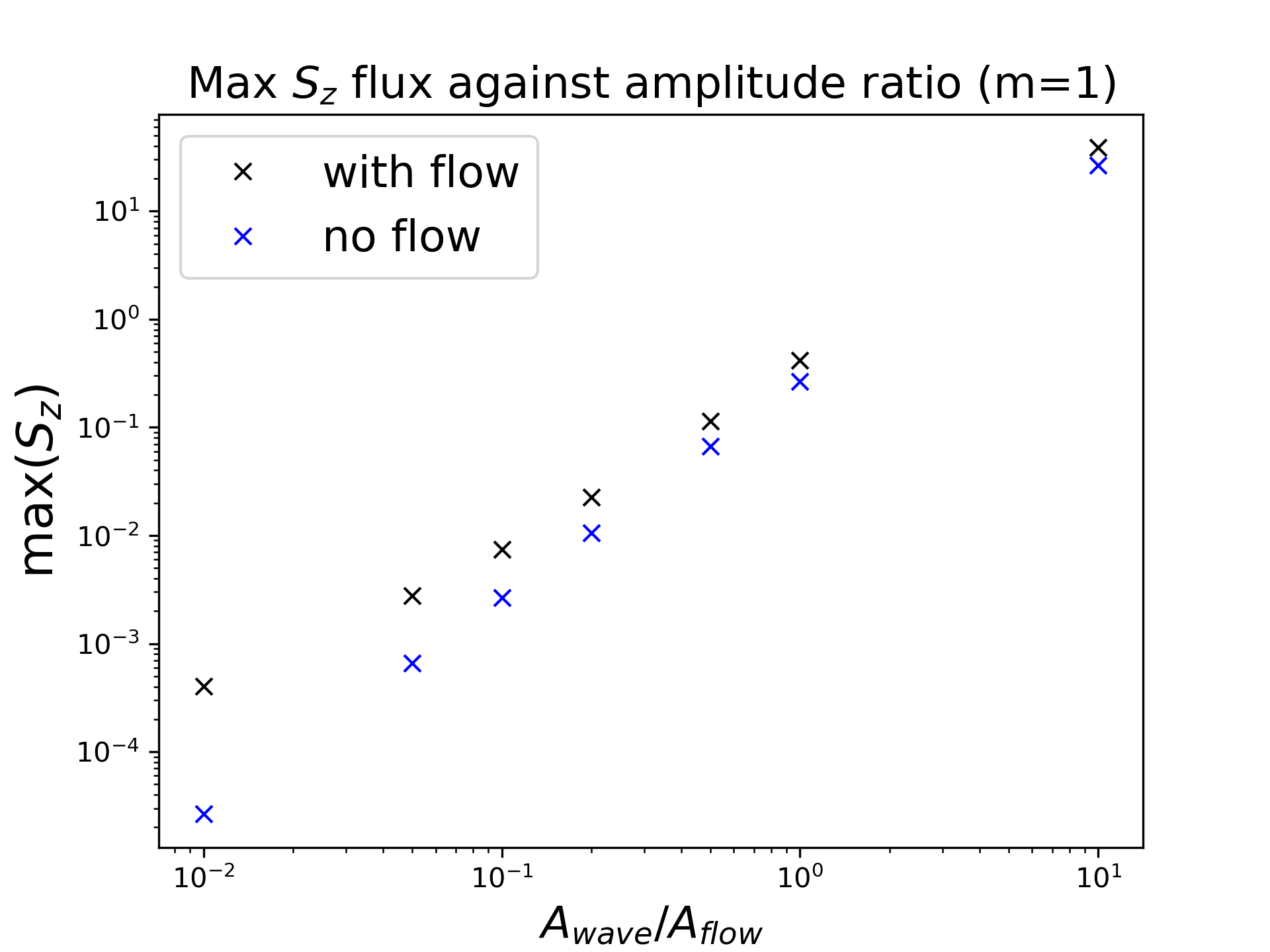}{0.33\textwidth}{(b)}          
          \fig{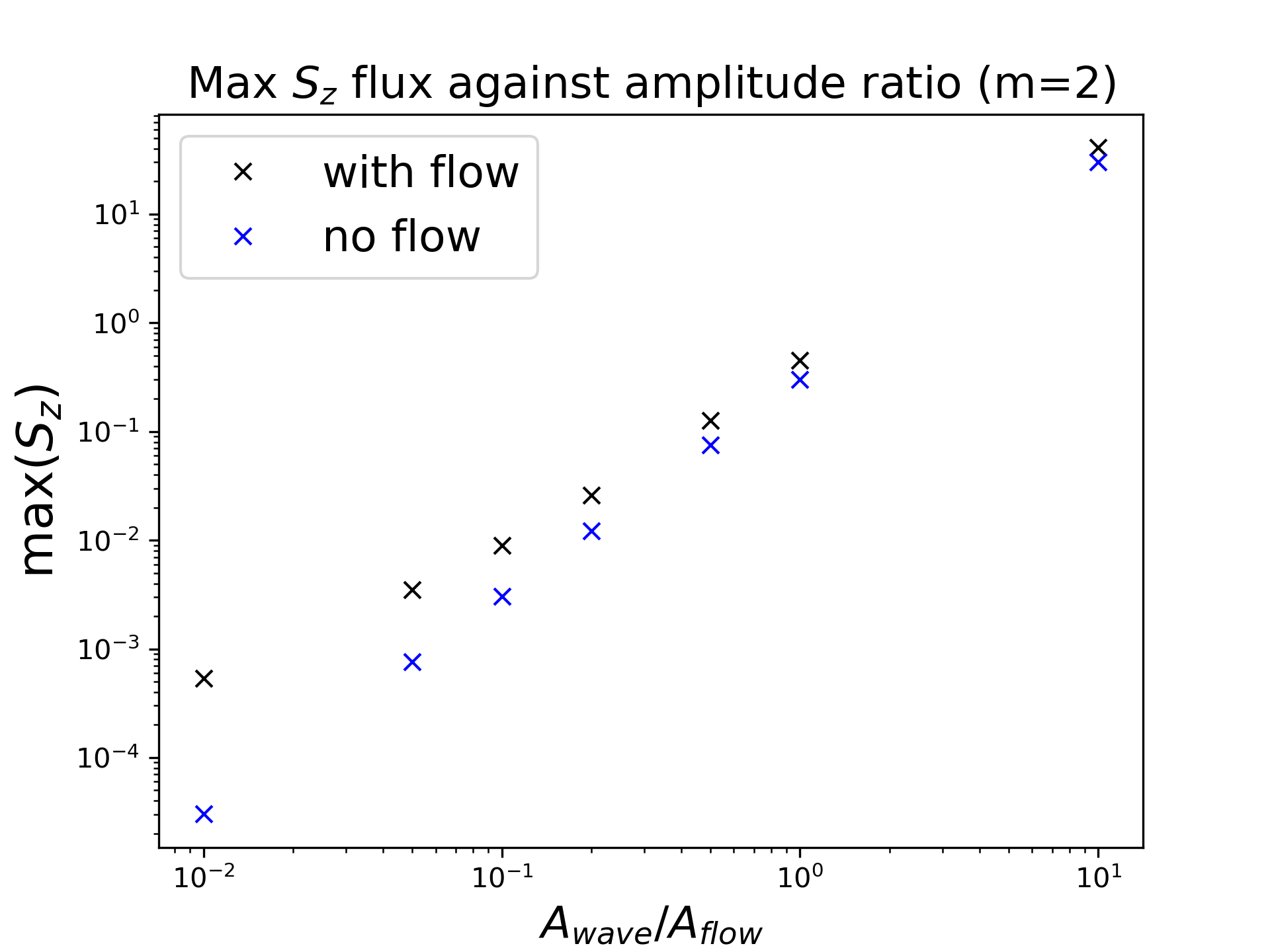}{0.33\textwidth}{(c)}
          }
\caption{Maximum value of $S_z$ computed using Equation (\ref{S_z_equation_withflow}) for (a) the sausage mode $m=0$, (b) the kink mode $m=1$ and (c) fluting mode $m=2$ both with (black crosses) and without (blue crosses) the presence of a background rotational flow.
\label{fig:Sz_flow_noflow_comparison}}
\end{figure*}

Figure \ref{fig:Sz_flow_noflow_comparison} displays the computed maximum values of $S_z$ for different MHD modes with and without the inclusion of a background rotational plasma flow in the model for the same pair of eigenvalues for each mode. Figure \ref{fig:Sz_flow_noflow_comparison}(a) demonstrates that there is no significant effect on the magnitude of $S_z$ for the sausage mode ($m=0$), except for the regime where the strength of the background flow is much greater than the strength of the perturbation where the Poynting flux associated with the wave increases by one order of magnitude. However, it can be seen in Figure \ref{fig:Sz_flow_noflow_comparison}(b) and Figure \ref{fig:Sz_flow_noflow_comparison}(c) that the presence of a background rotational flow appears to increase the magnitude of the magnetic energy transported by non-axisymmetric MHD modes for varying amplitude ratios. For all modes, it is clear that the Poynting flux is increased by larger amounts in the regime where the strength of the flow is much greater than the strength of the perturbation.
\begin{figure}
    \centering
    \includegraphics[width=0.43\textwidth]{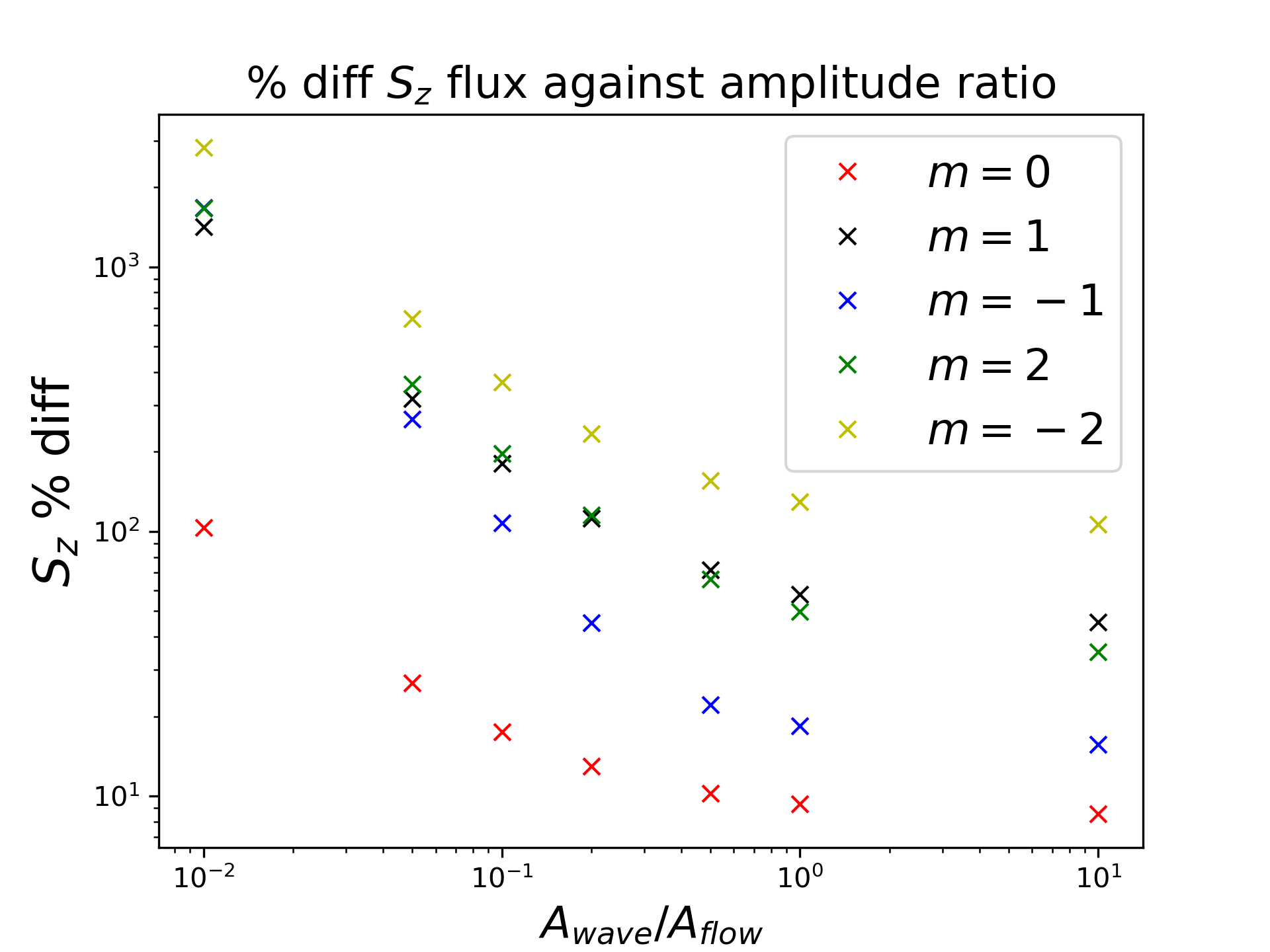}
    \caption{The percentage difference of the maximum value of $S_z$ for different modes in the presence of a background rotational flow when compared to the static case. Red, black, blue, green and yellow crosses correspond to modes $m=0, 1, -1, 2, -2$, respectively. \label{fig:Sz_perc_diff}}
\end{figure}
This is more evident in Figure \ref{fig:Sz_perc_diff} which shows the percentage difference between the maximum $S_z$ value for each mode when the background rotational flow is included, opposed to the static case of the Poynting flux associated with each mode without the background flow. Note that because $S_z$ associated with MHD modes is increased when background rotational flow is included, that the percentage difference displayed in Figure \ref{fig:Sz_perc_diff} represents a percentage increase from the case of a magnetic flux tube with no rotational flow. When the strength of the background flow is much greater than the perturbation, the $S_z$ associated with the MHD modes can be increased by more than $1000$\% for the non-axisymmetric modes, albeit that this is a large increase of a small number. However, the cumulative effect of the MHD waves in a sufficient number of solar vortex tubes present on the Sun at any given moment, originating from the intergranular lanes, may transport a significant amount of $S_z$ to the upper solar atmosphere.

\section{Discussion \& Conclusions}\label{sec:conclusions}
In this paper, we have explored how the vertical component of the Poynting flux, $S_z$, associated with MHD modes, is affected when in the presence of a background rotational flow. This is instructive when analysing both numerical and observational data of the dynamics of MHD waves in, for example, solar tornadoes. We have derived an analytical expression, given in Equation (\ref{S_z_equation_withflow}), for $S_z$ transported by MHD modes for any azimuthal wavenumber $m$, which depends solely on the plasma and wave quantities. This expression reduces to previously obtained analytical formulas for the Poynting flux associated with MHD modes in magnetic flux tubes without background rotational flows. Furthermore, we have presented a fundamental result that, for any untwisted magnetic flux tube, there is zero vertical Poynting flux associated with the background rotating equilibrium. In order for vortex tubes to self generate vertical Poynting flux in the solar atmosphere, they must posses magnetic twist in addition to background flows. Therefore, for magnetic flux tubes with weak pitch angles, whereby the axial component of the magnetic field dominates over the azimuthal component, any vertical Poynting flux can be associated with perturbations only, for example the propagation of MHD waves. This result has important consequences for interpreting the wealth of observational and numerical data of rotating structures in the solar atmosphere, suggesting that solar vortex tubes, solar tornadoes and magnetic swirls, amongst others see e.g. \citet[][]{Tziotziou2023SSRv}, can act as conduits for MHD waves in the solar atmosphere.

Exploiting the derived expression for $S_z$, we produced 2D visualisations of $S_z$ for different MHD modes under varying regimes of amplitude ratios. This was achieved through varying the amplitude of the obtained wave eigenfunctions with respect to the strength of the background plasma flow. We found that the presence of a background flow had little effect on the spatial distribution of $S_z$ for the axisymmetric sausage mode $m=0$. On the other hand, the presence of a background rotational flow has an effect on the non-axisymmetric modes ($|m| > 0$), whereby increased sub-structuring in the $S_z$ signal is apparent for higher order modes, and a notable swirling pattern appears for the superposition of the modes. It may be possible to calculate $S_z$ from observations of the lower solar atmosphere as both the velocity and magnetic fields are readily measurable, and it has been shown that MHD is a good approximation for $S_z$ in the solar photosphere \citep{Tilipman2023}. Techniques such as Local Correlation Tracking \citep[LCT;]{November1988}, Fourier Local Correlation Tracking \citep[FLCT;]{Fisher2008} and machine learning techniques such as Deepvel \citep{AsensioRamos2017}, provide information regarding the velocity field, whereas the magnetic field may be measured using spectroscopic techniques and Stokes inversions. The difficulty will then be separating the measured $S_z$ into the background and perturbed components in order to determine the contribution from MHD waves to the measured $S_z$ signal. However, the contribution from these different MHD modes to the total $S_z$ signal in rotating solar magnetic flux tubes may be retrievable using wave analysis techniques such as Proper Orthogonal Decomposition and Dynamic Mode Decomposition \citep{Albidah2021, Albidah2023, Jafarzadeh2024}.

Finally, we computed the maximum value of $S_z$ determined using Equation (\ref{S_z_equation_withflow}) for each MHD mode in the presence of a background rotational flow and compared this against the case of a static magnetic flux tube. We find that the presence of a background rotational flow increases the maximum value of $S_z$ for all modes, and that the increase is greater for higher azimuthal wavenumbers and for stronger background flows. To convert the results presented in this work into physical units, we can take numerical simulations of magnetic tornadoes as an example. \citet{Kuniyoshi2023} found that the Poynting flux at the transition region in a 3D radiative MHD simulation was 420\% greater in the presence of a magnetic tornado when compared to a region where the magnetic tornado was absent, and that this increase is roughly $3\times 10^5$ erg cm$^2$ s$^{-1}$. That increase would correspond to an amplitude ratio $v_{0,\varphi} = 10-20\hat{f}$, however, it is unclear the contribution from MHD waves to this increase. Going forward, it will be crucial to separate the Poynting fluxes associated with background flows (the rotating structures themselves) and any wave perturbations, in order to get a better understanding of the energy transport of MHD waves in solar vortices and their physical contribution to the energy budget of the solar atmosphere. The results of this study may be useful for interpreting the ratio between the amplitudes of background flows and MHD wave perturbations in numerical and observational data. In this work we have focused solely on the magnetic energy associated with MHD waves in rotating flux tubes, however, when modelling a plasma with non-zero plasma pressure, it is possible that the thermal energy associated with these waves is significant. An investigation into the thermal energy associated with MHD waves in solar vortex tubes should be the focus of future work.

\acknowledgements
SJS, VF, SSA and GV are grateful to the Science and Technology Facilities Council (STFC) grants ST/V000977/1, ST/Y001532/1.  VF, SSA and GV thank The Royal Society, International Exchanges Scheme, collaboration with  Instituto de Astrofisica
de Canarias, Spain (IES/R2/212183), Institute for Astronomy,
Astrophysics, Space Applications and Remote Sensing, National
Observatory of Athens, Greece (IES/R1/221095), and Indian
Institute of Astrophysics, India (IES/R1/211123) for the support provided. VF and SSA would like to thank the International Space Science Institute (ISSI) in Bern, Switzerland, for the hospitality provided to the members of the teams on `The Nature and Physics of Vortex Flows in Solar Plasmas' and 'Tracking Plasma Flows in the Sun’s Photosphere and Chromosphere: A Review \& Community Guide'. VF and GV are grateful to the Institute for Space-Earth Environmental Research (ISEE, International Joint Research Program, Nagoya University, Japan) for the support provided.

\bibliography{ref}{}
\bibliographystyle{aasjournal}

\end{document}